\documentclass[twocolumn]{aa} 

\usepackage{graphicx}
\usepackage{dcolumn}
\usepackage{txfonts}
\usepackage{natbib}
\bibpunct{(}{)}{;}{a}{}{,} 

\def\arcsec{\ifmmode^{\prime\prime}\;\else$^{\prime\prime}\;$\fi}
\def\arcmin{\ifmmode^{\prime}}
\def\deg{\hbox{$^\circ$}}

\begin{document}

\title{A study of a sample of high rotation measure AGNs through multifrequency single dish observations}
\titlerunning{A single dish study of high RM AGNs}
\authorrunning{Pasetto et al.}

   \subtitle{}
   \author{
          }

   \author{ Alice Pasetto \inst{1}\fnmsep\thanks{
  \email{apasetto@mpifr-bonn.mpg.de}\newline},
   		Alex Kraus \inst{1}, 
		Karl-Heinz Mack \inst{2},
          	Gabriele Bruni \inst{1,}\inst{2},
		Carlos Carrasco-Gonz\'alez \inst{3}
                 }

   \institute{Max-Planck Institut f\"ur Radioastronomie (MPIfR), 
              Auf dem H\"ugel 69, D-53121 Bonn, Germany
              \and
             Istituto di Radioastronomia-INAF (IRA-INAF), Via Gobetti 101, I-40129 Bologna, Italy
             \and
             Instituto de Radioastronom\'ia y Astrof\'isica (IRAf-UNAM), Antigua Carretera a P\'atzcuaro 8701, 58089 Morelia, Michoac\'an, M\'exico
             }

   \date{Received   ; accepted }


  \abstract
   {We characterised and studied, in the radio band, a sample of candidates of high Rotation Measure (RM). These point-like objects show a strong depolarisation at 21cm. This feature suggests the presence of a very dense medium surrounding them in a combination of a strong magnetic field.
}
   {This work aims at selecting and studying a sample of radio sources with high RM, thus to study their physical conditions and their status with respect to their surrounding medium. We want to understand if any connection is present between the AGN hosting galaxy medium with some evolutionary track and/or some restarting phase of the AGN itself.}
   {Multifrequency single-dish observations were performed with the 100-m Effelsberg telescope to define the initial sample, to characterise the spectral energy distribution (SED) of the final sample (30 targets) and to determine their RM in the 11 to 2 cm wavelength range. }
   {From the observations, the SED together with polarisation information, i.e. the fractional polarisation and the polarisation angle, have been determined. Three different object types were revealed from the SEDs analysis: \textit{Older}, \textit{GPS-like} and \textit{Mixed}. For each of the targets, the rotation measure has been found and the depolarisation has been modelled. No significant correlation have been found between the depolarisation behaviours and the SEDs, while a correlation has been found between sources with mixed SED (with an old component at low frequency and compact components at high frequencies) and high values of the rotation measure (with values in the rest frame larger than 1000 rad/m$^{2}$).   }
   {This work helps us to define and identify a sample of sources with high RM. From the analysis we can conclude that the sources showing a restarting phase at high frequency (with a \textit{Mixed} SED), are characterised by a really dense and/or a magnetised medium that strongly rotates the polarisation angle at the different frequencies, leading to a high RM. 
}


   \maketitle
%

\section{Introduction}

Active Galactic Nuclei (AGNs) are the most powerful objects in the Universe. Hosting a Supermassive Black Hole (SMBH), AGNs have a spectrum that emits at all wavelengths, from radio to gamma ray. Thanks to this, they are well studied but still our understanding about these objects is not completely clear.
The unification scheme for radio loud AGN, where their appearance strongly depend on their orientation \citep{OB82, UP95}, is by now accepted by most of the scientific community, but still several are the open questions. How the medium of the hosting galaxy in the vicinity of the SMBH is characterised and how strong their magnetic field is, are important elements to understand the jets ejection mechanism.
It is also important to understand whether AGNs are characterised, in the radio band, by some periodic activity phase \citep{Ma06, Saikia09, Cz09}. Besides, thanks to studies of samples of compact AGNs, \cite{ODeaBaum97} identified the existence of an anti-correlation between the linear size and the peak frequency in their spectra as the result of the synchrotron self absorption (SSA) mechanism: the higher the peak frequency, the smaller the source. 
A possible reason could be found in their early evolutionary stage, that leads to very compact radio sources the High Frequency Peakers (HFP) \citep{Dallacasa00}, with peak frequency around 8-10 GHz, to the Compact Steep-spectrum Sources (CSS) \citep{Saikia88, Fanti90}, with a peak frequency around hundred of MHz, passing through a state of Giga-Hertz Peaked-spectrum sources (GPS) \citep{Gopal-Krishna83, Stanghellini90, ODea91}, where the peak frequency appears around few GHz. However, whether AGNs are really experiencing the above evolution from young quasars to large-scale radio sources \citep{ODea98}, is still in doubt.

The study and the analysis of the polarisation information in the radio band, is a powerful tool that can help to clarify these questions.
The Stokes parameters (I,Q,U and V) are the observable quantities needed to obtain information on the linear polarization.
The polarization flux density (S$_{Pol}$) is defined as:

\begin{equation}
	S_{Pol}=\sqrt{Q^2+U^2}
\end{equation}
\noindent
The fractional polarization is thus the ratio between the polarized flux density (S$_{pol}$) with the total flux density, expressed with the Stokes I: 

\begin{equation}
	m = \frac{S_{pol}}{I}
\end{equation}
\noindent
The observed electric vector polarisation angle (EVPA) can be also expressed using the observable Stokes parameters Q and U as:

\begin{equation}
	\chi = \frac{1}{2} \cdot \arctan \frac{U}{Q}\qquad [\rm{rad}] .
\end{equation}
\noindent
From this quantities it is possible to study two important aspects of this project: the Faraday rotation and the Faraday depolaization.
They can probe the interstellar medium (ISM) and the strength of the magnetic field of the host galaxy. In order to perform this kind of studies, observations in a wide wavelengths range are necessary.

The Faraday rotation is the rotation of an electromagnetic wave that occurs when it passes through a magnetised plasma. Different unresolved regions of polarized emission will likely experience different amount of Faraday rotation. Therefore, following the explanation proposed by \cite{Burn66}, we can express the Faraday rotation using the Faraday depth:

\begin{equation}
	\phi = \frac{e^3}{2\pi m_e^2 c^4} \int_d^0 n_{e} B_{\parallel}dl \qquad[\rm{rad/m^2}],
\end{equation}
\noindent
where $n_{e}$ is the electron density of the medium (expressed in cm$^{-3}$), B$_{\parallel}$ is the component of the magnetic field along the line of sight (expressed in $\mu$G) and \textit{l} is the geometrical depth of the medium along the line of sight (expressed in parsecs).
The integral is performed along the line of sight from the source (at distance d) to the observer.
In the simplest case of an homogeneous medium, $\phi$ is equal to the Rotation Measure (RM), defined as the gradient of the polarization angle $\chi$ with the observed $\lambda^{2}$,:

\begin{equation}
	\chi(\lambda) = RM \lambda^{2} + \chi(0) \qquad [\rm{rad}]. 
\end{equation}
\noindent
\noindent
If the medium is inhomogeneous or unresolved, the RM can change within the source and a deviation, more or less strong, from the $\lambda^{2}$ law occurs \citep{Burn66, Vallee80, SaSa88}. This can be an indication of the presence of multiple RM components thus a hint of the complexity of the source.

Together with the RM, it is important to study the fractional polarisation (\textit{m}) behaviour as a function of the observed wavelength.
When the Faraday rotation causes the reduction of the fractional polarisation, the source is subjected to depolarisation. This can be internal or external; the first occurs when the Faraday rotating component is intermixed to the radio emitting region and the latter when several Faraday screens are somewhere between the radio source and the observer. Several depolarisation and repolarisation (where the fractional polarization increases at longer wavelengths) models have been developed \citep{Burn66, Tribble91, Homan02, Rossetti09, Mantovani09, Hovatta12}.

The analysis of these two features, the RM and the depolarisation, can give information about the density distribution of the ISM that surrounds the radio source, its clumpiness and the strength of the magnetic field.
The importance of connecting the polarisation properties, like a very high RM value and strong depolarisation, with the ambient medium, is evident since years \citep{Burn66, Laing84, Tribble91, Rossetti08}. Some observational works have found sources with very high RM with single dish and interferometric technique \citep{Kato87, Benn05, Trippe12, Kravchenko15} and also with studies with higher resolution VLBI technique \citep{Zavala04, Attridge05, Jorstad07}.
However, a deep study of a relationship between the RM and the ambient medium is very difficult since it would require the study of a large sample in a large range of frequencies and with simultaneous observations. 

In this work we present an homogeneous observational study performed with the 100-m Effelsberg telescope of a medium-sized sample of radio sources. We made a selection of sources from the full northern sky based on strong depolarisation at a low frequency which defined a sample of high RM candidates. These sample was then observed at several frequencies in the 2-15 GHz range with the single dish telescope. We want to study whether any connection is present between the AGN hosting galaxy medium with some evolutionary track and/or some periodic activity phase of the AGN itself. Further observations on our sample, using  the JVLA, EVN and VLBA interferometers, will be reported in future papers.

\section{Sample selection criteria}

Our sample was created by selecting sources from the NRAO VLA Sky Survey (NVSS) \citep{Condon98}, a survey at 1.4 GHz, which contains a total of 1773484 entries. From these, we considered only sources matching the following criteria:  
\begin{itemize}

\item Flux density S$_{NVSS}$ ${\geq}$ 300 mJy;
\vspace{0.1cm}
\item Unresolved with major axis $\theta^{maj}_{NVSS}$ ${\leq}$ 45${\arcsec}$;
\item Declination $\delta$ ${\geq}$ --10$\deg$ ;
\vspace{0.1cm}
\item Polarisation flux density S$^{pol}_{NVSS}$ ${\le}$ 0.87 mJy, i.e. unpolarized sources;

\end{itemize}
The minimum flux density of 300 mJy at 1.4 GHz was chosen in order to be able to detect enough total flux density at higher frequencies (roughly S=70 mJy at 10.45 GHz) to perform polarisation studies, assuming that the majority of the AGNs could be characterised by a steep radio spectrum with $\alpha$= --0.7.
The last point is essential as we are interested in studying sources suffering for a strong depolarisation at 1.4 GHz, hint of a possible high RM.
The value 0.87 mJy represents the 3$\sigma$$^{pol}_{1.4}$ (where the rms fluctuation level $\sigma$$^{pol}_{1.4}$ = 0.29 mJy/beam, for the NVSS survey) \citep{Condon98}.
The result of this selection is a list of 2890 point-like sources with no detected polarisation flux density (thus $\sim$20\% of the brightest sources in the NVSS do not have polarisation detection).

As a second step, we cross correlated the obtained sample with the Faint Images of the Radio Sky at Twenty-cm (FIRST) catalog \citep{White97}. Having a better angular resolution, only unresolved sources, i.e. with a angular size $\leq$5$\arcsec$, were selected. The objective of this was to increase the probability to select possibly compact and/or high redshift candidates.
The result of the cross correlation is a list of 537 bright, point-like and unpolarised sources. 

Given the strong dependence of the Faraday depolarisation effect with the observing frequency (the lower the frequency, the stronger the depolarisation becomes), we observed the entire cross correlated list with the 100-m Effelsberg telescope at 10.45 GHz (see section 4.1), in order to search for polarised flux density, thus suggesting a strong depolarisation at 1.4 GHz. The final sample of high RM candidates is composed of 30 sources ($\sim$ 6\% of the initial cross correlated sample). We checked that the targets are isolated sources, extracting images from the FIRST catalogue using an image size of 5$\arcmin$, corresponding to the size of the Effelsberg beam at 11cm (2.64 GHz). The final sample have been finally observed at several frequencies (see section 4.2) in order to determine their SEDs and their RM (see section 6.2 and 6.4 respectively).

\section{Possible selection biases}
The selection criteria we followed could result in two possible biases present in our sample. The first is the lack of sensitivity of our observations for sources weaker than 150 mJy at 10.45 GHz. This could exclude a fraction of steep spectrum sources, for which the polarised flux density drops below the minimum 3 mJy limit detectable in our observations. In fact, starting from the original list (537 unpolarised sources at 1.4 GHz), 30 sources show significant polarisation at 10.45 GHz, the remaining 507 not. Setting a 3 mJy upper limit (3$\sigma$, assuming the Effelsberg rms=1 mJy/beam) to their polarised flux density and assuming a typical detectable fractional polarisation being larger than 2\% \citep{Condon98}, we can only assert that the unpolarised sources are 77 targets with no polarisation detection at 10.45 GHz and S$_{10.45}$ $\gtrsim$ 150 mJy.
Thus for the remaining 430 targets we cannot discern if they are polarised or not and, as a consequence, we can not include them into our study.

In contrast, our flux density selection criterium (S$_{1.4GHz}$$\ge$ 300 mJy) should exclude from our analysis potential sources with synchrotron spectra peaking at higher frequencies, thus potential GPS-HFP targets. These sources, peaking in the range between 2 and 10 GHz, are optically thick at 1.4 GHz. Then, at this frequency their emission increases with a spectral index of 2.5. Therefore, while they can be very bright at high frequencies, we are excluding them from our analysis due to their weakness at 1.4 GHz. 

\section{Observations}
In the following, we describe the observational campaign carried out for this work with the 100-m Effelsberg single dish telescope.
 
\subsection{10.45 GHz observations}
In order to identify high RM candidates, we initially selected unpolarised sources at 1.4 GHz in the NVSS catalogue, assuming that the non-polarisation in some cases is caused by in-band depolarisation due to a very high RM. Therefore the strategy we adopted was to observe at higher frequency searching for polarised flux density. Using the 100-m Effelsberg telescope, observations at 10.45 GHz on the initial sample of 537 targets were performed during the winter semester 2012-13. 
The 10.45 GHz receiver has a bandwidth of 300 MHz and a system temperature (Tsys) of $\sim$50 K on the sky (zenith) and a FWHM of 69$\arcsec$. The system delivers Right Circular Polarisation (RCP) and Left Circular Polarisation (LCP) and is connected to an IF-polarimetry that provides full polarisation information, thus giving the Stokes parameters: I, Q, U and theoretically also V. Due to the expected weakness of Stokes V in extragalactic sources, that has not been studied.
All 537 sources are point-like compared to the Effelsberg beam.
Therefore they were observed in cross-scan mode, along azimuth and elevation with a scan length of 6$\arcmin$ and a scanning speed of $\sim$7$\arcsec$/sec, using a total of twelve subscans (6 scans for each direction).  
Since the sources are bright enough, focus and pointing were checked by the targets themselves.
3C286, 3C48 and 3C161 have been observed as flux density calibrators, the flux densities of which were based on the scale of \cite{Baars77} and counter-checked with more recent data coming from the 100-m Effelsberg calibrators monitoring program \citep{Kraus03}.
The polarisation information (the fractional polarisation \textit{m} and the polarisation angle $\chi$) of the mentioned calibrators, were checked using the recent and continuously updated values available in the Effelsberg wiki-page\footnote{https://eff100mwiki.mpifr-bonn.mpg.de/doku.php "Calibrators and their polarisation" section in the Effelsberg User Guide}.
NGC7027 and 0951+69 have been chosen as unpolarised calibrators to determine the leakage terms, the instrumental polarisation, which is of the order of 1\%.
This first observational session at high frequency leaded to a list of 30 sources which became the high RM candidates (see Tab.1).
 
\subsection{Follow-up programme}
For the sources with detected polarisation at 10.45 GHz, a follow-up programme was performed in order to determine their RM value.
The 30 sources have been observed in cross-scan mode at 2.64 GHz (FWHM of 275$\arcsec$), 4.85 GHz (FWHM of 146$\arcsec$), 8.35 GHz (FWHM of 80.6$\arcsec$), 10.45 GHz (FWHM of 69$\arcsec$) and 14.60 GHz (FWHM of 50.8$\arcsec$) (see the Effelsberg wiki-page\footnote{https://eff100mwiki.mpifr-bonn.mpg.de/doku.php; the section "Receiver and calibration" in the Effelsberg User Guide} for more details on the receivers).
The scan length and the scan velocity were chosen following the technical parameters of the different receivers.
As previously done with the observations at 10.45 GHz, suitable sources, such as 3C286, 3C48 and 3C161, were observed as flux densities calibrators and, also, to check for the polarisation information ($\textit{m}$ and $\chi$).
Again NGC7027 and 0951+69 were observed for the determination of the instrumental polarisation.
The total intensity and the polarisation information for all the 30 targets were collected and they are presented in section 6.
The sample have been observed quasi-simultaneously with time between observations from few hours to few days (in few cases in order to repeat some bad observations).
Tab. 1 -- 6 in Appendix A, contain the values of total intensity and the polarisation information of the high RM candidates collected during this follow-up campaign.

\subsection{Time variability}
Flux density and polarization variability over the entire electromagnetic spectrum  is a common phenomenon in AGN \citep{Peterson01} and could therefore also influence our observations. However, strong variations are usually seen only at frequencies > 10 GHz and mainly in blazar sources (radio-loud AGN seen at small angles to the axis of the jet) with variations of several factors in the total flux density (e.g. by a factor of $\sim$4 for the sources BL Lac and 3C 273 at 15.0 GHz within 20 years) \citep{Lister09}. 

During the follow-up observations, the high RM candidates have been observed nearly simultaneously at the various frequencies; the time between the individual scans was in most cases only a few minutes (except for few cases where bad observations had to be repeated). From the obtained SEDs (presented in section 6.2) of our targets, we can assert that the high RM sample is mainly composed by quasars, and we can most likely exclude contamination from blazar type source. Besides, most of our observations are performed at frequencies $\leq$10 GHz, hence, strong variations are rather unlikely. Furthermore, our repeated observations at 10.45 GHz (for the follow-up campaign; section 4.2) didn't reveal any strong variations - at the most, small oscillations (of the order of few \%) have been seen on a time scale of few months.

As one observes at lower frequencies, time variability is less strong. A considerable time window for this project is given by the NVSS and low-frequencies surveys, for which, as said, should not give any variability issues. Therefore, since our sources seem to be not strongly variable at the hi??gher observed frequency (10.45 GHz), we can expect that the combination of the new Effelsberg data and the data taken from literature, is not strongly affected by a time variability significant for the purposes of this work.
Therefore, we can safely assume that neither our RM determination nor the SEDs are significantly influenced by source variability.

\subsection{Data reduction}
Data reduction has been done using the TOOLBOX package for single dish data (Max Planck Institute for Radio Astronomy; see the Effelsberg wiki-page\footnote{https://eff100mwiki.mpifr-bonn.mpg.de/doku.php section "Using the Toolbox to inspect cross-scans" in the Effelsberg User Guide}).
For coherency, the same data reduction procedure has been followed for all frequencies like the correction for the opacity, determined by a fit between the Tsys versus airmass distribution, and the pointing offsets, where the offsets in longitude are applied to the latitude data and vice versa. Moreover a baseline subtraction, an averaging of all the subscans and few other standard adjustments were applied. 
Flux calibration has been done using the [Jy/K] factor calculated from the flux calibrators opportunely observed during the different observational sessions.
The leakage terms have been adequately obtained through observations on unpolarised point like calibrators.
\noindent
The errors have been calculated through error propagation taking into account uncertainties from the Gaussian fit to the cross-scanned data, pointing correction and various calibration uncertainties, e.g. changes in the noise diode or changes in the focus.

\section{Comparison samples}
We compare our results with two comparison samples.
We used the 77 bona fide unpolarised sources at 10.45 GHz, listed in Tab. 1, to compare their spectral indices with those of our high RM candidates.

\begin{table*}

\centering
\tiny
\caption{Table of the 77 bona fide unpolarised sources at 10.45 GHz. The coordinates are taken from the FIRST survey and the flux density are taken from the NVSS survey. It follows the flux densities at 10.45 GHz and the upper limits on the polarization flux density at 10.45 GHz measured with Effelsberg and finally the spectral index between the two frequncies.}
\rotatebox{90}{
\centering
\begin{tabular}{|llrrrrr|}
\hline
  \multicolumn{1}{|c}{Name} &
  \multicolumn{1}{c}{RA} &
  \multicolumn{1}{c}{DEC} &
  \multicolumn{1}{c}{S$_{1.4}$} &
  \multicolumn{1}{c}{S$_{10.45}$} &
  \multicolumn{1}{c}{P$_{10.45}$} &
  \multicolumn{1}{c|}{$\alpha$$^{10.45}_{1.4}$} \\
  \multicolumn{1}{|c}{} &
  \multicolumn{1}{c}{[J2000]} &
  \multicolumn{1}{c}{[J2000]} &
  \multicolumn{1}{c}{[mJy]} &
  \multicolumn{1}{c}{[mJy]} &
  \multicolumn{1}{c}{[mJy]} &
  \multicolumn{1}{c|}{} \\
\hline\hline
  1327+4326 & 13:27:20.964 & 43:26:27.90 & 660   $\pm$ 20  & 340  $\pm$ 30   & $<$ 30 & -0.32 $\pm$ 0.10\\
  1333+1649 & 13:33:35.771 & 16:49:04.18 & 390   $\pm$ 10  & 410  $\pm$ 10   & $<$ 9  &   0.02 $\pm$ 0.04\\
  1339+6328 & 13:39:23.766 & 63:28:58.13 & 480   $\pm$ 10  & 183  $\pm$  3   & $<$ 9  &  -0.47 $\pm$ 0.03\\
  1347+1835 & 13:47:23.484 & 18:35:37.82 & 360   $\pm$ 10  & 270  $\pm$ 10   & $<$ 9  &  -0.14 $\pm$ 0.054\\
  1357+4353 & 13:57:40.584 & 43:53:59.73 & 690   $\pm$ 10  & 290  $\pm$ 10   & $<$ 9  &  -0.43 $\pm$ 0.04\\
  1358+4737 & 13:58:40.667 & 47:37:58.12 & 690   $\pm$ 20  & 260  $\pm$ 10   & $<$ 6  &  -0.48 $\pm$ 0.05\\
  1407+2827 & 14:07:00.394 & 28:27:14.78 & 820   $\pm$ 20  & 920  $\pm$ 10   & $<$ 9  &   0.05 $\pm$ 0.03\\
  1410+3647 & 14:10:43.043 & 36:47:21.83 &1230   $\pm$ 40  & 220  $\pm$ 10   & $<$ 9  &  -0.85 $\pm$ 0.06\\
  1413+1509 & 14:13:41.645 & 15:09:39.65 & 470   $\pm$ 10  & 170  $\pm$ 10   & $<$ 9  &  -0.50 $\pm$ 0.07\\
  1442+4044 & 14:42:59.305 & 40:44:28.79 & 960   $\pm$ 30  & 153  $\pm$  3   & $<$ 9  &  -0.91 $\pm$ 0.04\\
  1450+0910 & 14:50:31.184 & 09:10:28.03 & 330   $\pm$ 10  & 390  $\pm$ 10   & $<$ 6  &   0.08 $\pm$ 0.04\\
  1451+1343 & 14:51:31.498 & 13:43:24.07 & 690   $\pm$ 20  & 181  $\pm$  3   & $<$ 9  &  -0.66 $\pm$ 0.03\\
  1458+3542 & 14:58:43.413 & 35:42:57.50 & 710   $\pm$ 20  & 190  $\pm$  3   & $<$ 9  &  -0.65 $\pm$ 0.03\\
  1504+3249 & 15:04:07.545 & 32:49:21.16 & 340   $\pm$ 10  & 190  $\pm$ 10   & $<$ 9  &  -0.28 $\pm$ 0.06\\
  1507+5857 & 15:07:47.370 & 58:57:27.71 & 540   $\pm$ 20  & 180  $\pm$ 10   & $<$ 30 & -0.54 $\pm$ 0.07\\
  1509+4726 & 15:09:19.830 & 47:26:56.31 &1270   $\pm$ 40  & 180  $\pm$ 10   & $<$ 9  &  -0.97 $\pm$ 0.07\\
  1511+2208 & 15:11:05.568 & 22:08:06.70 & 410   $\pm$ 10  & 173  $\pm$  3   & $<$ 9  &  -0.42 $\pm$ 0.03\\
  1528+3738 & 15:28:27.922 & 37:38:09.48 & 780   $\pm$ 20  & 151  $\pm$  4   & $<$ 9  &  -0.81 $\pm$ 0.04\\
  1539+6113 & 15:39:48.117 & 61:13:56.36 & 500   $\pm$ 20  & 211  $\pm$  3   & $<$ 9  &  -0.42 $\pm$ 0.04\\
  1545+4751 & 15:45:08.526 & 47:51:54.69 & 690   $\pm$ 20  & 220  $\pm$ 10   & $<$ 6  &  -0.56 $\pm$ 0.06\\
  1606+3124 & 16:06:08.526 & 31:24:46.41 & 660   $\pm$ 20  & 550  $\pm$ 10   & $<$ 9  &  -0.09 $\pm$ 0.04\\
  1630+2131 & 16:30:11.240 & 21:31:34.38 & 510   $\pm$ 20  & 240  $\pm$  4   & $<$ 30 & -0.37 $\pm$ 0.04\\
  1640+1144 & 16:40:58.869 & 11:44:04.18 & 330   $\pm$ 10  & 220  $\pm$ 10   & $<$ 9  &  -0.20 $\pm$ 0.06\\
  1644+1305 & 16:44:41.199 & 13:05:19.69 &1340   $\pm$ 50  & 163  $\pm$  3   & $<$ 9  &  -1.04 $\pm$ 0.04\\
  1644+2536 & 16:44:59.061 & 25:36:30.98 & 730   $\pm$ 20  & 271  $\pm$  3   & $<$ 9  &  -0.49 $\pm$ 0.03\\
  1645+1113 & 16:45:54.692 & 11:13:52.59 & 460   $\pm$ 10  & 170  $\pm$  3   & $<$ 30 & -0.49 $\pm$ 0.03\\
  1647+1720 & 16:47:41.833 & 17:20:11.76 &2130   $\pm$ 70  & 490  $\pm$ 10   & $<$ 9  &  -0.73 $\pm$ 0.04\\
  1711+3019 & 17:11:19.937 & 30:19:17.67 &1040   $\pm$ 30  & 240  $\pm$ 10   & $<$ 9  &  -0.72 $\pm$ 0.05\\
  1735+5049 & 17:35:48.985 & 50:49:11.68 & 430   $\pm$ 13  & 920  $\pm$ 10   & $<$ 9  &   0.37 $\pm$ 0.03\\
  2126--0119 & 21:26:32.768 & --01:19:32.37& 340   $\pm$ 10  & 220  $\pm$ 10 & $<$ 9  &  -0.21 $\pm$ 0.06\\
  2145+0431 & 21:45:17.756 & 04:31:32.05 &1430   $\pm$ 40  & 234  $\pm$  4   & $<$ 9  &  -0.90 $\pm$ 0.03\\
  2151+0552 & 21:51:37.876 & 05:52:12.87 & 680   $\pm$ 20  & 500  $\pm$ 10   & $<$ 9  &  -0.15 $\pm$ 0.04\\
  2153+1241 & 21:53:04.651 & 12:41:05.19 & 430   $\pm$ 10  & 210  $\pm$ 10   & $<$ 30 & -0.35 $\pm$ 0.06\\
  2322+0812 & 23:22:36.097 & 08:12:01.66 &1180   $\pm$ 40  & 430  $\pm$ 20   & $<$ 9  &  -0.50 $\pm$ 0.06\\
  2331+0705 & 23:31:55.513 & 07:05:42.08 & 540   $\pm$ 20  & 181  $\pm$  4   & $<$ 9  &  -0.54 $\pm$ 0.04\\
  2333--0903 & 23:33:47.282 & --09:03:04.17&1000   $\pm$ 30  & 170  $\pm$ 10 & $<$ 9  &  -0.88 $\pm$ 0.07\\
  2341+0018 & 23:41:06.908 & 00:18:33.56 & 430   $\pm$ 10  & 190  $\pm$ 10   & $<$ 9  &  -0.40 $\pm$ 0.06\\
  2354--0019 & 23:54:09.171 & --00:19:47.89& 350   $\pm$ 10  & 291  $\pm$  4 & $<$ 9  &  -0.09 $\pm$ 0.03\\
    &   &   &    &     &   &  \\
\hline
\end{tabular}
}

\centering
\tiny
\rotatebox{90}{
\centering
\begin{tabular}{|llrrrrr|}
\hline
  \multicolumn{1}{|c}{Name} &
  \multicolumn{1}{c}{RA} &
  \multicolumn{1}{c}{DEC} &
  \multicolumn{1}{c}{S$_{1.4}$} &
  \multicolumn{1}{c}{S$_{10.45}$} &
  \multicolumn{1}{c}{P$_{10.45}$} &
  \multicolumn{1}{c|}{$\alpha$$^{10.45}_{1.4}$} \\
  \multicolumn{1}{|c}{} &
  \multicolumn{1}{c}{[J2000]} &
  \multicolumn{1}{c}{[J2000]} &
  \multicolumn{1}{c}{[mJy]} &
  \multicolumn{1}{c}{[mJy]} &
  \multicolumn{1}{c}{[mJy]} &
  \multicolumn{1}{c|}{} \\
\hline\hline
  0132--0804 & 01:32:41.129 & --08:04:04.83 &310   $\pm$ 10  & 181  $\pm$  2 & $<$  6 &  -0.27 $\pm$ 0.04 \\
  0134+0003 & 01:34:12.700 & 00:03:45.29 & 920   $\pm$ 30  & 263  $\pm$  1   & $<$  6 &  -0.62 $\pm$ 0.04 \\
  0249+0619 & 02:49:18.010 & 06:19:51.85 & 500   $\pm$ 20  & 620  $\pm$ 10   & $<$  9 &   0.11 $\pm$ 0.05 \\
  0323+0534 & 03:23:20.254 & 05:34:11.93 & 2790  $\pm$ 80  & 330  $\pm$  3   & $<$  6 &  -1.06 $\pm$ 0.03 \\
  0706+4647 & 07:06:48.082 & 46:47:56.39 & 1590  $\pm$ 50  & 270  $\pm$ 10   & $<$  9 &  -0.88 $\pm$ 0.06 \\
  0737+6430 & 07:37:58.988 & 64:30:43.25 & 420   $\pm$ 10  & 350  $\pm$ 10   & $<$  9 &  -0.09 $\pm$ 0.04 \\
  0805+2106 & 08:05:38.530 & 21:06:51.92 & 930   $\pm$ 30  & 430  $\pm$ 30   & $<$  12 &  -0.38 $\pm$ 0.09 \\
  0808+2646 & 08:08:36.756 & 26:46:36.73 & 450   $\pm$ 10  & 250  $\pm$ 10   & $<$  9 &  -0.29 $\pm$ 0.05 \\
  0830+2323 & 08:30:21.693 & 23:23:25.72 & 1100  $\pm$ 30  & 173  $\pm$  3   & $<$  6 &  -0.92 $\pm$ 0.03\\
  0902+4310 & 09:02:30.934 & 43:10:14.07 & 340   $\pm$ 10  & 530  $\pm$ 20   & $<$  9 &   0.22 $\pm$ 0.05\\
  0923+3849 & 09:23:14.443 & 38:49:39.74 & 380   $\pm$ 10  & 320  $\pm$  4   & $<$  9 &  -0.08 $\pm$ 0.03\\
  0945+4636 & 09:45:42.096 & 46:36:50.60 & 470   $\pm$ 10  & 290  $\pm$  4   & $<$  9 &  -0.24 $\pm$ 0.02\\
  0952+2828 & 09:52:06.089 & 28:28:32.37 & 1360  $\pm$ 40  & 260  $\pm$  4   & $<$  6 &  -0.82 $\pm$ 0.03\\
  0954+2639 & 09:54:39.795 & 26:39:24.56 & 310   $\pm$ 10  & 170  $\pm$  3   & $<$  30 & -0.29 $\pm$ 0.04\\
  1006+1713 & 10:06:31.755 & 17:13:17.15 & 570   $\pm$ 20  & 230  $\pm$ 10   & $<$  12 &  -0.45 $\pm$ 0.06\\
  1028+3844 & 10:28:44.304 & 38:44:36.67 & 660   $\pm$ 20  & 180  $\pm$  4   & $<$  6 &  -0.64 $\pm$ 0.04\\
  1033+3935 & 10:33:22.051 & 39:35:51.12 & 400   $\pm$ 10  & 280  $\pm$ 10   & $<$  9 &  -0.17 $\pm$ 0.04\\
  1035+5628 & 10:35:07.058 & 56:28:46.81 & 1800  $\pm$ 50  & 620  $\pm$ 10   & $<$  9 &  -0.53 $\pm$ 0.03\\
  1047+1456 & 10:47:32.403 & 14:56:46.57 & 372   $\pm$ 10  & 590  $\pm$ 30   & $<$  30 &  0.22 $\pm$ 0.06\\
  1057+0012 & 10:57:15.781 & 00:12:03.74 & 890   $\pm$ 30  & 190  $\pm$  3   & $<$  6 &  -0.76 $\pm$ 0.04\\
  1058--0309 & 10:58:10.991 & --03:09:26.81 & 450  $\pm$ 10  & 180  $\pm$  4 & $<$  12 &  -0.45 $\pm$ 0.03\\
  1101+3904 & 11:01:30.074 & 39:04:32.78 & 340   $\pm$ 10  & 200  $\pm$  4   & $<$  6 &  -0.26 $\pm$ 0.04\\
  1110+6028 & 11:10:13.085 & 60:28:42.29 & 430   $\pm$ 10  & 210  $\pm$ 10   & $<$  9 &  -0.35 $\pm$ 0.06\\
  1124+1919 & 11:24:43.869 & 19:19:29.53 & 880   $\pm$ 30  & 190  $\pm$ 10   & $<$  9 &  -0.76 $\pm$ 0.07\\
  1135+4258 & 11:35:55.999 & 42:58:44.64 & 1450  $\pm$ 40  & 170  $\pm$  2   & $<$  9 &  -1.06 $\pm$ 0.03\\
  1143+1834 & 11:43:26.063 & 18:34:38.40 & 310   $\pm$ 10  & 220  $\pm$  4   & $<$  9 &  -0.17 $\pm$ 0.04\\
  1148+0752 & 11:48:30.779 & 07:52:07.54 & 610   $\pm$ 20  & 200  $\pm$  3   & $<$  6 &  -0.55 $\pm$ 0.04\\
  1148+5924 & 11:48:50.352 & 59:24:56.68 & 480   $\pm$ 10  & 450  $\pm$ 10   & $<$  9 &  -0.03 $\pm$ 0.03\\
  1155+4555 & 11:55:10.998 & 45:55:39.85 & 610   $\pm$ 20  & 170  $\pm$  4   & $<$  9 &  -0.63 $\pm$ 0.04\\
  1204+5202 & 12:04:18.615 & 52:02:17.83 & 960   $\pm$ 30  & 152  $\pm$  4   & $<$  6 &  -0.91 $\pm$ 0.04\\
  1208+5413 & 12:08:27.495 & 54:13:19.74 & 440   $\pm$ 10  & 200  $\pm$ 10   & $<$  9 &  -0.39 $\pm$ 0.06\\
  1215--0628 & 12:15:14.412 & --06:28:03.96& 360   $\pm$ 10  & 200  $\pm$ 10 & $<$  30 & -0.29 $\pm$ 0.06\\
  1220+2916 & 12:20:06.820 & 29:16:50.70 & 390   $\pm$ 10  & 173  $\pm$  4   & $<$  9 &  -0.40 $\pm$ 0.03\\
  1234+4753 & 12:34:13.330 & 47:53:51.40 & 360   $\pm$ 10  & 230  $\pm$ 10   & $<$  9 &  -0.22 $\pm$ 0.05\\
  1244+4048 & 12:44:49.200 & 40:48:06.35 &1340   $\pm$ 40  & 370  $\pm$ 10   & $<$  6 &  -0.64 $\pm$ 0.04\\
  1254+0859 & 12:54:58.953 & 08:59:47.57 & 670   $\pm$ 20  & 250  $\pm$  3   & $<$  9 &  -0.49 $\pm$ 0.03\\
  1313+5458 & 13:13:37.869 & 54:58:23.89 &1310   $\pm$ 40  & 264  $\pm$  4   & $<$  9 &  -0.79 $\pm$ 0.03\\
  1324+4048 & 13:24:12.067 & 40:48:11.58 & 350   $\pm$ 10  & 190  $\pm$ 10   & $<$  9 &  -0.30 $\pm$ 0.06\\
  1326+5712 & 13:26:50.572 & 57:12:06.85 & 520   $\pm$ 20  & 193  $\pm$  4   & $<$  9 &  -0.49 $\pm$ 0.05\\
  \hline
\end{tabular}
}

\end{table*}

We also compare our RM results with the sample of \cite{Farnes14}.
They present a catalogue of multi-wavelength linear polarisation and total intensity radio data for polarised sources from the NVSS. The result of this work is a catalogue of 951 sources with the SEDs in both total intensity and fractional polarisation.
 
\section{Results and discussion}

In Tab.2 we present information about the 30 high RM candidates: the sources coordinates taken from the FIRST catalogue (they are more precise, compared to the NVSS survey, thanks to the better angular resolution of the survey), their total flux density at 1.4 GHz from the NVSS catalogue, their total flux density measured with the Effelsberg telescope at 10.45 GHz (both expressed in mJy), the spectral index ($\alpha_{1.4-10.45}$), the observed RM value (RM$_{obs}$), the redshifts (when available from literature), the contribution of the RM of our Galaxy (RM$_{mw}$), the values of which have been taken from the most recent work by \cite{Oppermann15}, and finally the rest frame RM corrected by their redshifts (RM$_{rf}$). Where the redshift was not available in the literature, we assumed a mean value: z$_{mean}$ = 1.5 in order to have a rough idea on the intrinsic RM value. For the source 0845+0439 we receive its redshift value via private communication after spectroscopic measurements tests at the Large Binocular Telescope (LBT). Almost all the targets are QSO type.

In the following, we discuss about the spectral index distribution, their SEDs, and their polarisation information with a detailed explanation of the RM determination and the depolarisation behaviour.

\begin{table*}
\caption{Table listing the general information of the sources: the source name, the source coordinates right ascension and declination [J2000] taken from the FIRST catalogue, the flux density at 1.4 GHz from the NVSS catalogue, the measured flux density at 10.45 GHz with Effelsberg, the spectral index among these two frequencies $\alpha_{1.4-10.45}$, the observed RM (RM$_{obs}$), the contribution of the RM from the Milky Way (RM$_{mw}$), the redshifts of the sources taken from the literature (where this information was not available, its mean value was considered; marked here with an asterisk) and the rest frame RM (RM${_{rf}}$) }
\small
\centering
\begin{tabular}{lccrrrrcrr}
 \hline\hline                                                    
 Source & RA  &   DEC    &  S$^{NVSS}_{1.4GHz}$~~& S$^{Eff}_{10GHz}$~~&    $\alpha$$^{10.45}_{1.4}$ ~~~  &RM$_{obs}$~~~      & z~ &  RM${_{mw}}$~~ &  RM${_{rf}}$~~~~  \\
   &   [J2000]   & [J2000]  &      [mJy] ~~&   [mJy]   ~~      &     & [rad/m$^{2}$] ~ &  &  [rad/m$^{2}$] ~~  & [rad/m$^{2}$]~~~~\\                   
 \hline                                                  
  0239--0234 & 02:39:45.480  &  -02:34:40.98  & 300  $\pm$  10   &  723  $\pm$  6~~    &  0.44    $\pm$    0.04 &    --40    $\pm$   10~~   & 1.1~~   &  --100     $\pm$    100     &     --190   $\pm$  30~~~~ \\  
  0243--0550 & 02:43:12.464  &  -05:50:55.36  & 560  $\pm$  17   &  548  $\pm$  2~~    &  -0.01   $\pm$    0.04 &    600     $\pm$   100   & 1.8~~   &  --100     $\pm$    100     &     4500     $\pm$  400~~ \\ 
  0742+4900  & 07:42:02.763  &  +49:00:15.65  & 398  $\pm$  12   &  430  $\pm$  1~~    &  0.04    $\pm$    0.04 &    --200   $\pm$   30~~   & 2.3~~   &  30       $\pm$    10~~     &    --2170   $\pm$  370~~ \\ 
  0751+2716  & 07:51:41.492  &  +27:16:31.65  & 590  $\pm$  20   &  82  $\pm$  1~~    &  -0.98   $\pm$    0.04 &    500     $\pm$   100   & 3.2    &  --20     $\pm$    10~~     &     8800       $\pm$  900~~ \\  
  0845+0439  & 08:45:17.151  &  +04:39:46.64  & 380  $\pm$  10   &  682  $\pm$  6~~    &  0.30    $\pm$    0.04 &    1920    $\pm$   20~~   & 0.8~~   &  --100     $\pm$    100     &     6230    $\pm$  50~~~~  \\ 
  0925+3159  & 09:25:32.726  &  +31:59:52.86  & 551  $\pm$  17   &  97  $\pm$  4~~    &  -0.86   $\pm$    0.06 &    --100    $\pm$   100   & 1.5*    &  --10     $\pm$    10~~     &     --400    $\pm$  300~~ \\ 
  0958+3224  & 09:58:20.939  &  +32:24:02.16  &1250  $\pm$  40   &  660  $\pm$  6~~    &  -0.32   $\pm$    0.04 &    2200    $\pm$   100   & 0.5~~   &  --10     $\pm$    10~~     &     5200     $\pm$  200~~\\ 
  1015+0318  & 10:15:34.024  &  +03:18:50.06  & 416  $\pm$  13   &  91  $\pm$  2~~    &  -0.76   $\pm$    0.04 &    200     $\pm$   100   & 1.5*    &  --40     $\pm$    10~~     &     1600      $\pm$  300~~ \\  
  1043+2408  & 10:43:09.032  &  +24:08:35.45  & 320  $\pm$  10   & 1070  $\pm$  10     & 0.60    $\pm$    0.04 &    --60    $\pm$   10~~   & 0.6~    &  30       $\pm$    20~~     &     --150    $\pm$  20~~~~ \\ 
  1044+0655  & 10:44:55.921  &  +06:55:37.94  & 490  $\pm$  20   &  295  $\pm$  5~~    &  -0.25   $\pm$    0.06 &    --210   $\pm$   20~~   & 2.1~~   &  20       $\pm$    40~~     &    --2030   $\pm$  170~~ \\ 
  1048+0141  & 10:48:22.850  &  +01:41:47.46  & 380  $\pm$  10   &  328  $\pm$  3~~    &  -0.07   $\pm$    0.03 &    --2510  $\pm$   30~~   & 0.7~~   &  --100     $\pm$    100     &    --7160   $\pm$  90~~~~   \\ 
  1146+5356  & 11:46:44.186  &  +53:56:43.36  & 367  $\pm$  11   &  614  $\pm$  6~~    &  0.26    $\pm$    0.04 &    --450   $\pm$   10~~   & 2.2~~   &  10       $\pm$    10~~     &   --4570    $\pm$  90~~~~   \\ 
  1213+1307  & 12:13:32.146  &  +13:07:20.43  &1340  $\pm$  40   &  421  $\pm$  3~~    &  -0.60   $\pm$    0.04 &    20      $\pm$   2~~~~  & 1.1~~   &  20       $\pm$    20~~     &       90    $\pm$  10~~~~   \\ 
  1246--0730 & 12:46:04.231  &  -07:30:46.63  & 550  $\pm$  20   & 1040  $\pm$  10     & 0.32    $\pm$    0.04 &     880    $\pm$   10~~   & 1.2~~   &  10       $\pm$    40~~     &     4610     $\pm$  60~~~~ \\ 
  1311+1417  & 13:11:07.835  &  +14:17:46.69  & 734  $\pm$  22   &  207  $\pm$  1~~    &  -0.63   $\pm$    0.03 &    570     $\pm$   10~~   & 1.9~~   &  10       $\pm$    10~~     &     4940    $\pm$  100~~ \\ 
  1312+5548  & 13:12:53.193  &  +55:48:13.21  & 590  $\pm$  20   &  140  $\pm$  3~~    &  -0.72   $\pm$    0.04 &   --1000    $\pm$   200    & 1.5*    &  20       $\pm$    10~~    &    --6000   $\pm$  1000 \\ 
  1351+0830  & 13:51:16.926  &  +08:30:39.82  & 350  $\pm$  10   &  308  $\pm$  5~~    &  -0.06   $\pm$    0.05 &    70      $\pm$   20~~   & 1.4~~   &  50       $\pm$    30~~     &      410    $\pm$  120~~ \\ 
  1405+0415  & 14:05:01.113  &  +04:15:35.87  & 930  $\pm$  30   &  712 $\pm$   10    &  -0.13   $\pm$    0.04 &     1153   $\pm$   4~~~~  & 3.2~~   &  30       $\pm$    60~~     &    20420     $\pm$  80~~~~  \\ 
  1435--0414 & 14:35:39.884  &  -04:14:55.20  & 480  $\pm$  10   &  146  $\pm$  1~~    &  -0.60   $\pm$    0.03 &    1080    $\pm$   30~~   & 0.8~~   &  100       $\pm$    100    &     3500     $\pm$  80~~~~  \\ 
  1549+5038  & 15:49:17.447  &  +50:38:05.87  & 630  $\pm$  20   &  813  $\pm$  1~~    &  0.135   $\pm$    0.04 &    100     $\pm$   100   & 2.2~~   &  20       $\pm$    10~~     &     1400     $\pm$  500~~ \\  
  1616+0459  & 16:16:37.530  &  +04:59:31.96  & 330  $\pm$  10   & 1035 $\pm$  10     & 0.57    $\pm$    0.04 &    2530  $\pm$   40~~   & 3.2~~   &  150      $\pm$    10~~        &   44630      $\pm$  720~~  \\ 
  1616+2647  & 16:16:38.340  &  +26:47:01.39  &1480  $\pm$  50   &  294 $\pm$   5~~   &  -0.80   $\pm$    0.06 &    --140   $\pm$   30~~   & 1.5*    &  10       $\pm$    10~~     &     --880    $\pm$  190~~ \\ 
  1647+3752  & 16:47:25.735  &  +37:52:18.32  & 630  $\pm$  20   &  152 $\pm$   2~~   &  -0.70   $\pm$    0.04 &    80      $\pm$   10~~   & 1.5*    &  --20     $\pm$    10~~     &      490     $\pm$  80~~~~  \\ 
  1713+2813  & 17:13:25.930  &  +28:13:07.10  &1030  $\pm$  40   &  124 $\pm$   2~~   &  -1.05   $\pm$    0.05 &    84      $\pm$   2~~~~  & 1.5*    &  3        $\pm$    8~~~~    &      530     $\pm$  20~~~~  \\  
  1723+3417  & 17:23:20.801  &  +34:17:57.85  & 520  $\pm$  20   &  162 $\pm$   1~~   &  -0.58   $\pm$    0.04 &    83      $\pm$   2~~~~  & 0.2~~   &  --30     $\pm$    10~~     &      120     $\pm$  10~~~~  \\ 
  2050+0407  & 20:50:06.240  &  +04:07:49.22  & 565  $\pm$  17   &  566 $\pm$   2~~   &  0.01    $\pm$    0.06 &    60      $\pm$   10~~   & 1.5*    &  100      $\pm$    100    &      380       $\pm$  60~~~~\\ 
  2101+0341  & 21:01:38.833  &  +03:41:31.29  & 630  $\pm$  20   &  969  $\pm$  1~~    &  0.21    $\pm$    0.04 &    70      $\pm$   10~~   & 1.0~~   &  200      $\pm$    100     &      260     $\pm$  40~~~~  \\ 
  2147+0929  & 21:47:10.162  &  +09:29:46.63  & 930  $\pm$  30   &  760  $\pm$  5~~    &  -0.10   $\pm$    0.04 &    1400    $\pm$   20~~   & 1.1~~   &  --10     $\pm$    40~~     &     6260    $\pm$  110~~ \\ 
  2200+0708  & 22:00:57.607  &  +07:08:29.01  & 896  $\pm$  32   &  113  $\pm$  1~~    &  -1.03   $\pm$    0.04 &    --15    $\pm$   2~~~~  & 1.5*    &  --20     $\pm$    100     &      --90    $\pm$  20~~~~ \\ 
  2245+0324  & 22:45:28.284  &  +03:24:08.71  & 480  $\pm$  10   &  379  $\pm$  7~~    &  -0.12   $\pm$    0.04 &    --800   $\pm$   100   & 1.3~    &  --40     $\pm$    100      &    --4400    $\pm$  500~~ \\

 \hline\hline                                                    

\end{tabular}
\end{table*}%

\subsection{Spectral index distribution}

From the first observation run we were able to determine the spectral index (defined here as S $\propto$ $\nu^{\alpha}$) between 1.4 GHz (from the NVSS catalogue) and 10.45 GHz. This allows us to have a qualitative idea on the source type presented in the sample. 
As mentioned in section 5, we compared the 30 high RM candidates, with the 77 unpolarised sources at 10.45 GHz.

In Fig.1 the spectral index distributions for the targets considered to have a high RM (red colours) together with the unpolarised targets (black colours) and its cumulative plot distribution are shown. Both the histograms have been normalised with the number of the sources. We chose the bin dimension larger than the 3$\sigma$ of the larger error of the spectral index in our sample (in this case 3$\sigma_{\alpha}$= 0.18).
From the cumulative plot (Fig.1$\textit{b}$), the two distributions seem to follow a similar tendency up to a value of $\alpha$$\sim$--0.6, then they separate.
In fact, the spectral index distribution of the unpolarized sources (black histogram Fig.1$\textit{a}$) peaks at $\alpha_{peak}$$\simeq$ --0.5, thus we see mainly steep radio spectra sources.
On the other hand, the spectral index distribution of the high RM candidates (red histogram Fig.1$\textit{a}$) shows three different groups of objects. The three distributions peak, indeed, at different spectral indices revealing us the following classes of spectra:
\begin{itemize}
\item steep spectrum radio sources with $\alpha_{peak}$$\simeq$ --0.8;
\item flat spectrum radio sources with $\alpha_{peak}$$\simeq$ --0.1;
\item inverted spectrum radio sources with $\alpha_{peak}$$\simeq$ +0.2;
\end{itemize}

The distribution of radio spectral indices can be symptomatic of the orientations of radio sources \citep{OB82}. The first group could be associated to "lobe dominated", objects where large-scale structures dominate the radio synchrotron emission and where the steepness of the spectrum is due to radiative loss of the relativistic electrons. Flat spectrum radio sources could be associated with objects for which the lines of sight are closer to the radio axis. In these objects, we would see a superposition of several features, thus a continuous injection of relativistic electrons that become opaque at widely different frequencies.
The latter kind of objects could be representative of "core dominated" objects where the dominant synchrotron component is very compact and where it is possible that the synchrotron self absorption occur. If a turnover peak is revealed in the spectra of these objects, it is possible to associate them to the GPS source which are supposed to be AGN in an early phase of activity \citep{ODea98}.

From the comparison of the histograms in Fig.1, we can assert that, although the low-number in our sample, the high RM candidates do not seem to be represented by a particular class of targets. Instead, the three main types of spectra seem to represent equally these peculiar objects.
On the contrary, the other sources without detected polarisation at high frequency seem to be dominated by very steep spectra. This could be an indication that the objects without high RM could be dominated by very extended, and probably old, synchrotron components while high RM values could be found in different objects where the combined contribution of electron density and magnetic field is strong.

\begin{figure}[h!]
\begin{center}
\includegraphics[width=0.5\textwidth]{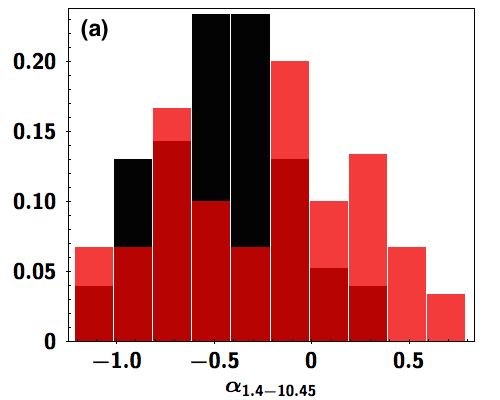}
\includegraphics[width=0.5\textwidth]{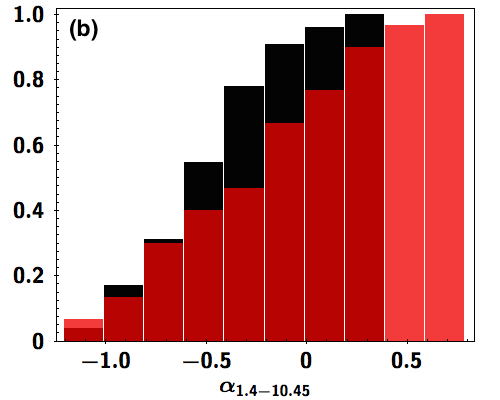}
\caption{Spectral index distributions. (a) comparison of the spectral index distribution for the high RM candidates (30 sources; red histogram) and the unpolarised sources (77 sources; black histogram). The distribution of the unpolarised sources peaks at $\alpha_{peak}$$\sim$ --0.5.  The high RM distribution shows 3 different types of objects: steep ($\alpha_{peak}$$\sim$ --0.8), flat ($\alpha_{peak}$$\sim$ --0.1) and inverted ($\alpha_{peak}$$\sim$ +0.2) spectrum radio sources. (b) the same but cumulative histograms.}
\label{}
\end{center}
\end{figure}

\subsection{Spectral energy distribution}
The wide frequency coverage obtained with our follow-up observations, plus data from literature, allows us to study the SED shape of these sources all characterised by synchrotron emission.
We, indeed, extended the SEDs to frequencies lower than 1.4 GHz, using the following surveys: VLSS (74 MHz with a resolution of 80$\arcsec$) \citep{Cohen07}, 7C (151 MHz with a resolution of 70$\arcsec$) \citep{Hales07}, WENSS (325 MHz with a resolution of 54$\arcsec$$\times$ 54$\arcsec$cosec$\delta$) \citep{Rengelink97} and TEXAS (365 MHz with a resolution of $\sim$50$\arcsec$) \citep{Douglas96}. Their values are listed in Tab. A.1 -- 6. As commented in section 2, we checked that our targets are isolated up to 5$\arcmin$. Therefore, the different beam sizes of each frequency should not be affected by back- or foreground sources. Moreover, all the sources were selected to be compact in the FIRST survey. Therefore, effects on the flux density due to extended structures are neither affecting our analysis.

For each source in our sample, we fit several models representing its SEDs:
\begin{itemize}
\item A power law, representing a purely optically thin synchrotron spectrum with a slope $\alpha_{thin}$:
\begin{equation}
S^{pl}_\nu \propto \nu^{\alpha_{thin}}
\end{equation}
\item A power law with a break (symptomatic of an ageing of the radio source) at frequency $\nu_{b}$:
\begin{equation}
S^{plb}_\nu= S^{pl}_\nu \left(1-\exp\left(\left(\frac{\nu}{\nu_b}\right)^{\alpha_{break}-\alpha_{thin}}\right)\right) 
\end{equation}
where $\alpha_{thin}$ is the spectral index at frequencies lower than $\nu_{b}$ and $\alpha_{break}$ is the spectral index at higher frequencies.
\item A single synchrotron self-absorption component:
\begin{equation}
S^{s}_{\nu} \propto \nu^{2.5}   \left( 1 - \exp\left(- \left(\frac{\nu}{\nu_{0}}\right)^{\alpha_{thin}-2.5}\right)\right)
\end{equation}
where $\nu_0$ is the frequency where the emission changes from optically thick, with a spectral index of 2.5,  to optically thin with a spectral index $\alpha_{thin}$.
\item A single synchrotron component with a break at frequency $\nu_{b}$:
\begin{equation}
S^{sb}_{\nu} = S^{s}_{\nu} \left(1-\exp\left(\left(\frac{\nu}{\nu_b}\right)^{\alpha_{break}-\alpha_{thin}}\right)\right) 
\end{equation}
\item A combination of several synchrotron components ($S^{s+}_{\nu} $) with fixed $\alpha_{thin}$= --0.7
\item A combination of a power law with one or two synchrotron components ($S^{pls}_{\nu} $) with $\alpha_{thin}$= free or $\alpha_{thin}$= --0.7.
\end{itemize}

\noindent
For each source, we selected the best model according to the lowest residual value. Tab. B.1 shows the chosen parameters for the different fitting functions used for the SEDs study. Tab. 3 is a quick look on the values for $\alpha_{thick}$ and $\alpha_{thin}$ we decided to use for each model.

Based on the fitted SEDs we noticed 3 main groups. We will refer to them from now on with:
\begin{itemize}
\item $\textit{Older}$: sources fitted with a power law (with or without a break) and sources with one synchrotron component with a break (the presence of a frequency break is an indication of an ageing of the source).
\item $\textit{GPS-like}$: SEDs with several synchrotron components peaking at frequencies $\ge$ 100 MHz.
\item $\textit{Mixed}$: the combination of the two above, thus a combination of a power law or a synchrotron component peaked at low frequency (with peak frequency $\le$ 100 MHz) and one or more synchrotron components at higher frequencies (with peak at frequencies $\ge$ 100 MHz). 
\end{itemize}

\begin{table}[h]
\caption{Quick look of the spectral indices $\alpha_{thick}$ and $\alpha_{thin}$ used for each model.}
\begin{center}
\begin{tabular}{lcc}
\hline
Model & $\alpha_{thick}$ &$\alpha_{thin}$ \\ 
\hline\hline
$S^{pl}_\nu$ & -- &  free\\
$S^{plb}_\nu$ & -- &free\\
$S^{s}_{\nu}$& 2.5 & free\\
$S^{sb}_{\nu}$ & 2.5 & free\\
$S^{s+}_{\nu} $ & 2.5 &--0.7 \\
$S^{pls}_{\nu} $ & 2.5 & free or --0.7 \\
 \hline
\end{tabular}
\end{center}
\label{default}
\end{table}%

\noindent
In Fig.2, we show three examples of the three object types we identified. Together with the total intensity SED, information on the polarisation flux density, the fractional polarisation ($\textit{m}$) and the polarisation angle ($\textit{$\chi$}$) are shown for each of the targets. An explanation and a discussion about the polarisation characteristics is presented in subsections 6.4 and 6.5. Similar plots for all the sources are available in Appendix C.

\begin{figure}[!h]
\begin{center}
\includegraphics[width=0.45\textwidth]{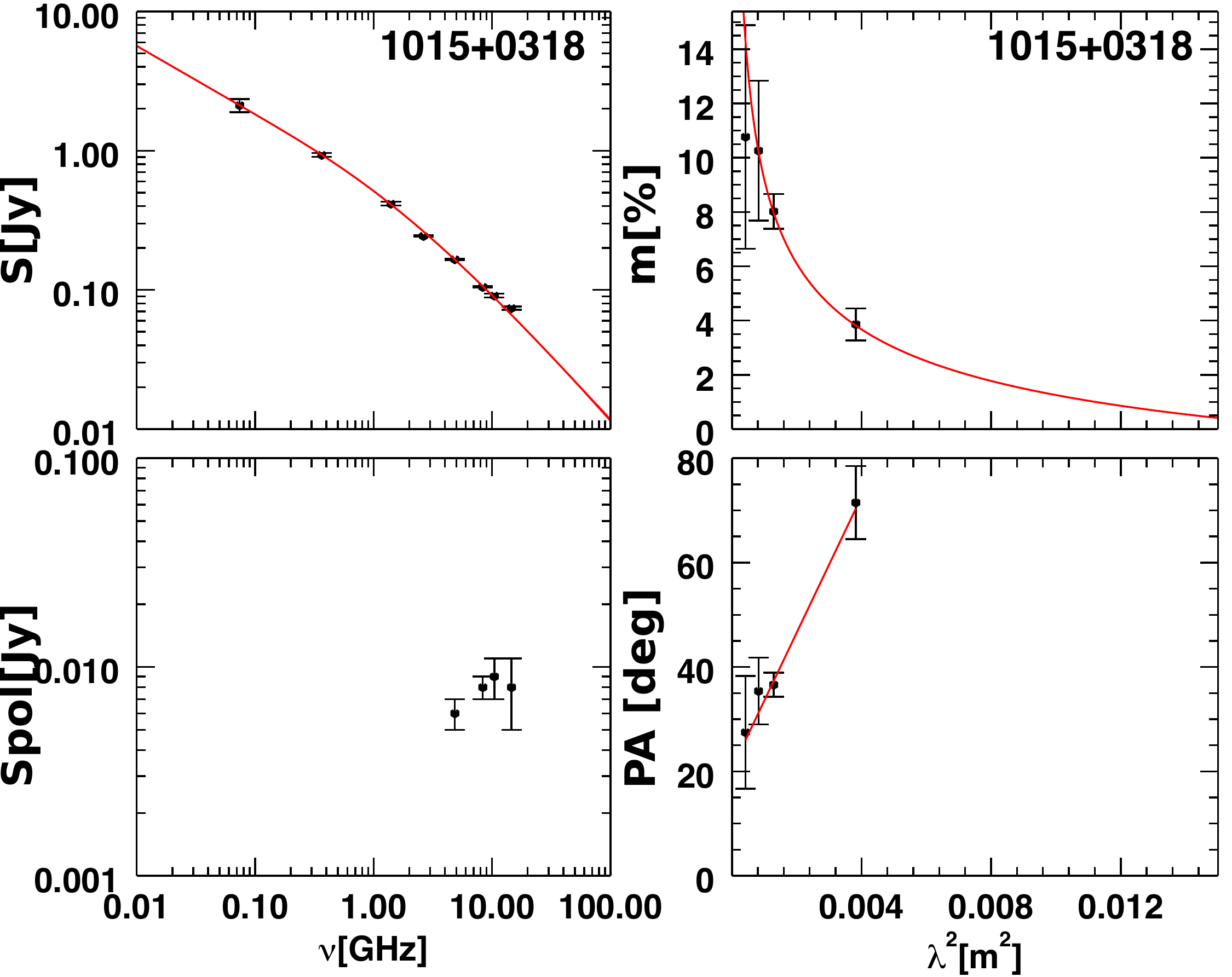}
\includegraphics[width=0.45\textwidth]{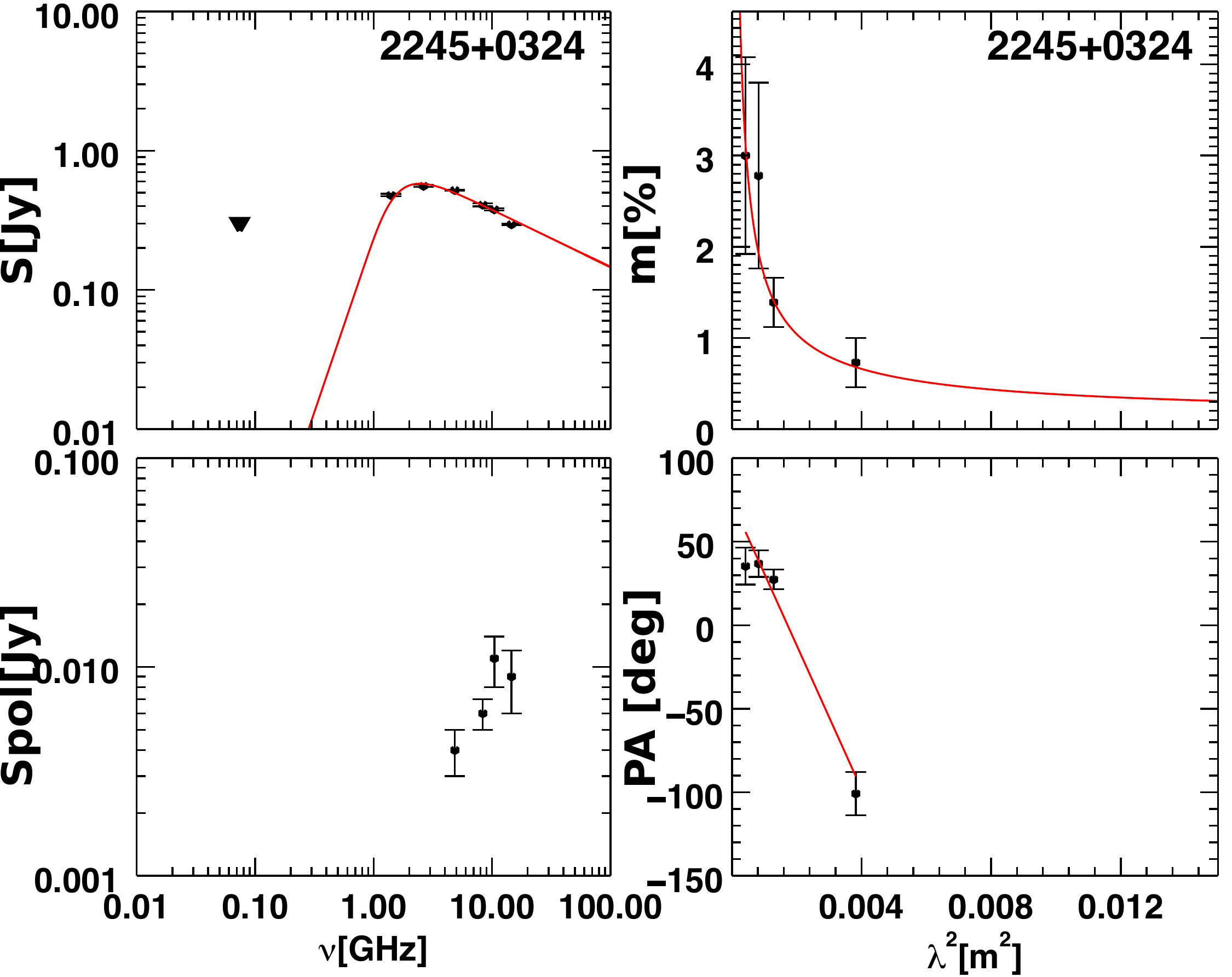}
\includegraphics[width=0.45\textwidth]{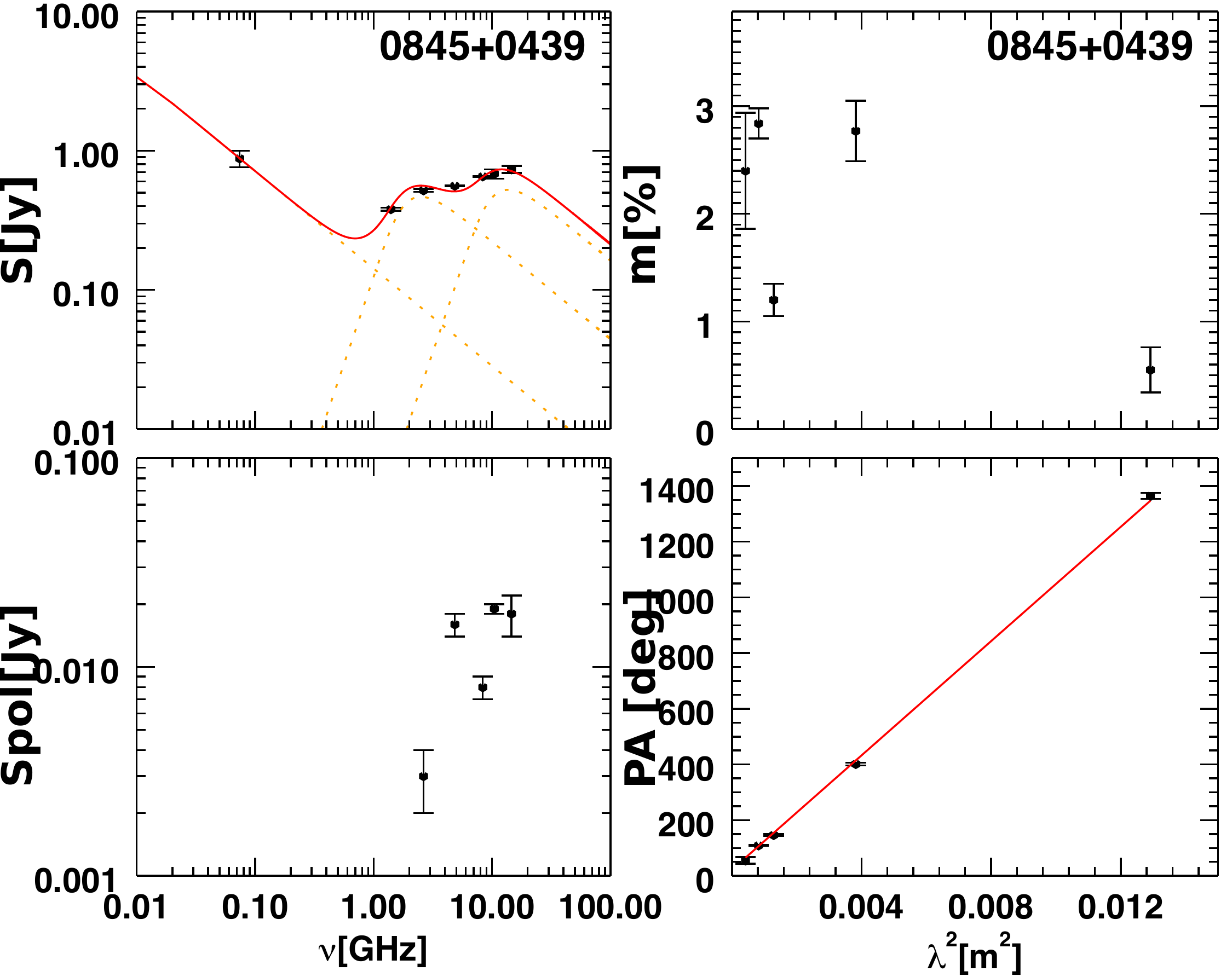}

\caption{SEDs for the sources: 1015+0318 classified as $\textit{Older}$, 2245+0324 classified as $\textit{GPS-like}$ and 0845+0439 classified as $\textit{Mixed}$. Black dot points are the Effelsberg and low frequency data from literature; 3-sigmas upper limits are drawn as triangles. Where present, various synchrotron components are plotted with orange dashed line. The fit of the spectra is the red straight line.
Together with the SEDs, polarisation information is presented.}

\label{default}
\end{center}
\end{figure}

From the analysis of the radio spectra we find that the sample splits into three (equal sized) parts:
\begin{itemize}
\item 1/3 of the sample can be considered as $\textit{Older}$ sources. Since for these sources the synchrotron peak is not visible or it peaks at very low MHz value, it is possible to assert that these sources have an extended and probably old synchrotron component; 
\item 1/3 of the sample is characterised by a $\textit{GPS-like}$ SED. These targets could have a more compact and probably early phases synchrotron components. 
\item 1/3 of the sample can be considered to have $\textit{Mixed}$ spectrum, i.e. a combination of $\textit{Older}$ and $\textit{GPS-like}$ features. This behaviour could be an indication of sources in which a restart radio emission activity occurs.
\end{itemize}

These results suggest that the high RM candidates are mainly (66\%) sources with compact high frequency components, thus probably new growing radio components. We can associate these targets with objects in a particular compact young phase (the $\textit{GPS-like}$) or in a reactivated activity phase ($\textit{Mixed}$).

For the $\textit{Older}$ sources a possible explanation for a high RM value could be  a very dense intervening material surrounding the already extended radio component.
From these considerations, in all the cases, the contribution inferred by the medium in which the source is embedded and/or the several Faraday screens that the radiation passes through (see subsections 6.4 and 6.5 for details) is very important.

\subsection{Magnetic Field Estimation}
\begin{table*}[h]
\small
\caption{Estimation of the magnetic field for those targets with one or more synchrotron components in their SEDs. We give lower limits of the magnetic field B[$\mu$G] considering their lower limit in angular size [mas] (calculated from the Inverse Compton relation). These estimations are corrected for the redshifts. Where the redshift was not available from the literature, we used the mean value of $\textit{z}$ from the sample: z$_{mean}$=1.5, marked in the table with an asterisk.  }
\centering
\begin{tabular}{clccccccc}

\hline\hline
 Name  &  z&	SEDfit  &  $\theta$ [mas]  & B [$\mu$G]  \\
\hline
2050+0407& 1.5* & $S^{s}_{\nu}$  &   0.8  & 20.0\\  
2245+0324& 1.3 & $S^{s}_{\nu}$   &   0.3  & 47.9 \\ 

\hline\hline
 Name     &  z  &  SEDfit        &  $\theta$ [mas]  & B [$\mu$G] \\    
\hline
0751+2716 & 3.2 & $S^{sb}_{\nu}$ &     7.0          & 2.2  \\
1311+1417 & 1.9 & $S^{sb}_{\nu}$ &     0.8          & 17.8 \\
1435-0414 & 0.8 & $S^{sb}_{\nu}$ &    16.3          & 1.7  \\

\hline\hline
 Name &  z &  	SEDfit   &   $\theta_{1}$ [mas]   & B$_1$ [$\mu$G] &  $\theta_{2}$ [mas]  & B$_2$ [$\mu$G] \\
\hline
1351+0830 &1.4 & $S^{s+}_{\nu} $     &  0.5   & 26.4   &  0.1  &  173.3  \\
1549+5038 &2.2 & $S^{s+}_{\nu} $     &  1.4   & 10.2   &  0.1  &  106.1  \\
1616+2647 &1.5*  & $S^{s+}_{\nu} $   &  12.2  & 1.9    &  2.2  &  11.2   \\
2101+0341 &1.0  & $S^{s+}_{\nu} $    &  0.5   & 19.4   &  0.1  &  220.8  \\

\hline\hline
 Name  &  z&  	SEDfit  &  $\theta_{1}$ [mas]  & B$_1$ [$\mu$G]  & $\theta_{2}$[mas]   & B$_2$ [$\mu$G] & $\theta_{3}$[mas]   & B$_3$ [$\mu$G]\\
\hline
0239-0234 & 1.1&  $S^{s+}_{\nu} $    &  1.2  & 11.0  &  0.2  &  82.0   &  0.1  &  247.6 \\
0742+4900 & 2.3&  $S^{s+}_{\nu} $    &  0.9  & 7.7   &  0.3  &  36.1   &  0.04 &  186.5 \\
1043+2408 & 0.6&  $S^{s+}_{\nu} $    &  1.7  & 12.0  &  0.2  &  121.8  &  0.1  &  437.4 \\
1044+0655 & 2.1&  $S^{s+}_{\nu} $    & 13.0  & 1.5   &  0.2  &  45.4   &  0.03 &  183.4\\
1048+0141 & 0.7&  $S^{s+}_{\nu} $    & 12.5  & 2.5   &  0.2  &  71.5   &  0.03 &  261.8 \\
1146+5356 & 2.2&  $S^{s+}_{\nu} $    &  2.9  & 3.4   &  0.3  &  40.3   &  0.1  &  137.1 \\
1246-0730 & 1.3&  $S^{s+}_{\nu} $    & 12.5  & 2.3   &  0.2  &  84.7   &  0.1  &  292.6 \\
2147+0929 & 1.1&  $S^{s+}_{\nu} $    & 18.0  & 2.0   &  0.3  &  56.0   &  0.1  &  227.0 \\

\hline\hline
 Name &  z &  	SEDfit  &  $\theta$[mas]   & B [$\mu$G] \\
\hline
1312+5548& 1.5*& $S^{pls}_{\nu} $  & 1.4 & 12.0 \\

\hline\hline
 Name   &  z & SEDfit  & $\theta_{1}$[mas]   & B$_1$ [$\mu$G] &   $\theta_{2}$[mas]  &   B$_2$ [$\mu$G]\\
\hline
0243-0550 &1.8& $S^{pls}_{\nu} $  & 0.6  & 19.0   & 0.1  & 105.3  \\
0845+0439 &0.8& $S^{pls}_{\nu} $  & 0.3  & 62.0   & 0.1  & 338.0  \\
1405+0415 &3.2& $S^{pls}_{\nu} $  & 1.0  & 10.1   & 0.1  &  87.6  \\
1616+0459 &3.2& $S^{pls}_{\nu} $  & 0.3  & 35.0   & 0.1  &  85.3  \\

\hline\hline

\end{tabular}
\end{table*}%

Together with the characterisation of the type of the targets, considerations on the magnetic field have been done. 
Lower limits on the magnetic field have been computed from observable quantities. 
Assuming that the spectral peaks seen on the SEDs are due to SSA, the magnetic field B of an homogeneous synchrotron component can be derived using the following relation \cite[see e.g.,][]{Kellermann81}:

\begin{equation}
	B \sim \theta^{4} \nu_{max}^{5} S_{max}^{-2}(1+z)^{-1}, \qquad [\rm{G}]
\end{equation}

\noindent
where $\theta$ is the angular dimension [mas], S$_{max}$ and $\nu_{max}$ are the peak flux density [Jy] and the peak frequency [GHz] of the synchrotron component and \textit{z} is the redshift.

Since all our targets are unresolved for the FIRST beam, we do not have information on their angular size. We obtained lower limits for the angular sizes by considering the inverse Compton limit for which the brightness temperature (T$_{B}$) is assumed not exceeding its maximum value T$_{B}$ $\le$ 10$^{12}$ K \citep{Kellermann81}:

\begin{equation}
	T_{B} = 1.22\times10^{12}\frac{ S_{max}}{\theta^{2}\cdot \nu_{max}^{2}} \le 10^{12} \qquad [\rm{K}] ,
\end{equation} 

hence:

\begin{equation}
	\theta = 1.1\cdot \frac{\sqrt{S_{max}}}{ \nu_{max}} \qquad [\rm{mas}] ,
\end{equation} 
\noindent
From the lower limits in the angular size, we obtained lower limits for the magnetic field strength [$\mu$G] for those targets with synchrotron component/s in their SEDs. For the sources with no information of the redshift \textit{z}, the mean value of the sample (z$_{mean}$=1.5) was used in order to have a rough estimation on their magnetic field strength.
The lower limit values are listed in Tab. 4 and are, for most of the sources, in the range of 1-100 $\mu$G.

\subsection{Rotation Measure}
The polarisation angle $\chi$ for all the targets was calculated and, by combining several frequencies, their Rotation Measure (RM) estimated. In order to determine the RM value we fit a straight line (linear regression fit) to the plot of the EVPA versus $\lambda$$^{2}$.  The data at our disposal cover a wide range in $\lambda$ but with large gaps in the $\lambda$$^{2}$ coverage, thus the resulting RM values suffer from \textit{n$\pi$} ambiguity. The strategy we adopted was to trust the observed polarisation angle values at the highest frequency, which suffer less from Faraday rotation and to apply, where necessary, wrapping by some integer multiple of 180 degrees to the lower frequencies data (mostly to the 2.64 GHz and to the 4.85 GHz data). The maximum number of wraps we decided to apply had been fixed to five. 

In Tab. 2 the values of the observed RM (RM$_{obs}$) together with the rest frame RM (RM$_{rf}$ ), both given in rad/m$^{2}$, are listed. As known, our Milky Way introduces a RM contribution that varies with the galactic latitude. The high RM targets are mainly at latitudes |b| > 30$\deg$, thus above the galactic plane, but following pioneer works \citep{Kronberg76, Kronberg77}, the Galactic contribution should be subtracted to the observed RM value. Looking at the most recent foreground galactic RM (RM$_{mw}$) map by \cite{Oppermann15}, we noticed that the RM$_{mw}$ contributions at the positions of our sources, are very small compared to the observed RM$_{obs}$ of our targets. However, the uncertainties of the RM$_{mw}$ are quite large. Therefore, since the correction for the galactic contribution in our case would only increase the uncertainty of the measures without a significant correction on the RM value, we decided to not apply it. For the same reason we decided to ignore also the cosmic-web contribution that is only a few rad/m$^2$ too \citep{Akahori11}.
The source rest frame RM$_{rf}$ was then calculated following the relation:

\begin{equation}
	RM_{rf} = RM_{obs} \times (1+z)^{2} \qquad [\rm{rad/m^2}]
\end{equation}
\noindent
For those sources with unknown z, we used the approximate mean value of our sample (z$_{mean}$=1.5) in order to have at least an indication on the intrinsic RM of the source. 
In Appendix C all the plots showing the RM fit for the 30 targets, together with the SEDs and other polarisation information are shown.

The fitted data points for the RM almost always showed a linear regression with the lowest $\chi^2$ fit, expected for the goodness of the fit. 
However, for 10 sources ($\sim$ 33\% of the sample; 0239--0234, 0742+4900, 1043+2408, 1213+1307, 1616+2647, 1713+2813, 2050+0407, 2101+0341, 2200+0708, 2245+0324) we noticed a deviation from the $\lambda$$^{2}$ law. Among these, 6 sources ($\sim$ 60\%), are \textit{GPS-like}. This suggests that these compact sources are characterised by several Faraday screens intervening the medium. Indeed if the radiation passes through different magnetised plasma, the latter could rotate the polarisation angle differently, with the result of a non linear behaviour of the data. This explains, in general, the low-RM (observed) associated for the majority of these 10 sources ($\sim$ 90\%).

In Fig. 3 we show the distribution of the fitted RM (RM$_{obs}$) for our 30 high RM sources (red histogram) compared with the targets from the Farnes catalogue \citep{Farnes14} (blue histogram) and its cumulative plot. Both the histograms have been normalised with the number of the sources of each sample: 30 sources for our sample and 951 for the Farnes sample. For a better visibility of the whole distribution, the histograms are shown in logarithmic scale. \cite{Farnes14} shows a RM$_{obs}$ distribution that is centred to zero with a $\sigma$$\sim$100 rad/m$^2$. From the two histograms in Fig. 3 it is evident that the distribution of the |RM$_{obs}$| of our sample is different with respect to the sources chosen from the Farnes catalogue. The cumulative plot underlines that 80\% of the Farnes targets have a |RM$_{obs}$| value below 20 rad/m$^2$, while 80\% of our targets have a |RM$_{obs}$| value of $\gtrsim$ 100 rad/m$^2$.
In order to check the discrepancy, we also run the Kolmogorov-Smirnov test to the distributions and it gives a probability that the two distributions are comparable of 3$\times$10$^{-6}$; the two distributions are different at a confidence level > 95\%.

We also compared the RM for the three main object types we have found analysing the SEDs. Their cumulative plot of the |RM$_{obs}$| and the |RM$_{rf}$| are shown in Fig. 4. In Fig. 4a it is clear that the three groups are different and it seems that the $\textit{Mixed}$ targets are those with the largest values of |RM$_{obs}$|. Indeed, after the correction of the observed RM to its value in the rest frame (Fig. 4b), we can assert that all the $\textit{Mixed}$ targets show |RM$_{rf}$| higher than 1000 rad/m$^2$, while for the two others only 50\% are above 500 rad/m$^2$, still 5$\sigma$ away from the Farnes distribution. This seems to suggest that sources that show a mixed SED, with an old component at low frequency and compact components at high frequencies, i.e. radio sources that are restarting their activity, are related to high values of the RM. 

\begin{figure}[]
\begin{center}
\includegraphics[width=0.5\textwidth]{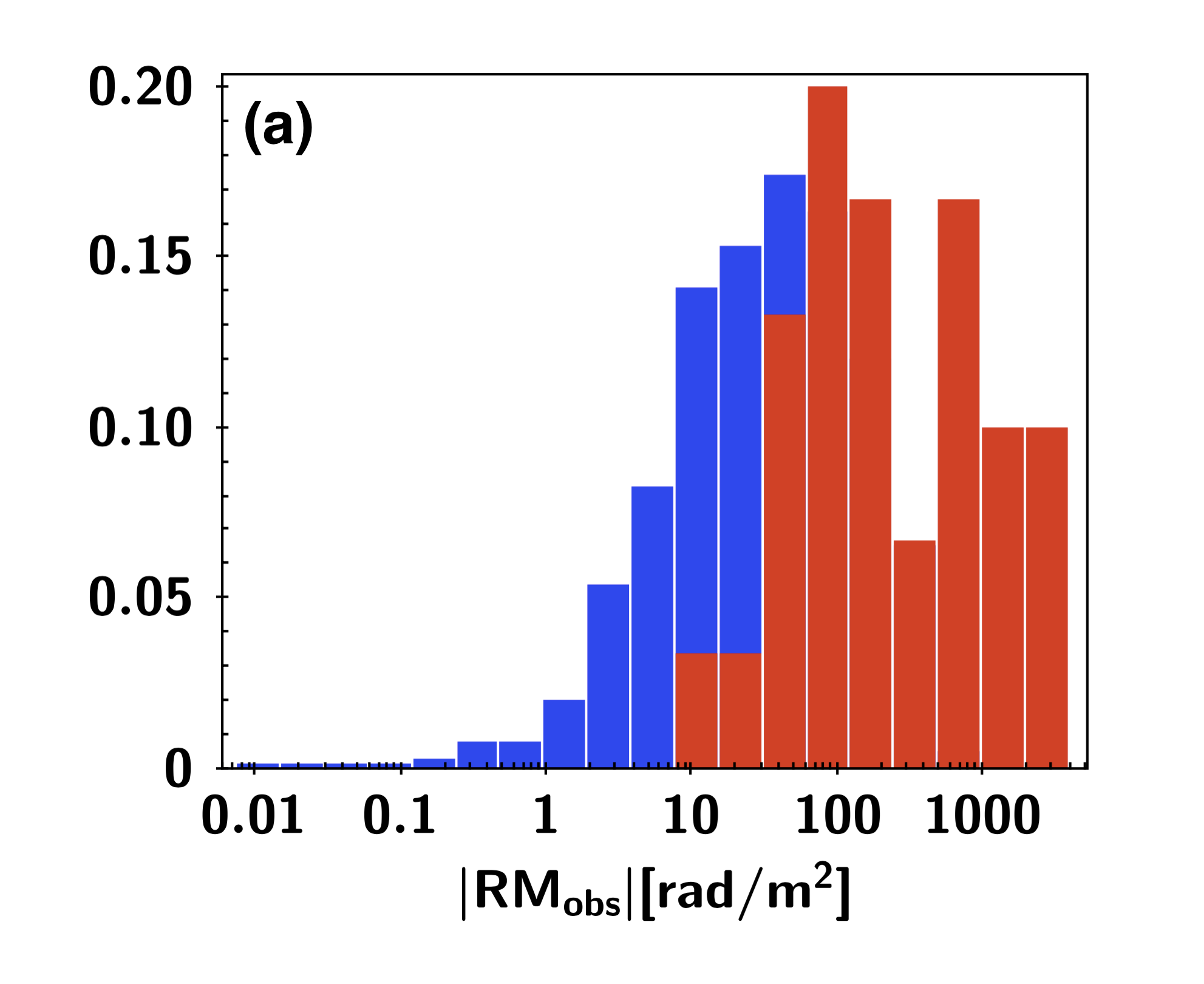}
\includegraphics[width=0.5\textwidth]{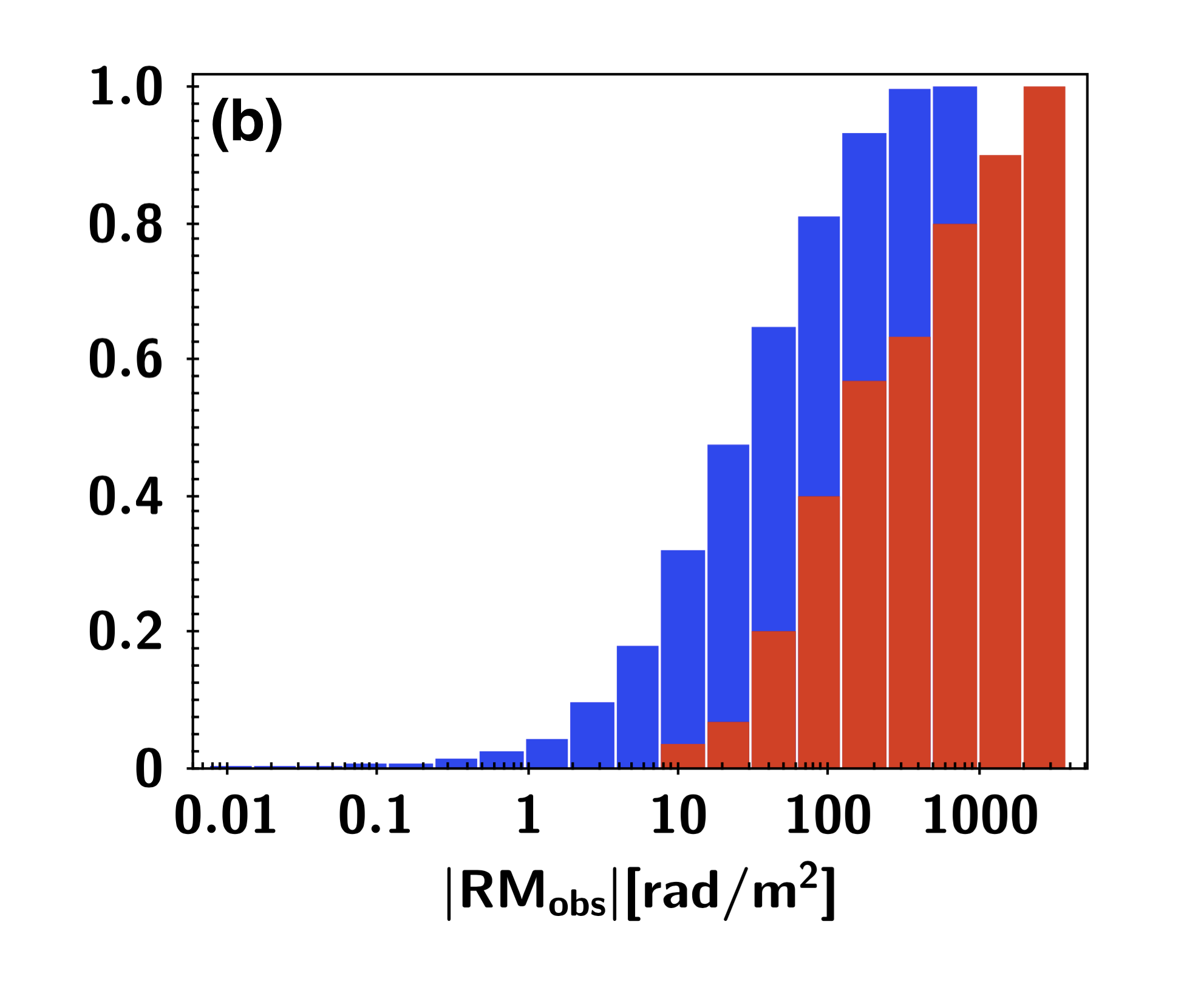}
\caption{a): histogram showing the distribution of the fitted rotation measure of our targets (red histogram) together with the Farnes catalogue (blue histogram). b): the same but cumulative.  
}
\label{default}
\end{center}
\end{figure}

\begin{figure}[]
\begin{center}
\includegraphics[width=0.5\textwidth]{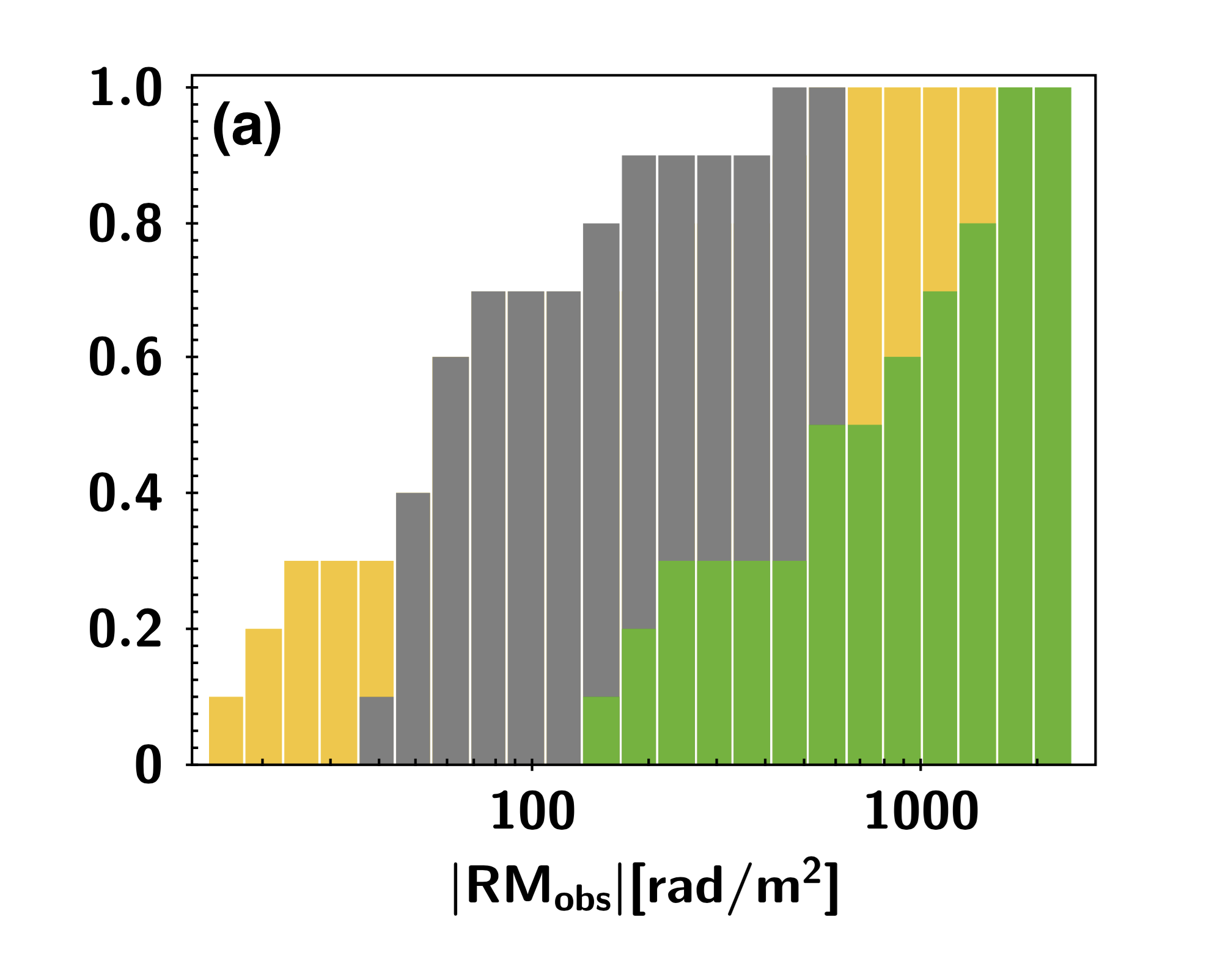}
\includegraphics[width=0.5\textwidth]{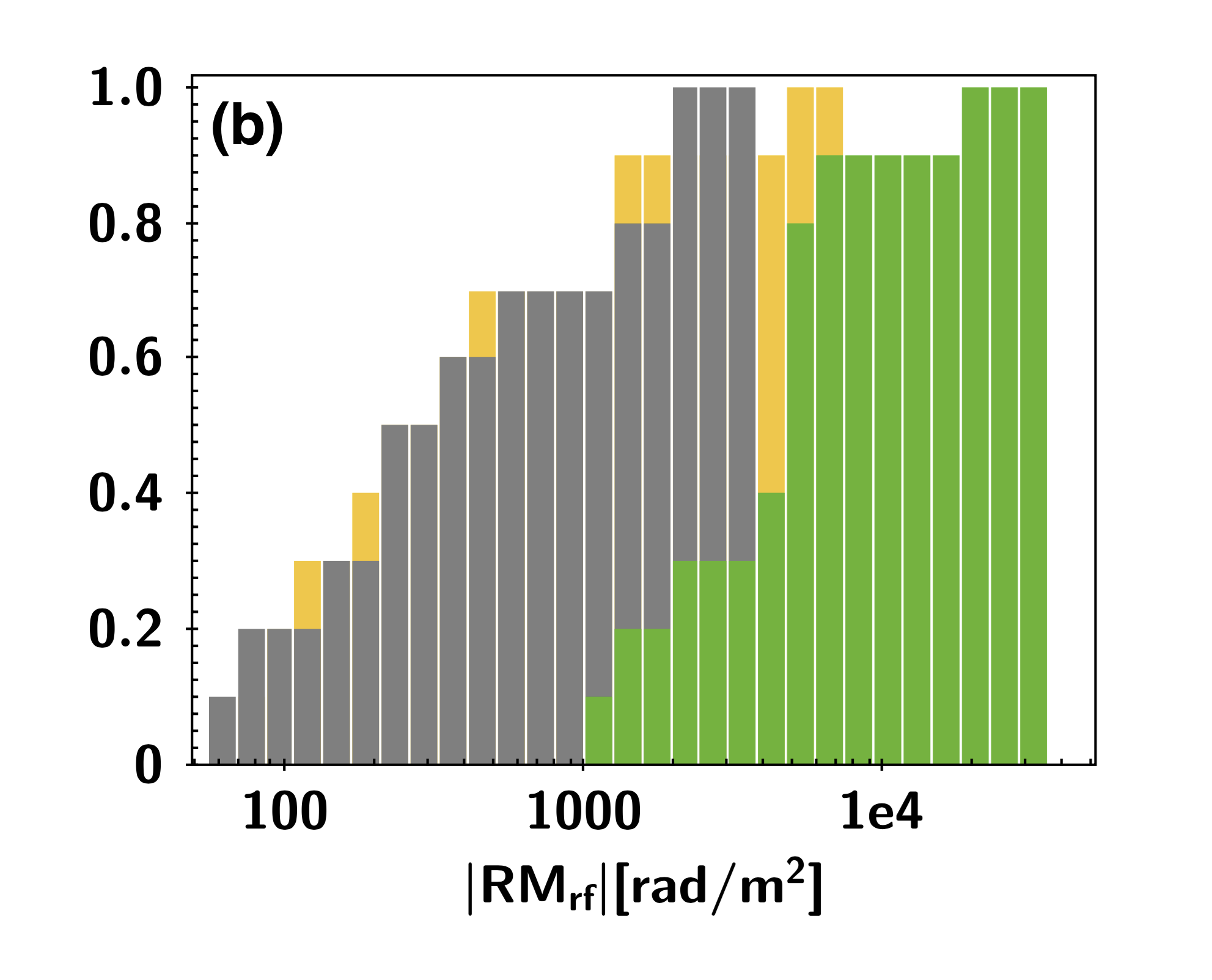}
\caption{(a) cumulative plot of the RM$_{obs}$ for the three object type. Yellow: $\textit{Older}$ type; grey: $\textit{GPS-like}$ type; green: $\textit{Mixed}$ type. (b) cumulative plot of the RM$_{rf}$ for the three objects types.}

\end{center}
\end{figure}

\subsection{Fractional polarisation, depolarisation and repolarisation}
The analysis of the fractional polarisation (\textit{m}), as a function of $\lambda^{2}$ indicates that the majority of the sources have a fractional polarisation that decreases with increasing wavelength (Saikia \& Salter 1988). This is indicative of a non-homogeneous medium that is present between the source and the observer.  

Several depolarisation (DP) models have been developed in order to explain the physical processes behind this behaviour.
The main models that one need to consider are: the \textit{Slab} model \citep{Burn66}, the \textit{Tribble} model \citep{Tribble91}, the \textit{Rossetti-Mantovani} model \citep{Rossetti08, Mantovani09} and the \textit{Repolariser} model \citep{Homan02, Mantovani09, Hovatta12}. Most of these models assume an optically thin emitting region and all of them make the assumption that we detect the same emitting region at each frequency. We saw from our SEDs that the majority of the sources have different synchrotron components contributing to the shape of the total intensity distribution. Thus, our unresolved sources can have several overlaps of optically thick and optically thin components together. This can, in the end, modifies the polarisation behaviour from the simplified way described from the models. 

As we cannot be sure that the polarised emitting region comes from the same emitting region within the frequency range and, due to the lack of a complete coverage along the bandwidth, we decided to follow a similar approach to that adopted by \cite{Farnes14}: we choose three models that are just a mathematical generalisation that can mimic the wavelength dependence of polarisation of the various physical models: a Gaussian (DP$_{Gauss}$), a power law (DP$_{PL}$) and a Gaussian with a constant term (DP$_{Gauss+}$).

\begin{equation}
	DP_{Gauss} = m_{0}\times \exp(-(\lambda-c_{1}){^2}/2c_{2}^{2})
\end{equation}

\begin{equation}
	DP_{PL} = m_{0}\times \lambda^{c_{3}}+b
\end{equation}

\begin{equation}
	DP_{Gauss+} = m_{0}\times \exp(-(\lambda-c_{1}){^2}/c_{2}^{2})+b
\end{equation}

\noindent
where $\lambda$ is in units of centimetres, m$_{0}$ in \%,  c$_{1}$ and c$_{2}$ in centimetres, c$_{3}$ is unit-less and b is a constant.
The Gaussian model is representative of the \textit{Slab} model, the power law model represents the \textit{Tribble} or, thanks to its flexibility, the \textit{Repolariser} model and finally the Gaussian with the constant term can provide the \textit{Rossetti-Mantovani} model.

Due to the different number of degrees of freedom for the various models the Gaussian and the power law models have been used whenever three or more data points were available and the Gaussian with constant model have been adopted when four or more data points were available. Proper fitting constraints were applied during the fitting process to ensure that only values of c$_{i}$ corresponding to physical solutions were obtained. 
Repolarised sources have been identified thanks to the flexibility of the power law model, which provides a polarisation spectral index c$_{3}$> 0. The determination of the residuals, together with a visual inspection, provides a measure for the goodness of the fit. See the Appendix C for the plots of the depolarisation models.   

Some of the sources ($\sim$ 30\% of the sample; see 0845+0439 as an example, Fig. 2) show a complex behaviour with a rising and decaying of the fractional polarisation; for them none of the three models fit properly the data. The behaviour of these sources can be better studied with better frequency sampled data and/or higher angular resolution observations.
Still from this first analysis it is possible to note that:
\begin{itemize}
\item 1/3 of the sample follow a model of a Gaussian, DP$_{Gauss}$, or a Gaussian with a constant term, DP$_{Gauss+}$.
\item 1/3 of the sample follow a power law model, DP$_{PL}$.
\item 1/3 of the sample cannot be fitted by any of the three models because of their complex shape, thus we can assert the existence of a complex model DP$_{complex}$ not proposed here, due to the lack of points for the fitting.
\end{itemize} 

We did not find any correlation between the DP models and the SEDs shapes or the RM values. 
Thus from these first single dish analysis on the polarisation fraction is not possible to indicate a possible trend that these objects are following.

\section{Summary and conclusions}
In this work, we presented the search for sources with a very high RM, as well as some follow-up studies of appropriate candidates. Assuming that a high RM causes strong in-band depolarisation, we observed a sample for unpolarised sources from the NVSS with the 100-m Effelsberg telescope at 10.45 GHz. After the identification of 30 potential high RM sources, we performed observations at 2.64, 4.85, 8.35, 10.45 and 14.60 GHz as well, in order to determine the SED and the RM of the sources. In some cases, exceptionally high RM were found, e.g. in 1616+0459 where its RM$_{obs}$ of $\sim$ 2550 rad/m$^{2}$ corresponds to a RM$_{rf}$ of $\sim$ 44300 rad/m$^{2}$. Our main conclusions can be summarised as follows.
\begin{itemize}
\item From our statistical study it turns out that the high RM candidates are not characterised by a specific object type. Indeed the spectral index distribution of the 30 high RM candidates shows three peaks representing all the possible object type (steep, flat and inverted spectrum radio sources). 

\item SEDs have been characterised and three groups (\textit{Older}, \textit{GPS-like}, \textit{Mixed}) have been identified.
high RM candidates are mainly (66\%) sources with compact high frequency components, probably new growing radio components, thus objects in a particular compact young phase (the $\textit{GPS-like}$) or in a reactivated activity phase ($\textit{Mixed}$).
The $\textit{Older}$ sources showing high RM could be surrounded by a very dense intervening material.

\item Lower limits for the magnetic field strength have been calculated for those sources with well defined synchrotron component/s in their SEDs. Due to the lack of angular resolution, no evidence of strong magnetic fields have been detected.
 
\item The behaviour of the fractional polarisation \textit{m} with $\lambda^{2}$ have been fitted using simpler mathematical representations of the main physical depolarisation models:  the \textit{Slab} model, the \textit{Tribble} model, the \textit{Repolariser} and the \textit{Rossetti-Mantovani} model.
Three groups have been detected: sources following a Gaussian or Gaussian with a constant term model, sources following a power law model and sources having a complex behaviour for which we cannot fit any of the model because of their complex shape. No correlation between the SED type and the fractional polarisation behaviour has been found. 

\item The RM has been determined for the all the sources of the sample, among these, 11 sources show a deviation from a simple linear behaviour. This feature could be an indication that several Faraday screens within the medium are affecting differently the polarisation angle distribution. 
Due to the lack of well sample data, we could not reconstruct their real behaviour. 

\item A strong correlation between the \textit{mixed} SEDs and a high value of the RM has been found. Indeed the \textit{mixed} sources have all a RM$_{rf}$ larger than 1000 rad/m$^{2}$. This could be an indication that these particular sources, showing a restarting phase at high frequency, are characterised by a really dense and/or a magnetised medium that strongly rotates the polarisation angle. 

\end{itemize}
Extreme cases, i.e. all the sources with |RM$_{obs}$|$\geq$500 rad/m$^2$, are being studied by us with interferometric technique, using the JVLA, VLBA and EVN interferometers. The results will be published in future papers.  

\begin{acknowledgements}
Based on observations with the 100-m telescope of the MPIfR (Max-Planck-Institut f\"ur Radioastronomie) at Effelsberg.
This research has made use of the NASA/IPAC Extragalactic Database (NED) which is operated by the Jet Propulsion Laboratory, California Institute of Technology, under contract with the National Aeronautics and Space Administration.
AP is a member of the International Max Planck Research School (IMPRS) for Astronomy and Astrophysics
at the Universities of Bonn and Cologne.
C.C-G. acknowledges support by UNAM-DGAPA-PAPIIT grant number IA101214.
We thank the useful help from the Effelsberg operators.
We acknowledge Roberto Fanti for useful discussions and suggestions and Jochen Heidt for providing us with spectral redshifts of some of the sources.
We thank the anonymous referee for careful revising of the paper and useful comments.
\end{acknowledgements}

\bibliographystyle{aa} 
\bibliography{EffhighRMpaper} 


\Online
\clearpage

\begin{appendix} 
\section{}
\begin{table*}[h]
\small
\centering
\caption{Table of values. For the source 0243-0550 we could provide fluxes only at 10.45 GHz and at 4.85 GHz. For the latter we split the bandwidth (500 MHz) into 3 sub-bands, each of them are $\sim$160 MHz wide. The Effelsberg data are those in the 2.64 to 14.60 GHz range. Values at 1.4 GHz are taken from the NVSS survey, values at 360 MHz cm are taken from the TEXAS survey, values at 320 MHz are taken from the WENSS survey, values at 150 MHz are taken from the 7C survey and values at 74 MHz are taken from the VLSS survey. Upper limits are indicated with the < symbol and the data non-available with the -- symbol.}
\begin{tabular}{|l|c|c|c|c|c|}
\hline
Name       &$\nu$ [GHz]&  S [mJy]          & S$_{Pol}$ [mJy]       &     m [\%]           & $\chi$ [deg]        \\    
\hline
           &  ~~14.60     &  650  $\pm$ 10   & 55  $\pm$ 4 &  8.4   $\pm$ 0.6  &    93.9  $\pm$   2.0 \\    
           &  ~~10.45     &  723  $\pm$  6   & 51  $\pm$ 1 &  7.0   $\pm$ 0.1  &    88.9  $\pm$   0.4 \\    
           &  ~~8.35      &  740  $\pm$  6   & 44  $\pm$ 1 &  6.0   $\pm$ 0.2  &    88.9  $\pm$   0.6 \\    
           &  ~~4.85      &  666  $\pm$  4   & 20  $\pm$ 1 &  3.0   $\pm$ 0.2  &    79.5  $\pm$   2.4 \\    
0239-0234  &  ~~2.64    &  548  $\pm$  7   & 20  $\pm$ 3 &  3.6   $\pm$ 0.6  &    62.8  $\pm$   4.4 \\    
           &  ~~1.40    &  300  $\pm$ 10   &   --        &  --    &     --     \\ 
           &  ~~~~0.36   &  250  $\pm$ 22   &   --        &  --    &     --     \\   
           &  ~~~~0.32   &      --          &   --        &  --    &     --     \\  
           &  ~~~~0.15   &      --          &   --        &  --    &     --     \\ 
           &  ~~~~0.07   &  <300            &   --        &  --    &     --     \\   
\hline
           & 10.45    &  548  $\pm$   2  &  9  $\pm$ 2 &  1.6   $\pm$ 0.4  &     1.0  $\pm$   9.9  \\    
           &  5.00    &  673  $\pm$  14  & 13  $\pm$ 4 &  1.9   $\pm$ 0.6  &   102.1  $\pm$  10.4  \\    
           &  4.80    &  691  $\pm$  15  & 13  $\pm$ 4 &  1.9   $\pm$ 0.6  &   109.3  $\pm$   9.5  \\    
           &  4.60    &  647  $\pm$   3  & 12  $\pm$ 1 &  1.8   $\pm$ 0.1  &   117.5  $\pm$   2.8  \\    
0243-0550  &  1.40    &  560  $\pm$  20  &   --     &  --    &     --     \\ 
           &  0.36   &  390  $\pm$  25  &   --     &  --    &     --     \\  
           &  0.32   &      --          &   --     &  --    &     --     \\  
           &  0.15   &      --          &   --     &  --    &     --     \\  
           &  0.07   &  990  $\pm$  160 &   --     &  --    &     --     \\  
\hline  
           &  14.60   &  422 $\pm$    3  & 15  $\pm$ 4 &  3.6   $\pm$ 0.8  &   -17.2  $\pm$   6.8 \\   
           &  10.45   &  430 $\pm$    1  & 16  $\pm$ 1 &  3.8   $\pm$ 0.2  &   -14.2  $\pm$   0.9 \\   
           &  8.35    &  432 $\pm$    2  & 12  $\pm$ 1 &  2.8   $\pm$ 0.2  &   -16.7  $\pm$   1.8 \\   
           &  4.85    &  416 $\pm$    2  &  6  $\pm$ 2 &  1.4   $\pm$ 0.4  &   -53.8  $\pm$   6.3 \\   
0742+4900  &  2.64    &  489 $\pm$    4  &   --     &  --    &     --      \\   
           &  1.40    &  398 $\pm$   12  &   --     &  --    &     --     \\ 
           &  0.36   &      --          &   --     &  --    &     --     \\ 
           &  0.32   &  127 $\pm$    4  &   --     &  --    &     --     \\ 
           &  0.15   & <78              &   --     &  --    &     --     \\ 
           &  0.07   & <300             &   --     &  --    &     --     \\ 
\hline
           &  14.60   &   57 $\pm$    6  & 14  $\pm$ 4 & 23.8   $\pm$ 7.8  &   -22.2  $\pm$   8.9 \\ 
           &  10.45   &   82 $\pm$    2  & 11  $\pm$ 3 & 13.3   $\pm$ 3.2  &   -21.9  $\pm$   6.9 \\ 
           &  8.35    &  106 $\pm$    1  &  9  $\pm$ 1 &  8.2   $\pm$ 0.7  &    -1.5  $\pm$   2.5 \\ 
           &  4.85    &  193 $\pm$    2  &  5  $\pm$ 1 &  2.5   $\pm$ 0.6  &    58.2  $\pm$   7.0 \\ 
0751+2716  &  2.64    &  325 $\pm$    3  &   --     &  --    &     --      \\ 
           &  1.40    &  590 $\pm$   20  &   --     &  --    &     --     \\
           &  0.36   & 1470 $\pm$   80  &   --     &  --    &     --     \\ 
           &  0.32   &        --        &   --     &  --    &     --     \\ 
           &  0.15   &        --        &   --     &  --    &     --     \\ 
           &  0.07   & 890  $\pm$  120  &   --     &  --    &     --     \\ 
\hline
           &  14.60   & 735  $\pm$  43   & 18  $\pm$ 4 &  2.4   $\pm$ 0.5  &    55.2  $\pm$  11.7  \\ 
           &  10.45   & 682  $\pm$  52   & 19  $\pm$ 1 &  2.8   $\pm$ 0.1  &   109.1  $\pm$   1.6  \\ 
           &  8.35    & 654  $\pm$   5   &  8  $\pm$ 1 &  1.2   $\pm$ 0.1  &   146.4  $\pm$   3.5  \\ 
           &  4.85    & 560  $\pm$   3   & 16  $\pm$ 2 &  2.7   $\pm$ 0.2  &   401.5  $\pm$   5.2  \\ 
0845+0439  &  2.64    & 519  $\pm$  13   &  3  $\pm$ 1 &  0.5   $\pm$ 0.2  &   1364.5 $\pm$   10.9 \\ 
           &  1.40    & 380  $\pm$  10   &   --     &  --    &     --     \\
           &  0.36   &       --         &   --     &  --    &     --     \\
           &  0.32   &      --          &   --     &  --    &     --     \\
           &  0.15   &      --          &   --     &  --    &     --     \\
           &  0.07   & 880  $\pm$  120  &   --     &  --    &     --     \\
\hline

\end{tabular}
\end{table*}%

\begin{table*}[h]
\small
\centering
\caption{Table of values. Continue. }
\begin{tabular}{|l|c|c|c|c|c|}
\hline
Name       &$\nu$ [GHz]&  S [mJy]          & S$_{Pol}$ [mJy]       &     m [\%]           & $\chi$ [deg]        \\    
\hline
           &  14.60   &  63  $\pm$   2   &   --     &  --    &    --      \\  
           &  10.45   &  97  $\pm$   4   & 6  $\pm$ 2 &  6.6   $\pm$ 2.4  &    41.1  $\pm$   8.0  \\  
           &  8.35    & 121  $\pm$   1   & 6  $\pm$ 1 &  5.0   $\pm$ 0.6  &    38.1  $\pm$   3.6  \\  
           &  4.85    & 181  $\pm$   2   & 7  $\pm$ 1 &  3.6   $\pm$ 0.6  &    31.1  $\pm$   6.1  \\  
0925+3159  &  2.64    & 316  $\pm$   4   &   --     &  --    &    --      \\ 
           &  1.40    & 551  $\pm$  17   &   --     &  --    &     --     \\
           &  0.36   & 1800 $\pm$  30   &   --     &  --    &     --     \\ 
           &  0.32   & 1970 $\pm$   4   &   --     &  --    &     --     \\ 
           &  0.15   & 3315 $\pm$  50   &   --     &  --    &     --     \\ 
           &  0.07   & 6160 $\pm$  650  &   --     &  --    &     --     \\ 
\hline
           &  14.60   & 584  $\pm$  15   &    --     &  --    &    --      \\  
           &  10.45   & 660  $\pm$   6   &  9  $\pm$ 3 &  1.2   $\pm$ 0.4  &    47.7  $\pm$ 122.8 \\   
           &  8.35    & 616  $\pm$   6   & 17  $\pm$ 5 &  2.8   $\pm$ 0.7  &   106.0  $\pm$   6.9 \\   
           &  4.85    & 830  $\pm$   3   & 22  $\pm$ 2 &  2.6   $\pm$ 0.2  &   396.7  $\pm$   5.1 \\   
0958+3224  &  2.64    & 1100 $\pm$  13   &    --     &  --    &    --      \\   
           &  1.40    & 1250 $\pm$  40   &   --     &  --    &     --     \\
           &  0.36   & 3630 $\pm$  70   &   --     &  --    &     --     \\  
           &  0.32   & 3880 $\pm$  10   &   --     &  --    &     --     \\  
           &  0.15   & 5060 $\pm$  700  &   --     &  --    &     --     \\  
           &  0.07   & 7000 $\pm$  740  &   --     &  --    &     --     \\  
\hline
           &  14.60   &   74 $\pm$    2  & 8  $\pm$ 2 & 10.7   $\pm$ 4.1  &   27.5   $\pm$ 10.8  \\  
           &  10.45   &   91 $\pm$    3  & 9  $\pm$ 2 & 10.2   $\pm$ 2.5  &   35.4   $\pm$  6.4  \\  
           &  8.35    &  105 $\pm$    1  & 8  $\pm$ 1 &  8.0   $\pm$ 0.6  &   36.6   $\pm$  2.3  \\  
           &  4.85    &  166 $\pm$    2  & 6  $\pm$ 1 &  3.8   $\pm$ 0.6  &   71.5   $\pm$  7.0  \\  
1015+0318  &  2.64    &  244 $\pm$    2  &   --     &  --    &    --      \\   
           &  1.40    &  416 $\pm$   13  &   --     &  --    &     --     \\
           &  0.36   &  933 $\pm$   30  &   --     &  --    &     --     \\  
           &  0.32   &      --          &   --     &  --    &     --     \\
           &  0.15   &      --          &   --     &  --    &     --     \\
           &  0.07   & 2120 $\pm$   230 &   --     &  --    &     --     \\  
\hline
           &  14.60   & 1170 $\pm$   10  & 60  $\pm$ 4 &  5.2   $\pm$ 0.3  &    92.9  $\pm$   1.8 \\  
           &  10.45   & 1070 $\pm$   10  & 67  $\pm$ 2 &  6.4   $\pm$ 0.2  &    93.7  $\pm$   1.4 \\  
           &  8.35    & 1050 $\pm$   10  & 42  $\pm$ 1 &  4.0   $\pm$ 0.1  &    82.1  $\pm$   0.5 \\  
           &  4.85    &  903 $\pm$    4  & 32  $\pm$ 1 &  3.5   $\pm$ 0.2  &    69.9  $\pm$   1.4 \\  
1043+2408  &  2.64    &  675 $\pm$    3  &  8  $\pm$ 2 &  1.1   $\pm$ 0.3  &    85.7  $\pm$  6.7  \\  
           &  1.40    &  320 $\pm$   10  &   --     &  --    &     --     \\
           &  0.36   &  460 $\pm$   76  &   --     &  --    &     --     \\
           &  0.32   &        --        &   --     &  --    &     --     \\
           &  0.15   &        --        &   --     &  --    &     --     \\
           &  0.07   &  <300            &   --     &  --    &     --     \\
\hline
           &  14.60   &  279 $\pm$   3  & 24  $\pm$ 4 & 8.5   $\pm$ 1.3  &    139.4 $\pm$    4.5\\   
           &  10.45   &  295 $\pm$   1  & 27  $\pm$ 1 & 9.3   $\pm$ 0.3  &    130.9 $\pm$    1.3\\   
           &  8.35    &  314 $\pm$   2  & 26  $\pm$ 1 & 8.4   $\pm$ 0.2  &    126.3 $\pm$    0.8\\   
           &  4.85    &  347 $\pm$   2  & 20  $\pm$ 1 & 5.6   $\pm$ 0.3  &    97.1  $\pm$   5.2 \\  
1044+0655  &  2.64    &  387 $\pm$   5  &  6  $\pm$ 2 & 1.4   $\pm$ 0.6  &    -9.2  $\pm$  12.8 \\  
           &  1.40    &  490 $\pm$   20 &   --     &  --    &     --     \\
           &  0.36   &  845 $\pm$   27 &   --     &  --    &     --     \\  
           &  0.32   &        --       &   --     &  --    &     --     \\
           &  0.15   &        --       &   --     &  --    &     --     \\
           &  0.07   & 1310 $\pm$160   &   --     &  --    &     --     \\  
\hline

\end{tabular}
\end{table*}%

\begin{table*}[h]
\small
\centering
\caption{Table of values. Continue. }
\begin{tabular}{|l|c|c|c|c|c|}
\hline
Name       &$\nu$ [GHz]&  S [mJy]          & S$_{Pol}$ [mJy]       &     m [\%]           & $\chi$ [deg]        \\    
\hline
           &  14.60   & 273  $\pm$   4  &  8  $\pm$ 1 &  2.7   $\pm$ 0.3  &   182.6  $\pm$   2.1 \\ 
           &  10.45   & 328  $\pm$   4  & 11  $\pm$ 2 &  3.2   $\pm$ 0.7  &   105.6  $\pm$   6.9 \\   
           &  8.35    & 342  $\pm$   3  &  7  $\pm$ 1 &  1.9   $\pm$ 0.3  &    71.4  $\pm$   4.5 \\   
           &  4.85    & 408  $\pm$   2  & 12  $\pm$ 1 &  2.8   $\pm$ 0.2  &  -282.4  $\pm$   5.7 \\ 
1048+0141  &  2.64    & 465  $\pm$   6  &   --     &  --    &    --      \\    
           &  1.40    & 380  $\pm$  10  &   --     &  --    &     --     \\ 
           &  0.36   & 554  $\pm$  43  &   --     &  --    &     --     \\   
           &  0.32   &       --        &   --     &  --    &     --     \\ 
           &  0.15   &       --        &   --     &  --    &     --     \\ 
           &  0.07   & 1050$\pm$150    &   --     &  --    &     --     \\   
\hline
           &  14.60   & 552  $\pm$   4   & 11  $\pm$ 4 &  2.1   $\pm$ 0.7  &    61.3  $\pm$   8.9 \\   
           &  10.45   & 614  $\pm$  13   & 15  $\pm$ 1 &  2.4   $\pm$ 0.1  &    45.3  $\pm$   1.0 \\   
           &  8.35    & 631  $\pm$   3   & 15  $\pm$ 1 &  2.4   $\pm$ 0.1  &    34.7  $\pm$   1.5 \\   
           &  4.85    & 630  $\pm$   2   & 18  $\pm$ 1 &  3.0   $\pm$ 0.2  &   -17.5  $\pm$   5.2 \\   
1146+5356  &  2.64    & 578  $\pm$   7   & 12  $\pm$ 2 &  2.1   $\pm$ 0.4  &  -244.5  $\pm$   5.2 \\  
           &  1.40    & 367  $\pm$  11   &   --     &  --    &     --     \\
           &  0.36   & 340  $\pm$  20   &   --     &  --    &     --     \\ 
           &  0.32   & 398  $\pm$   4   &   --     &  --    &     --     \\ 
           &  0.15   & 286  $\pm$  20   &   --     &  --    &     --     \\ 
           &  0.07   & <300             &   --     &  --    &     --     \\
\hline
           &  14.60   & 332  $\pm$   3   & 13  $\pm$ 3 &  4.0   $\pm$ 1.0  &    51.0  $\pm$   7.4  \\  
           &  10.45   & 421  $\pm$   2   & 15  $\pm$ 1 &  3.5   $\pm$ 0.3  &    53.5  $\pm$   1.9  \\  
           &  8.35    & 487  $\pm$   2   & 18  $\pm$ 1 &  3.7   $\pm$ 0.1  &    55.7  $\pm$   1.0  \\  
           &  4.85    & 693  $\pm$   3   & 23  $\pm$ 2 &  3.3   $\pm$ 0.2  &    58.8  $\pm$   1.5  \\                                                                                                                                                                                      
1213+1307  &  2.64    & 976  $\pm$   3   & 26  $\pm$ 1 &  2.7   $\pm$ 0.0  &    67.2  $\pm$   0.8  \\  
           &  1.40    & 1340 $\pm$  40   &   --     &  --    &     --     \\
           &  0.36   & 2520 $\pm$  70   &   --     &  --    &     --     \\ 
           &  0.32   &      --          &   --     &  --    &     --     \\
           &  0.15   &      --          &   --     &  --    &     --     \\
           &  0.07   & 4890 $\pm$  510  &   --     &  --    &     --     \\ 
\hline
           &  14.60   & 1030 $\pm$  08   & 13  $\pm$ 1 &  1.2   $\pm$ 0.1  &   -4.7   $\pm$  6.7   \\ 
           &  10.45   & 1040 $\pm$  23   & 12  $\pm$ 1 &  1.1   $\pm$ 0.1  &    9.1   $\pm$  4.6   \\ 
           &  8.35    & 965  $\pm$   8   &  8  $\pm$ 1 &  0.8   $\pm$ 0.1  &   29.1   $\pm$  3.6   \\ 
           &  4.85    & 917  $\pm$   4   &  9  $\pm$ 1 &  1.0   $\pm$ 0.2  &  180.3   $\pm$  6.4   \\ 
1246-0730  &  2.64    & 686  $\pm$   9   & 13  $\pm$ 3 &  1.8   $\pm$ 0.4  &  576.8   $\pm$  6.2   \\ 
           &  1.40    & 550  $\pm$  20   &   --     &  --    &     --     \\
           &  0.36   & 948  $\pm$  37   &   --     &  --    &     --     \\
           &  0.32   &        --        &   --     &  --    &     --     \\
           &  0.15   &        --        &   --     &  --    &     --     \\
           &  0.07   & 1280 $\pm$  160  &   --     &  --    &     --     \\
\hline
           &  14.60   & 150  $\pm$   3   & 10  $\pm$ 2 &  6.8   $\pm$ 1.0  &  99.8    $\pm$ 3.8  \\  
           &  10.45   & 207  $\pm$   1   & 11  $\pm$ 1 &  5.1   $\pm$ 0.4  & 115.9    $\pm$ 3.4  \\  
           &  8.35    & 252  $\pm$   2   & 14  $\pm$ 1 &  5.4   $\pm$ 0.3  & 136.0    $\pm$ 2.0  \\  
           &  4.85    & 414  $\pm$   3   &  7  $\pm$ 2 &  1.6   $\pm$ 0.4  & 154.4    $\pm$ 6.6  \\  
1311+1417  &  2.64    & 614  $\pm$   8   &  4  $\pm$ 1 &  0.6   $\pm$ 0.1  & 494.5    $\pm$ 7.2  \\  
           &  1.40    & 734  $\pm$  22   &   --     &  --    &     --     \\  
           &  0.36   & 291  $\pm$  25   &   --     &  --    &     --     \\    
           &  0.32   &      --          &   --     &  --    &     --     \\  
           &  0.15   &      --          &   --     &  --    &     --     \\  
           &  0.07   & <300             &   --     &  --    &     --     \\    
\hline

\end{tabular}
\end{table*}%

\begin{table*}[h]
\small
\centering
\caption{Table of values. Continue. }
\begin{tabular}{|l|c|c|c|c|c|}
\hline
Name       &$\nu$ [GHz]&  S [mJy]          & S$_{Pol}$ [mJy]       &     m [\%]           & $\chi$ [deg]        \\    
\hline
           &  14.60   & 106  $\pm$   3   & 5  $\pm$ 1 &  4.5   $\pm$ 2.3  &     1.1  $\pm$  15.1\\  
           &  10.45   & 140  $\pm$   4   & 6  $\pm$ 1 &  4.0   $\pm$ 0.7  &    -8.1  $\pm$   0.1\\  
           &  8.35    & 164  $\pm$   1   & 4  $\pm$ 1 &  2.6   $\pm$ 0.4  &   -35.4  $\pm$   4.5\\  
           &  4.85    & 253  $\pm$   2   &   --     &  --    &    --      \\  
1312+5548  &  2.64    & 390  $\pm$   5   &   --     &  --    &    --      \\
           &  1.40    & 590  $\pm$  20   &   --     &  --    &     --     \\ 
           &  0.36   & 491  $\pm$  36   &   --     &  --    &     --     \\   
           &  0.32   & 410  $\pm$   4   &   --     &  --    &     --     \\   
           &  0.15   & 198  $\pm$  30	 &   --     &  --    &     --     \\  
           &  0.07   & <300             &   --     &  --    &     --     \\   
\hline
           &  14.60   & 283  $\pm$   4   & 5  $\pm$ 1 &  1.8   $\pm$ 0.5  &    93.0  $\pm$   8.3 \\ 
           &  10.45   & 308  $\pm$   5   & 9  $\pm$ 2 &  2.8   $\pm$ 0.7  &    92.1  $\pm$  10.7 \\ 
           &  8.35    & 317  $\pm$   2   & 8  $\pm$ 1 &  2.3   $\pm$ 0.3  &    94.4  $\pm$   3.6 \\ 
           &  4.85    & 297  $\pm$   2   & 8  $\pm$ 1 &  2.6   $\pm$ 0.4  &    99.1  $\pm$   6.5 \\ 
1351+0830  &  2.64    & 290  $\pm$   4   & 6  $\pm$ 2 &  2.0   $\pm$ 0.8  &   140.5  $\pm$  12.0 \\ 
           &  1.40    & 350  $\pm$  10   &   --     &  --    &     --     \\
           &  0.36   &      --          &   --     &  --    &     --     \\ 
           &  0.32   &      --          &   --     &  --    &     --     \\ 
           &  0.15   &      --          &   --     &  --    &     --     \\ 
           &  0.07   & <300             &   --     &  --    &     --     \\   
\hline
           &  14.60   & 723  $\pm$   5   &   --     &  --    &    --      \\  
           &  10.45   & 712  $\pm$  10   & 16  $\pm$ 5 &  2.3   $\pm$ 0.6  &   2.2    $\pm$ 7.3   \\ 
           &  8.35    & 772  $\pm$   3   & 13  $\pm$ 1 &  1.7   $\pm$ 0.1  &   9.3    $\pm$ 1.6   \\ 
           &  4.85    & 803  $\pm$   3   & 23  $\pm$ 1 &  2.8   $\pm$ 0.1  &   185.6  $\pm$   2.1 \\ 
1405+0415  &  2.64    & 893  $\pm$  11   & 33  $\pm$ 3 &  3.6   $\pm$ 0.2  &   728.3  $\pm$   2.0 \\ 
           &  1.40    & 930  $\pm$  30   &   --     &  --    &     --     \\ 
           &  0.36   &1240  $\pm$  30   &   --     &  --    &     --     \\   
           &  0.32   &        --        &   --     &  --    &     --     \\ 
           &  0.15   &        --        &   --     &  --    &     --     \\ 
           &  0.07   &3170  $\pm$  330  &   --     &  --    &     --     \\   
\hline
           &  14.60   & 112  $\pm$   2   & 13  $\pm$  4 & 11.8   $\pm$ 3.3  &   121.6  $\pm$   8.1 \\ 
           &  10.45   & 146  $\pm$   1   & 11  $\pm$  1 &  7.3   $\pm$ 0.3  &   126.2  $\pm$   0.1 \\ 
           &  8.35    & 175  $\pm$   1   &  9  $\pm$  1 &  5.2   $\pm$ 0.4  &   129.8  $\pm$   2.4 \\ 
           &  4.85    & 259  $\pm$   2   &  8  $\pm$  2 &  3.1   $\pm$ 0.6  &   134.3 $\pm$   4.4 \\ 
1435-0414  &  2.64    & 375  $\pm$   6   &   --     &  --    &    --      \\
           &  1.40    & 480  $\pm$  10   &   --     &  --    &     --     \\
           &  0.36   & 753  $\pm$  36   &   --     &  --    &     --     \\  
           &  0.32   &      --          &   --     &  --    &     --     \\
           &  0.15   &      --          &   --     &  --    &     --     \\
           &  0.07   & 1070 $\pm$ 160   &   --     &  --    &     --     \\  
\hline
           &  14.60   &  733 $\pm$    6  & 11  $\pm$ 4 &  1.5   $\pm$ 0.5  &    90.2  $\pm$   8.9\\ 
           &  10.45   &  813 $\pm$    1  & 14  $\pm$ 1 &  1.7   $\pm$ 0.1  &    99.5  $\pm$   2.1\\ 
           &  8.35    &  844 $\pm$    4  & 12  $\pm$ 1 &  1.3   $\pm$ 0.1  &   106.1  $\pm$   1.9\\ 
           &  4.85    &  812 $\pm$    3  &  5  $\pm$ 1 &  0.6   $\pm$ 0.2  &   117.1  $\pm$   8.6\\ 
1549+5038  &  2.64    &  648 $\pm$    8  &   --     &  --    &    --      \\ 
           &  1.40    &  630 $\pm$   20  &   --     &  --    &     --     \\ 
           &  0.36   &  397 $\pm$   35  &   --     &  --    &     --     \\ 
           &  0.32   &  348 $\pm$    4  &   --     &  --    &     --     \\ 
           &  0.15   & <69              &   --     &  --    &     --     \\ 
           &  0.07   & <300             &   --     &  --    &     --     \\ 
\hline

\end{tabular}
\end{table*}%

\begin{table*}[h]
\small
\centering
\caption{Table of values. Continue. }
\begin{tabular}{|l|c|c|c|c|c|}
\hline
Name       &$\nu$ [GHz]&  S [mJy]          & S$_{Pol}$ [mJy]       &     m [\%]           & $\chi$ [deg]        \\    
\hline
           &  14.60   &  885 $\pm$    3  & 23  $\pm$ 1 &  2.6   $\pm$ 0.1  & -55.0    $\pm$ 1.6  \\ 
           &  10.45   & 1035 $\pm$   10  & 16  $\pm$ 1 &  1.5   $\pm$ 0.1  &  -1.0    $\pm$ 1.4  \\ 
           &  8.35    & 1160 $\pm$    9  & 19  $\pm$ 1 &  1.6   $\pm$ 0.1  &  70.6    $\pm$ 1.5  \\ 
           &  4.85    & 1140 $\pm$    5  &   --     &  --    &    --      \\          
1616+0459  &  2.64    &  777 $\pm$    9  &   --     &  --    &    --      \\  
           &  1.40    &  330 $\pm$   10  &   --     &  --    &     --     \\
           &  0.36   &  301 $\pm$   50  &   --     &  --    &     --     \\ 
           &  0.32   &      --          &   --     &  --    &     --     \\
           &  0.15   &      --          &   --     &  --    &     --     \\
           &  0.07   &1060  $\pm$   140 &   --     &  --    &     --     \\ 
\hline
           &  14.60   & 216  $\pm$   1   & 15  $\pm$ 1 &  6.7   $\pm$ 0.4  &  97.0    $\pm$ 3.3  \\  
           &  10.45   & 294  $\pm$   5   & 19  $\pm$ 1 &  6.4   $\pm$ 0.1  &  96.0    $\pm$ 0.7  \\  
           &  8.35    & 375  $\pm$   3   & 20  $\pm$ 1 &  5.3   $\pm$ 0.2  &  93.4    $\pm$ 1.1  \\  
           &  4.85    & 618  $\pm$   3   &  6  $\pm$ 2 &  1.0   $\pm$ 0.2  &  65.7    $\pm$ 6.7  \\  
1616+2647  &  2.64    & 959  $\pm$  12   &   --     &  --    &    --      \\   
           &  1.40    & 1480 $\pm$  50   &   --     &  --    &     --     \\ 
           &  0.36   & 1710 $\pm$  36   &   --     &  --    &     --     \\    
           &  0.32   &      --          &   --     &  --    &     --     \\  
           &  0.15   &      --          &   --     &  --    &     --     \\  
           &  0.07   & 1140 $\pm$ 130   &   --     &  --    &     --     \\    
\hline
           &  14.60   & 117  $\pm$   1   &  13  $\pm$ 1 & 10.8   $\pm$ 1.1  &  89.0    $\pm$ 2.6  \\  
           &  10.45   & 152  $\pm$   2   &  17  $\pm$ 1 & 11.4   $\pm$ 0.3  &  90.1    $\pm$ 3.8  \\  
           &  8.35    & 181  $\pm$   1   &  19  $\pm$ 1 & 10.4   $\pm$ 0.3  &  92.7    $\pm$ 1.0  \\  
           &  4.85    & 276  $\pm$   2   &  17  $\pm$ 1 &  6.1   $\pm$ 0.4  & 100.6    $\pm$ 2.6  \\  
1647+3752  &  2.64    & 431  $\pm$   7   &   7  $\pm$ 2 &  1.5   $\pm$ 0.4  & 147      $\pm$ 9.1  \\  
           &  1.40    & 630  $\pm$  20   &   --     &  --    &     --     \\
           &  0.36   &       --         &   --     &  --    &     --     \\
           &  0.32   & 1480 $\pm$   5   &   --     &  --    &     --     \\
           &  0.15   & 1920 $\pm$  50	  &   --     &  --    &     --     \\
           &  0.07   & 2730 $\pm$ 290   &   --     &  --    &     --     \\
\hline
           &  14.60   &  080 $\pm$    2  &  10 $\pm$  3&  12.5  $\pm$ 4.2  &    73.2  $\pm$   9.7\\ 
           &  10.45   &  124 $\pm$    2  &  15 $\pm$  1&  11.7  $\pm$ 0.1  &    78.8  $\pm$   1.6\\ 
           &  8.35    &  159 $\pm$    1  &  18 $\pm$  1&  11.5  $\pm$ 0.4  &    77.0  $\pm$   1.0\\ 
           &  4.85    &  301 $\pm$    2  &  25 $\pm$  1&   8.2  $\pm$ 0.4  &    86.4  $\pm$   1.7\\ 
1713+2813  &  2.64    &  566 $\pm$    2  &  24 $\pm$  1&   4.3  $\pm$ 0.1  &   130.6  $\pm$   1.2\\ 
           &  1.40    & 1030 $\pm$   10  &   --     &  --    &     --     \\
           &  0.36   & 2530 $\pm$   50  &   --     &  --    &     --     \\
           &  0.32   &      --          &   --     &  --    &     --     \\ 
           &  0.15   &      --          &   --     &  --    &     --     \\ 
           &  0.07   &4980  $\pm$ 530   &   --     &  --    &     --     \\
\hline
           &  14.60   & 139  $\pm$    4  &   7 $\pm$  2&   4.8  $\pm$ 1.5  &    82.6  $\pm$   3.7 \\ 
           &  10.45   & 162  $\pm$    1  &   8 $\pm$  1&   5.1  $\pm$ 0.6  &    86.3  $\pm$   0.3 \\ 
           &  8.35    & 176  $\pm$    1  &   7 $\pm$  1&   4.0  $\pm$ 0.4  &    86.3  $\pm$   2.7 \\ 
           &  4.85    & 329  $\pm$    2  &   8 $\pm$  1&   2.3  $\pm$ 0.3  &    98.8  $\pm$   5.4 \\ 
1723+3417  &  2.64    & 716  $\pm$    2  &  24 $\pm$  2&   3.3  $\pm$ 0.2  &   140.5  $\pm$   1.4 \\ 
           &  1.40    & 520  $\pm$   20  &   --     &  --    &     --     \\
           &  0.36   &      --          &   --     &  --    &     --     \\
           &  0.32   & 4105 $\pm$    3  &   --     &  --    &     --     \\
           &  0.15   & <75              &   --     &  --    &     --     \\
           &  0.07   & <300             &   --     &  --    &     --     \\
\hline
\end{tabular}
\end{table*}%

\begin{table*}[h]
\small
\centering
\caption{Table of values. Continue. }
\begin{tabular}{|l|c|c|c|c|c|}
\hline
Name       &$\nu$ [GHz]&  S [mJy]          & S$_{Pol}$ [mJy]       &     m [\%]           & $\chi$ [deg]        \\    
\hline
           &  14.60   &  524 $\pm$    6  &  17 $\pm$   3&  3.0  $\pm$ 0.5   &-49.0    $\pm$  11.1  \\ 
           &  10.45   &  566 $\pm$    2  &  29 $\pm$   1&  5.1  $\pm$ 0.1   &-38.1    $\pm$  0.2  \\ 
           &  8.35    &  598 $\pm$    5  &  26 $\pm$   1&  4.2  $\pm$ 0.1   &-37.8    $\pm$  1.1  \\ 
           &  4.85    &  623 $\pm$    3  &  23 $\pm$   2&  3.7  $\pm$ 0.3   &-28.1    $\pm$  2.0  \\ 
2050+0407  &  2.64    &  607 $\pm$    7  &   9 $\pm$   3&  1.3  $\pm$ 0.4   &  2.6    $\pm$  8.1  \\ 
           &  1.40    &  565 $\pm$   17  &   --     &  --    &     --     \\
           &  0.36   &  410 $\pm$   26  &   --     &  --    &     --     \\ 
           &  0.32   &       --         &   --     &  --    &     --     \\
           &  0.15   &      --          &   --     &  --    &     --     \\
           &  0.07   & <300             &   --     &  --    &     --     \\ 
\hline
           &  14.60   &  918 $\pm$    9  &  32 $\pm$  4&  3.4  $\pm$ 0.4   &116.3    $\pm$  3.4  \\ 
           &  10.45   &  969 $\pm$    1  &  42 $\pm$  3&  4.3  $\pm$ 0.3   &116.5    $\pm$  2.1  \\ 
           &  8.35    &  878 $\pm$    7  &  39 $\pm$  1&  4.4  $\pm$ 0.1   &120.9    $\pm$  0.8  \\ 
           &  4.85    &  853 $\pm$    4  &  22 $\pm$  2&  2.5  $\pm$ 0.2   &125.7    $\pm$  1.7  \\ 
2101+0341  &  2.64    &  684 $\pm$    5  &   9 $\pm$  2&  1.3  $\pm$ 0.3   & 170.6   $\pm$  6.9 \\ 
           &  1.40    &  630 $\pm$   20  &   --     &  --    &     --     \\ 
           &  0.36   &      --          &   --     &  --    &     --     \\ 
           &  0.32   &      --          &   --     &  --    &     --     \\ 
           &  0.15   &      --          &   --     &  --    &     --     \\ 
           &  0.07   & <300             &   --     &  --    &     --     \\   
\hline
           &  14.60   &  719 $\pm$    7  &  15 $\pm$  4&   2.1  $\pm$ 0.5  &  -9.2   $\pm$  6.5 \\ 
           &  10.45   &  760 $\pm$    5  &  16 $\pm$  1&   2.1  $\pm$ 0.1  &  15.2   $\pm$  0.2 \\ 
           &  8.35    &  748 $\pm$    6  &  17 $\pm$  1&   2.2  $\pm$ 0.1  &  35.7   $\pm$  1.8 \\ 
           &  4.85    &  700 $\pm$    4  &  12 $\pm$  1&   1.7  $\pm$ 0.2  &  250.9  $\pm$  3.7\\ 
2147+0929  &  2.64    &  646 $\pm$    5  &   --     &  --    &    --      \\
           &  1.40    &  930 $\pm$   30  &   --     &  --    &     --     \\ 
           &  0.36   & 1120 $\pm$   30  &   --     &  --    &     --     \\   
           &  0.32   &      --          &   --     &  --    &     --     \\ 
           &  0.15   &      --          &   --     &  --    &     --     \\ 
           &  0.07   & 2160 $\pm$   230 &   --     &  --    &     --     \\   
\hline
           &  14.60   & 073  $\pm$     3 &   --     &  --    &    --      \\    
           &  10.45   & 113  $\pm$     1 &   7 $\pm$  1&   6.4  $\pm$ 0.2  &  69.3    $\pm$  1.6  \\   
           &  8.35    & 149  $\pm$     1 &   7 $\pm$  1&   5.0  $\pm$ 0.6  &  71.3    $\pm$  3.1  \\   
           &  4.85    & 282  $\pm$     2 &  11 $\pm$  1&   3.9  $\pm$ 0.4  &  73.6    $\pm$  3.8  \\   
2200+0708  &  2.64    & 515  $\pm$     1 &  12 $\pm$  1&   2.4  $\pm$ 0.1  &  60.3    $\pm$  1.3  \\   
           &  1.40    & 896  $\pm$    32 &   --     &  --    &     --     \\  
           &  0.36   & 2560 $\pm$    76 &   --     &  --    &     --     \\   
           &  0.32   &     --           &   --     &  --    &     --     \\  
           &  0.15   &     --           &   --     &  --    &     --     \\  
           &  0.07   & 5750  $\pm$ 600  &   --     &  --    &     --     \\   
\hline
           &  14.60   & 296  $\pm$     4 &   9 $\pm$  3&   3.0  $\pm$ 1.1  &  35.4    $\pm$ 11.1  \\   
           &  10.45   & 379  $\pm$     7 &  11 $\pm$  3&   2.7  $\pm$ 1.0  &  36.9    $\pm$ 7.9   \\  
           &  8.35    & 409  $\pm$    10 &   6 $\pm$  1&   1.3  $\pm$ 0.3  &  27.4    $\pm$ 5.9   \\  
           &  4.85    & 521  $\pm$     3 &   4 $\pm$  1&   0.7  $\pm$ 0.3  & -100.8   $\pm$ 12.9  \\ 
2245+0324  &  2.64    & 559  $\pm$    11 &   --     &  --    &    --      \\    
           &  1.40    & 480  $\pm$    10 &   --     &  --    &     --     \\ 
           &  0.36   &      --          &   --     &  --    &     --     \\ 
           &  0.32   &      --          &   --     &  --    &     --     \\ 
           &  0.15   &      --          &   --     &  --    &     --     \\ 
           &  0.07   &<300              &   --     &  --    &     --     \\   
\hline
\end{tabular}
\end{table*}%

\clearpage
\section{Parameters used for the SEDs fit} 

\begin{table*}
\small
\caption{Values of the SEDs best fit. In the table:  name of the source, the best fit model, $\alpha_{thin}$ (for the steep spectrum and also for the various syncrotron components), peak flux density (S$_{max}$), the peak frequency ($\nu_{max}$ ), and the $\chi^{2}$. The flux density \textit{S} is in Jy and the frequency $\nu$ is in GHz.
\newline
Explanation on the abbreviation I use in the Best fit column: $S^{pl}_\nu$ is linear fit; $S^{plb}_\nu$ is linear fit with a break; $S^{s}_{\nu}$ is one synchrotron component; $S^{sb}_{\nu}$ is one synchrotron component with a break;  $S^{s+}_{\nu} $ is two or more synchrotron components; $S^{pls}_{\nu} $ is linear fit at low frequency plus one or two synchrotron components at higher frequency.
}

\centering
\begin{tabular}{ccrrrrrrr}
\hline\hline
\textit{Name}   & 	\textit{type}     &      $\alpha_{thin}$    & $\chi^{2}$\\           
\hline
0925+3159 & $S^{pl}_\nu$ &   --0.9 & 8.0 \\ 
0958+3224 & $S^{pl}_\nu$ &   --0.5 & 1170.0 \\

\hline\hline
 \textit{Name}  &  	\textit{type}    &       $\alpha_{thin}$  & $\alpha_{break}$ &   $\nu_{break}$  & S$_{break}$ & $\chi^{2}$\\
\hline

1015+0318 &  $S^{plb}_\nu$    &  --0.5  &   --1.0  &    4.2    &  0.290  &   1.4 \\
1213+1307 &  $S^{plb}_\nu$    &  --0.4  &   --0.9  &   12.8    &  0.580  &   0.1\\
1647+3752 &  $S^{plb}_\nu$    &  --0.4  &   --1.0  &    4.5    &  0.466  &   3.0\\
1713+2813 &  $S^{plb}_\nu$    &  --0.4  &   --1.3  &    1.7    &  1.370  &   0.02\\
2200+0708 &  $S^{plb}_\nu$    &  --0.0  &   --1.0  &    0.2    &  6.263  &  29.0\\
\hline\hline
 \textit{Name}  &  	\textit{type}      &    $\alpha_{thin}$   & $\nu$   &    S$_{\nu}$ & $\chi^{2}$ \\
\hline
1723+3417  &  $S^{s}_{\nu}$   &       --0.9    &  2.0   &    0.550  &  0.6  \\
2050+0407  &  $S^{s}_{\nu}$   &       --0.1    &  1.1   &    0.700  & 5.0 \\
2245+0324  &  $S^{s}_{\nu}$   &       --0.4    &  2.5   &    0.580  & 0.5 \\

\hline\hline
 \textit{Name}  &  	\textit{type}    &     $\alpha_{thin}$     &  $\alpha_{break}$   &  $\nu$ & S$_{nu}$    &  $\nu_{break}$  & $\chi^{2}$\\    
\hline
0751+2716 &  $S^{sb}_{\nu}$ &  --0.0  &  --1.1  &    0.2 & 1.650   &  0.6 & 0.2 \\
1311+1417 &  $S^{sb}_{\nu}$ &  --0.1  &   --1.2  &    1.2   &   0.773   &  4.0 & 0.02 \\
1435--0414 &  $S^{sb}_{\nu}$ & --0.3  &   --1.0  &    0.1  &   1.071   &  6.2 & 0.003\\

\hline\hline
 \textit{Name}  &  	\textit{type}   &      $\nu{1}$ &   S$_{1}$ &  $\nu{2}$ &    S$_{2}$& $\chi^{2}$ \\
\hline
1351+0830  & $S^{s+}_{\nu} $     &     1.4   &    0.340  &    9.3 &    0.210 & 0.2\\
1549+5038  & $S^{s+}_{\nu} $     &     0.7   &    0.908  &    7.6 &    0.700 & 8.1 \\
1616+2647  & $S^{s+}_{\nu} $     &     0.1   &    1.320  &    0.6 &    1.620 & 5.1\\
2101+0341  & $S^{s+}_{\nu} $     &     1.9   &    0.720  &    9.9 &    0.700 & 5.0\\

\hline\hline
 \textit{Name}  &  	\textit{type}  &        $\nu{1}$ &    S$_{1}$  &   $\nu{2}$ &  S$_{2}$ &  $\nu{3}$ &   S$_{3}$& $\chi^{2}$ \\
\hline
0239--0234 &  $S^{s+}_{\nu} $    &    0.5   &     0.300   &   3.9  &   0.500  &   11.6   &  0.380 & 0.01\\
0742+4900 &  $S^{s+}_{\nu} $    &     0.6   &     0.200   &   2.7  &   0.400  &  13.8   &  0.241 & 0.2\\
1043+2408 &  $S^{s+}_{\nu} $    &     0.4   &     0.470   &   4.4  &   0.710  &  15.7   &  0.730 & 0.6 \\
1044+0655 &  $S^{s+}_{\nu} $    &     0.1  &     1.600   &   3.2  &   0.190  &  12.7   &  0.130 & 1.2\\
1048+0141 &  $S^{s+}_{\nu} $    &     0.1  &     1.130   &   2.7  &   0.320  &   10.0   &  0.110 & 0.1 \\
1146+5356 &  $S^{s+}_{\nu} $    &     0.2   &     0.400   &   2.9  &   0.470 &   10.0   &  0.350 & 1.2\\
1246--0730 &  $S^{s+}_{\nu} $    &    0.1   &     1.740   &   4.4  &   0.660  &  15.0   &  0.600 & 63.0 \\
2147+0929 &  $S^{s+}_{\nu} $    &     0.1  &     2.310   &   2.6  &   0.360  &   10.7   &  0.500 & 2.3\\

\hline\hline
 \textit{Name}  &  	\textit{type}   &  $\alpha_{thin}$  &  $\nu{2}$   &  S$_{2}$& $\chi^{2}$ \\
\hline
1312+5548 & $S^{pls}_{\nu} $  &   --0.7   & 0.7  & 0.750 & 0.3\\

\hline\hline
 \textit{Name}    & 	\textit{type}    &  $\nu{2}$  & S$_{2}$ &    $\nu{3}$ &   S$_{3}$ & $\chi^{2}$ \\
\hline
0243--0550 & $S^{pls}_{\nu} $     &   1.2  &    0.480  &   6.6   &  0.450  & 0.1\\
0845+0439 & $S^{pls}_{\nu} $      &   2.5  &    0.470  &  13.6   &  0.520  & 2.2 \\
1405+0415 & $S^{pls}_{\nu} $      &   1.0  &    0.770  &   8.2   &  0.500  & 25.0\\
1616+0459 & $S^{pls}_{\nu} $      &   3.3  &    0.610  &   8.0   &  0.720  & 0.2\\
\hline\hline

\end{tabular}
\end{table*}%

\clearpage

\section{Plots of the high RM candidates} 

\begin{figure}
\centering
\caption{Source 0239-0234, 0243-0550 and  0742+4900. For each source we present their SED $\textit{S}$[Jy], the polrization flux density $\textit{S$_{pol}$}$[Jy], the fractional polarization $\textit{m}$[\%] and the polarization angle PA[rad].}
\includegraphics[width=0.49\textwidth]{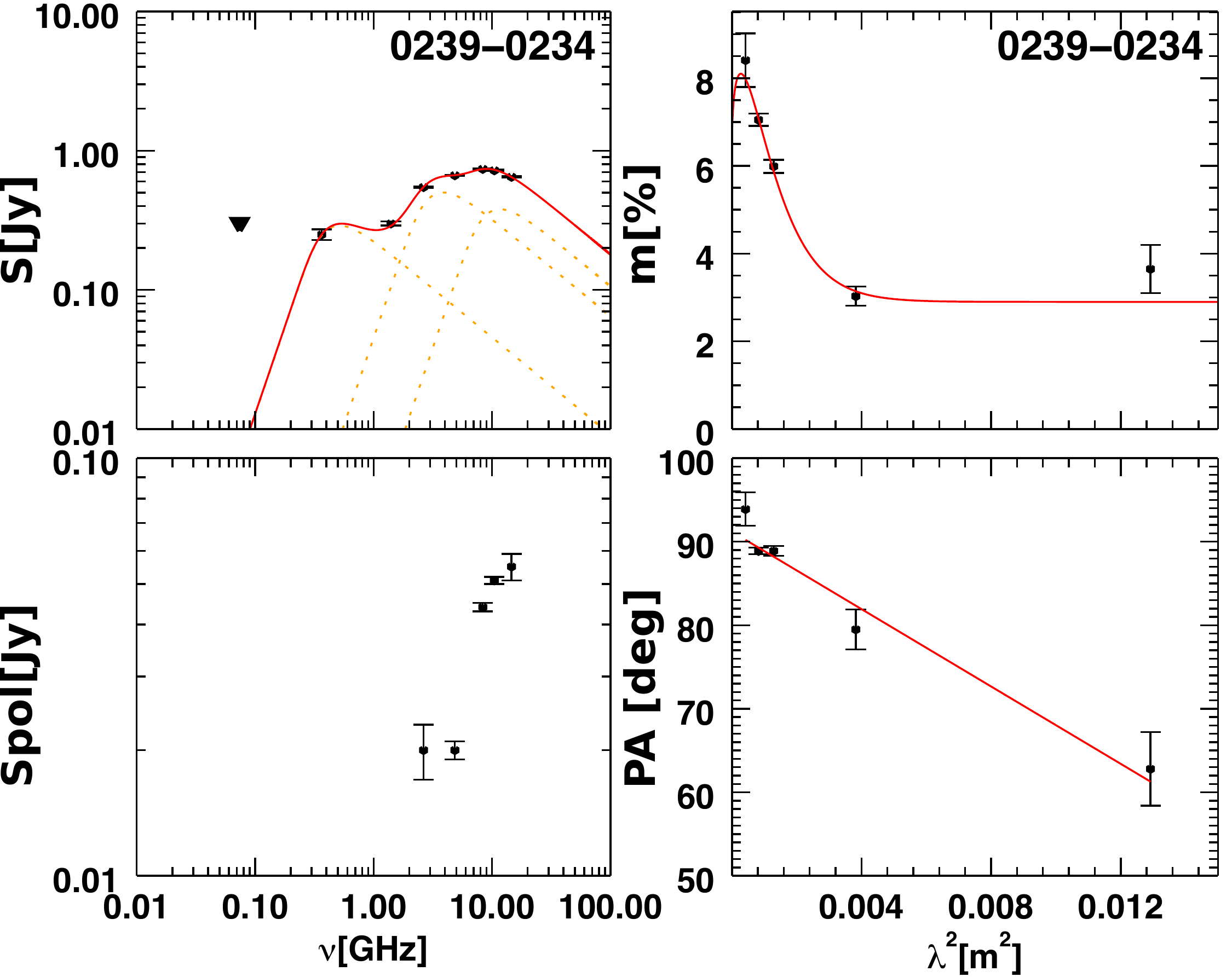}
\includegraphics[width=0.49\textwidth]{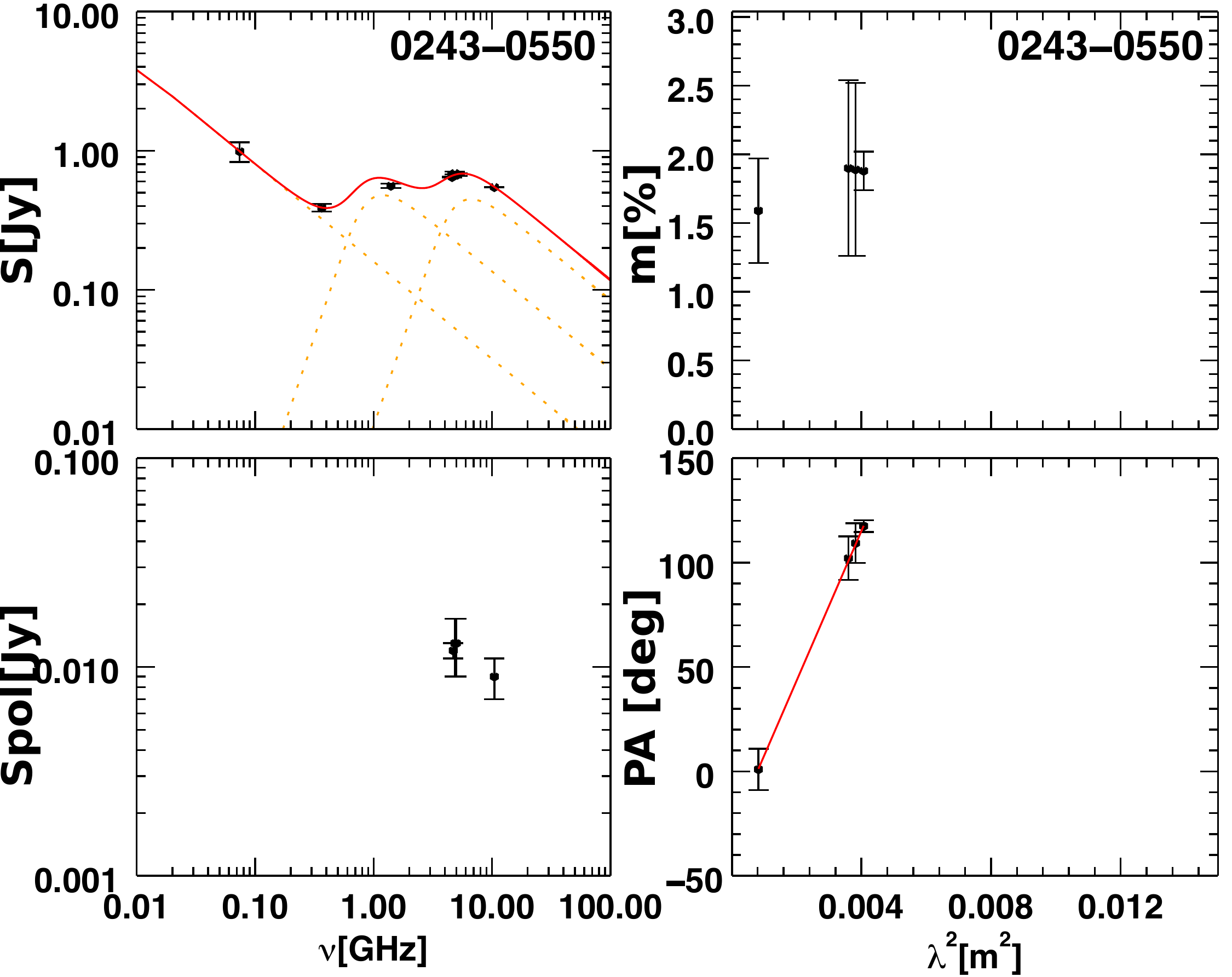}
\includegraphics[width=0.49\textwidth]{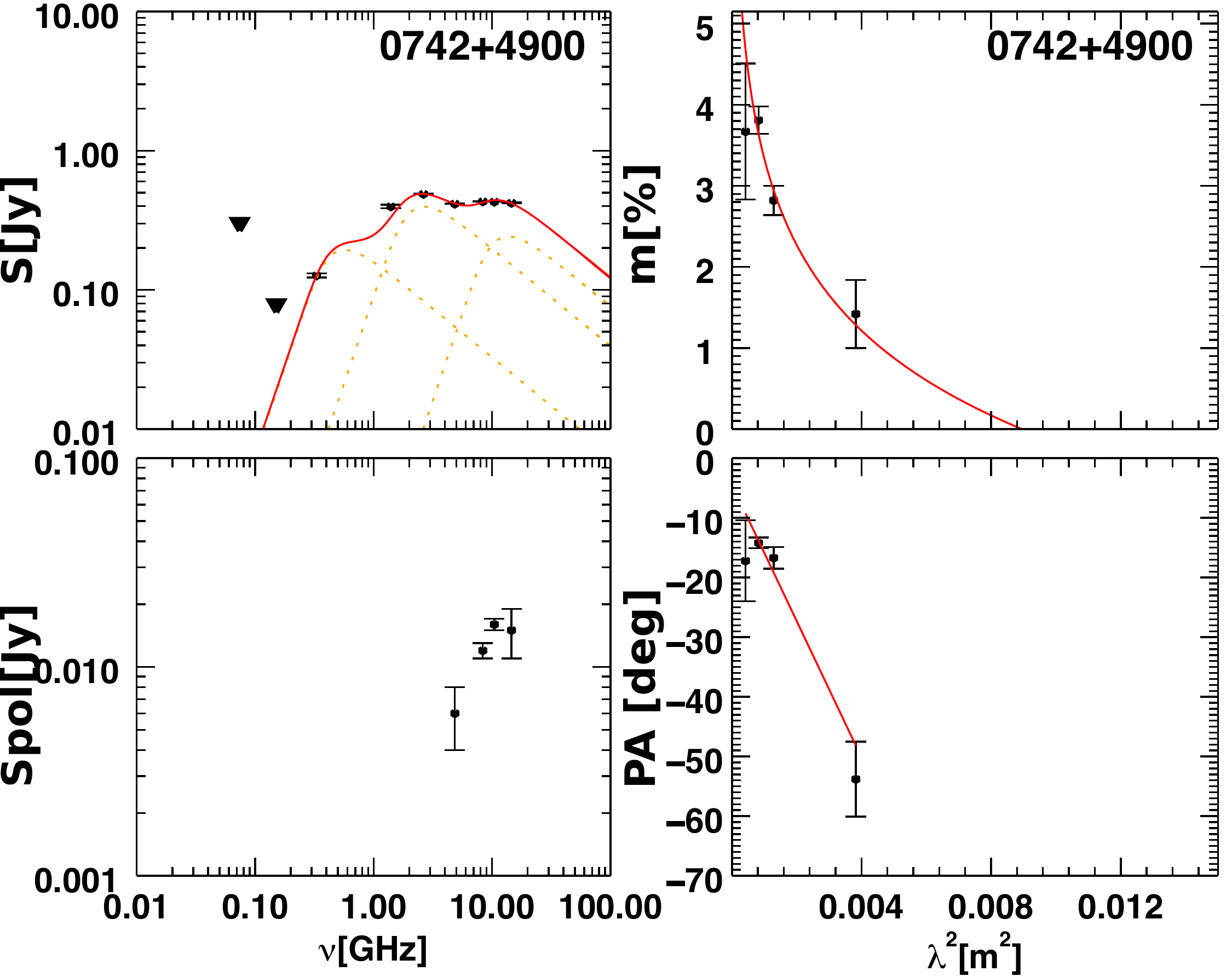}
\label{default}
\end{figure}

\begin{figure}
\centering
\caption{Source 0751+2716, 0845+0439 and 0925+3159. For each source we present their SED $\textit{S}$[Jy], the polrization flux density $\textit{S$_{pol}$}$[Jy], the fractional polarization $\textit{m}$[\%] and the polarization angle PA[rad]. }
\includegraphics[width=0.49\textwidth]{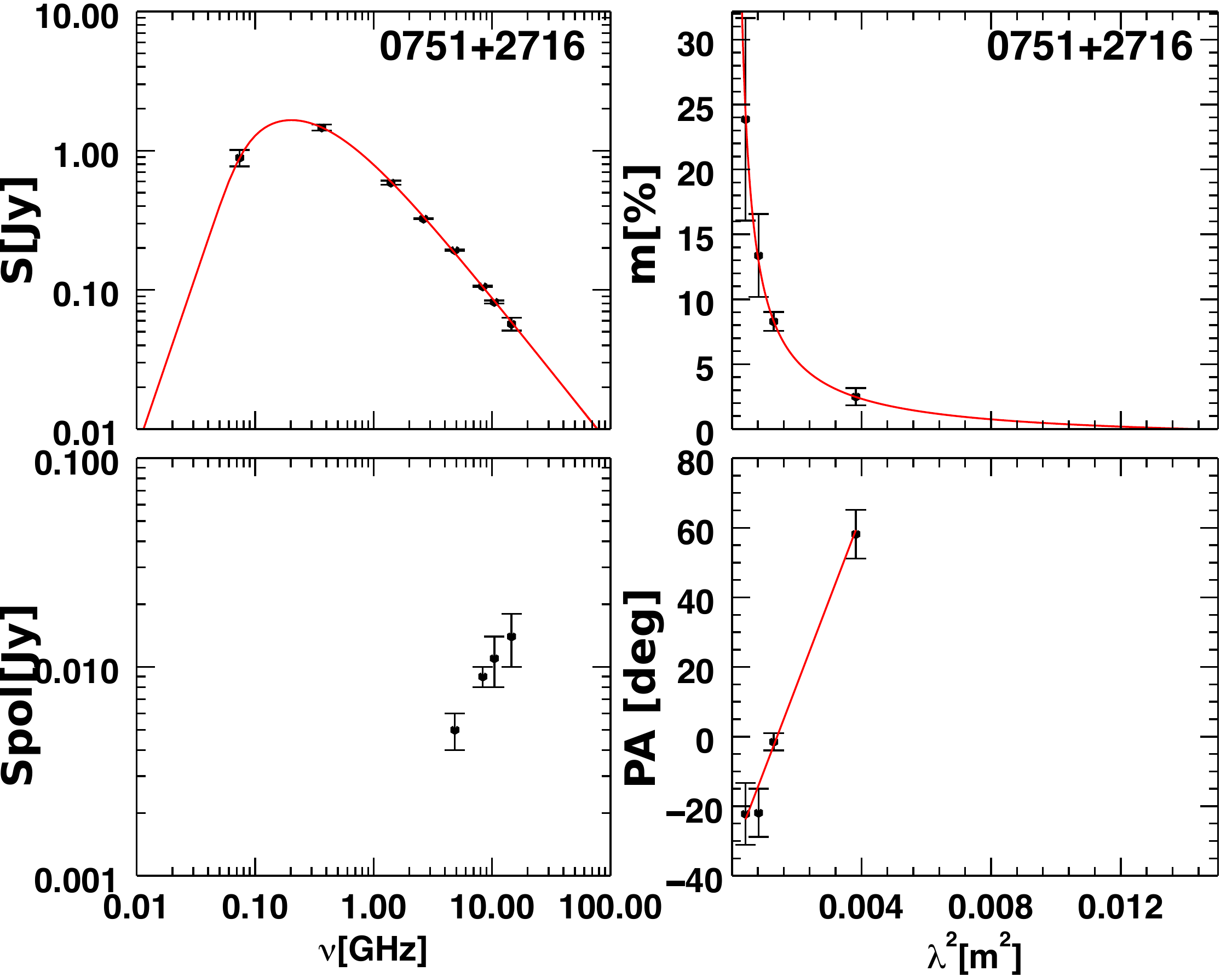}
\includegraphics[width=0.49\textwidth]{0845-eps-converted-to.pdf}
\includegraphics[width=0.49\textwidth]{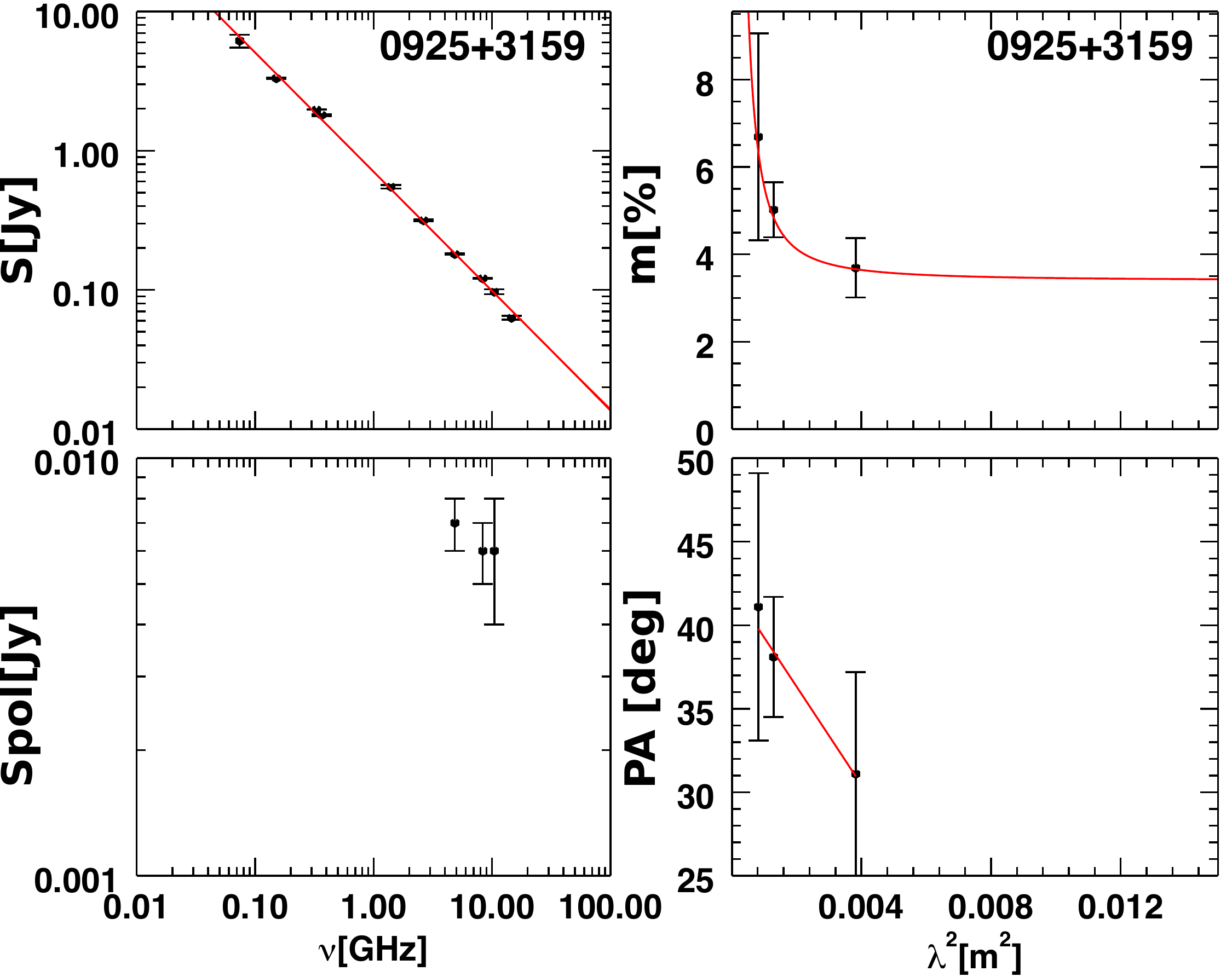}
\label{default}
\end{figure}

\begin{figure}[!h]
\begin{center}
\caption{Source 0958+3224, 1015+0318 and 1043+2408. For each source we present their SED $\textit{S}$[Jy], the polrization flux density $\textit{S$_{pol}$}$[Jy], the fractional polarization $\textit{m}$[\%] and the polarization angle PA[rad]. }
\includegraphics[width=0.49\textwidth]{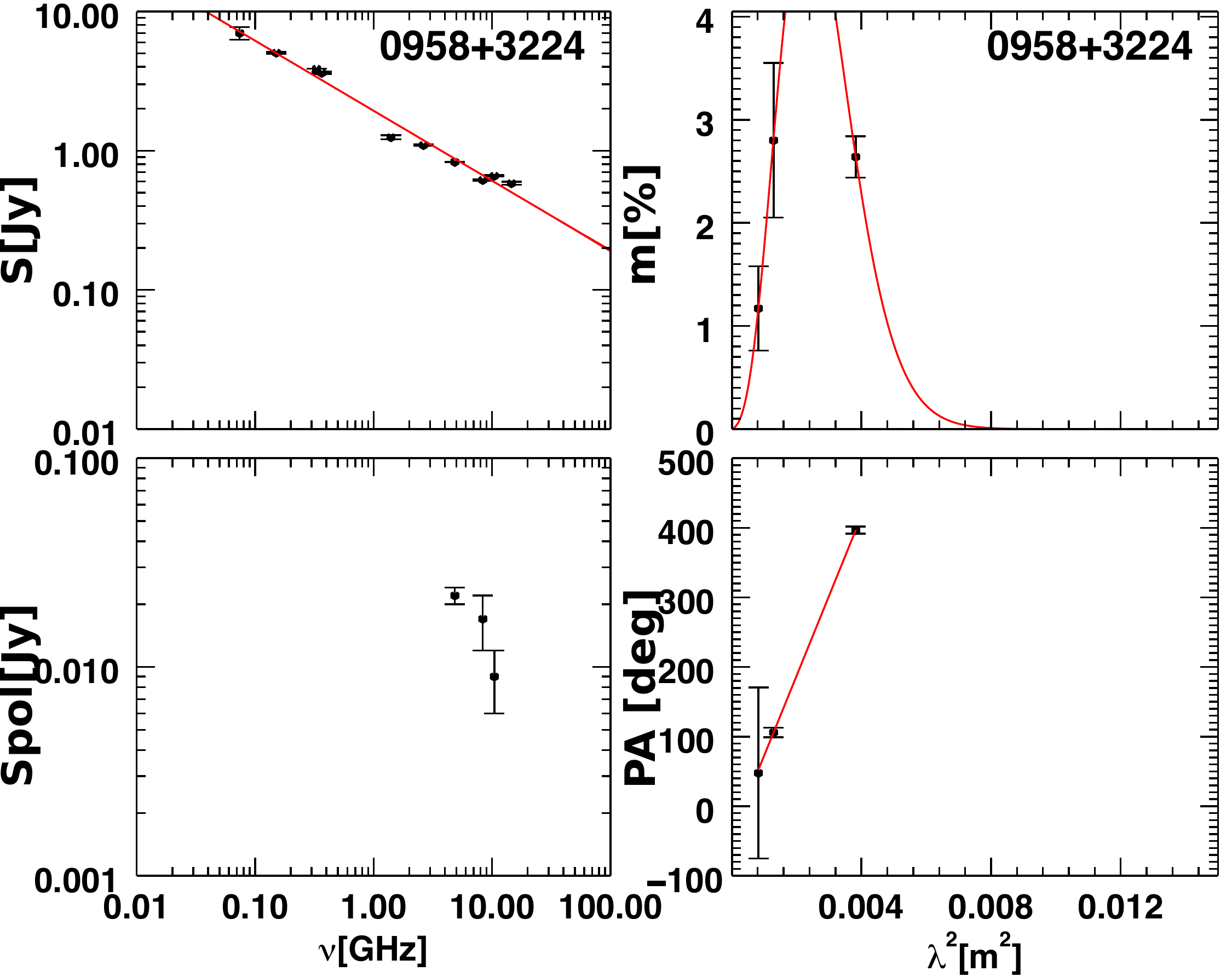}
\includegraphics[width=0.49\textwidth]{1015-eps-converted-to.pdf}
\includegraphics[width=0.49\textwidth]{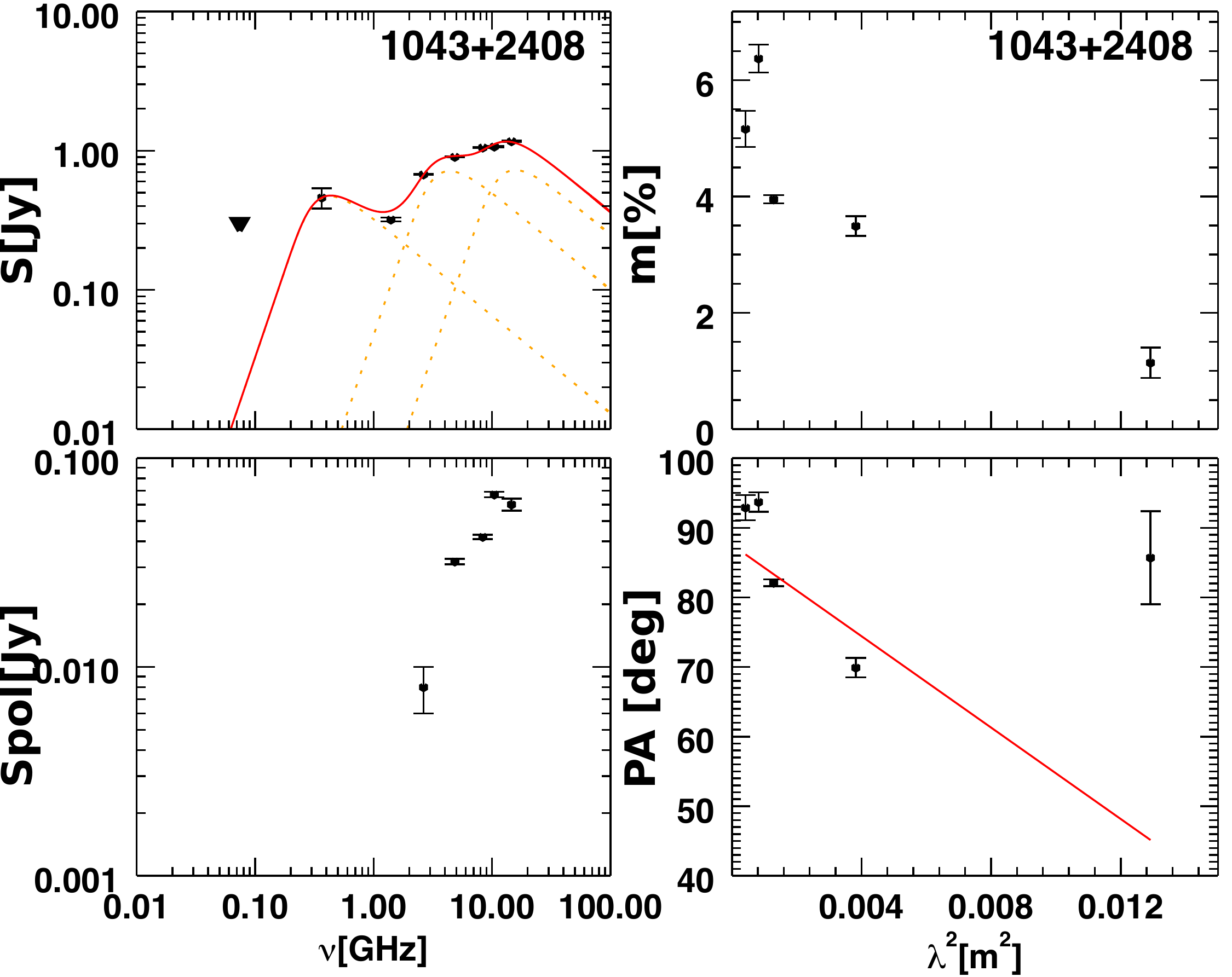}
\label{default}
\end{center}
\end{figure}

\begin{figure}[!h]
\begin{center}
\caption{Source 1044+0655, 1048+0141 and 1146+5356. For each source we present their SED $\textit{S}$[Jy], the polrization flux density $\textit{S$_{pol}$}$[Jy], the fractional polarization $\textit{m}$[\%] and the polarization angle PA[rad]. }
\includegraphics[width=0.49\textwidth]{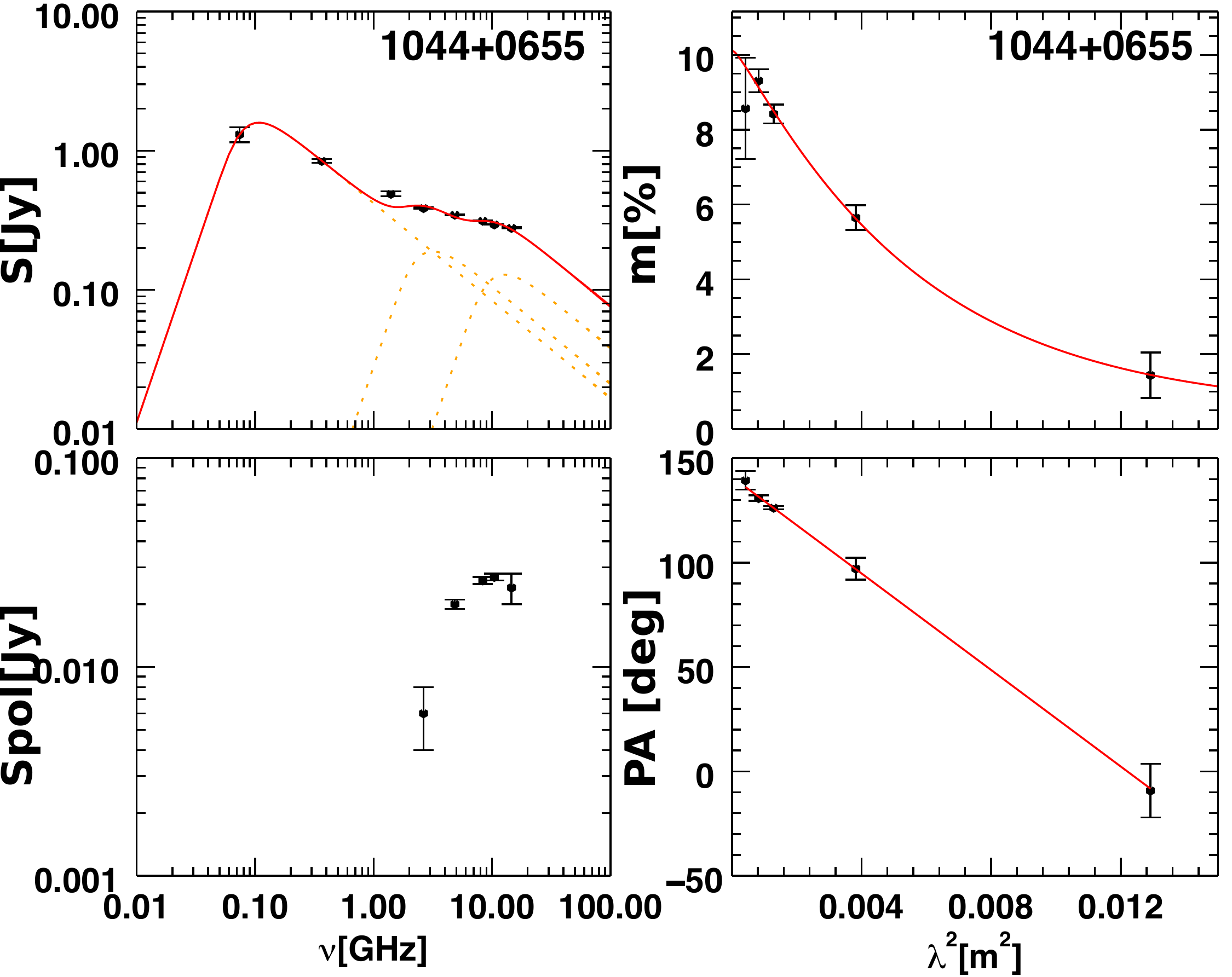}
\includegraphics[width=0.49\textwidth]{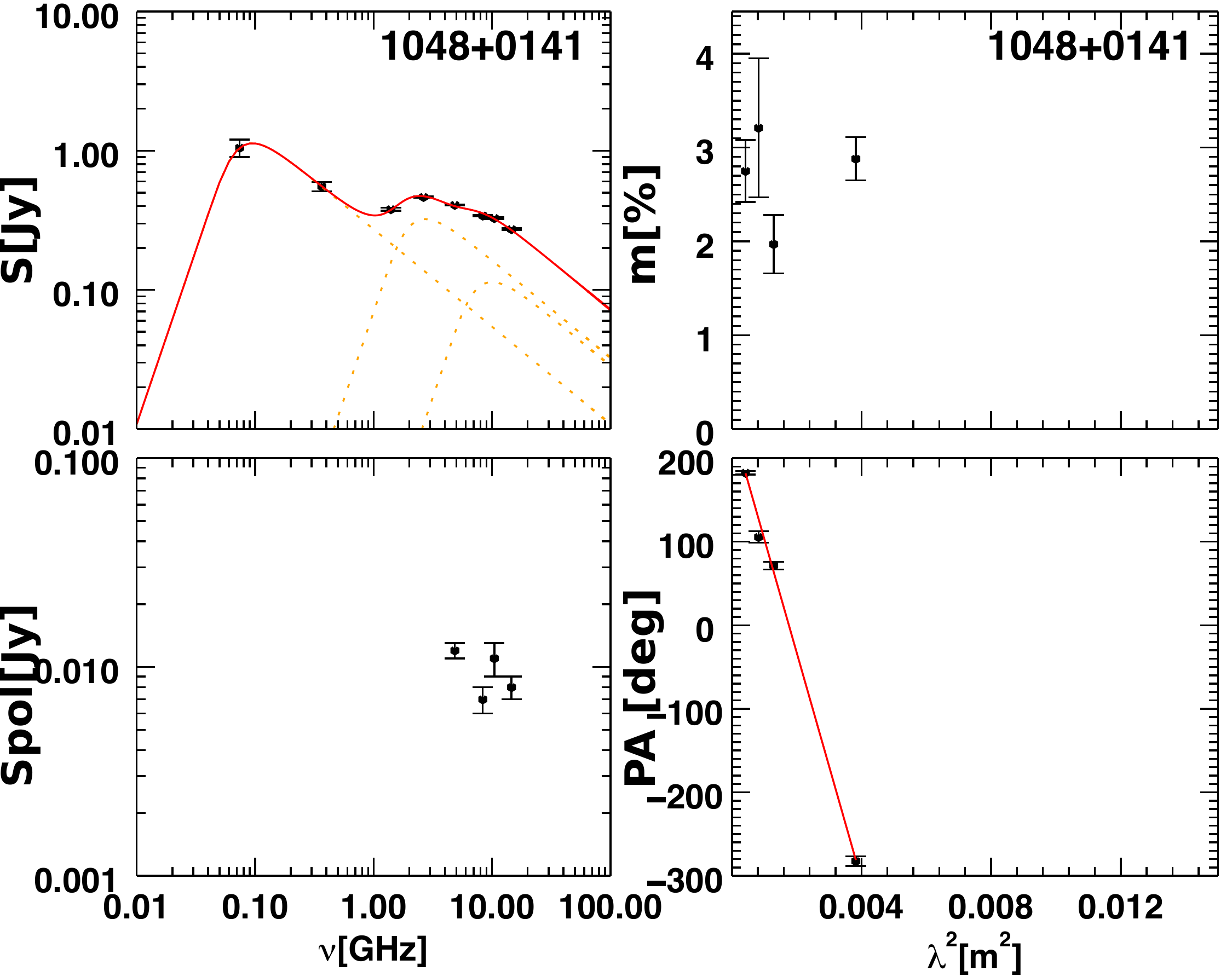}
\includegraphics[width=0.49\textwidth]{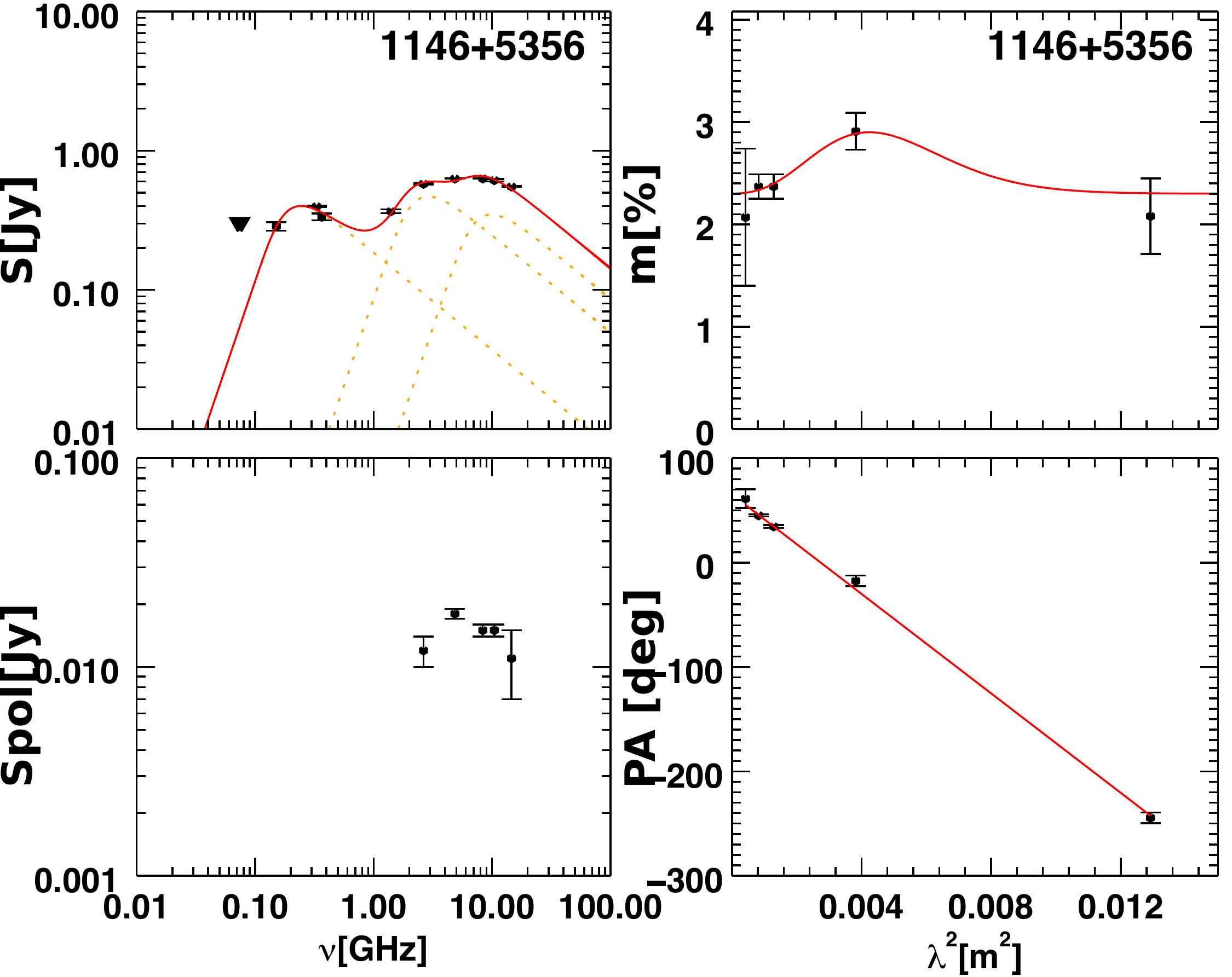}
\label{default}
\end{center}
\end{figure}

\begin{figure}[!h]
\begin{center}
\caption{Source 1213+1307, 1246-0730 and 1311+1417. For each source we present their SED $\textit{S}$[Jy], the polrization flux density $\textit{S$_{pol}$}$[Jy], the fractional polarization $\textit{m}$[\%] and the polarization angle PA[rad]. }
\includegraphics[width=0.49\textwidth]{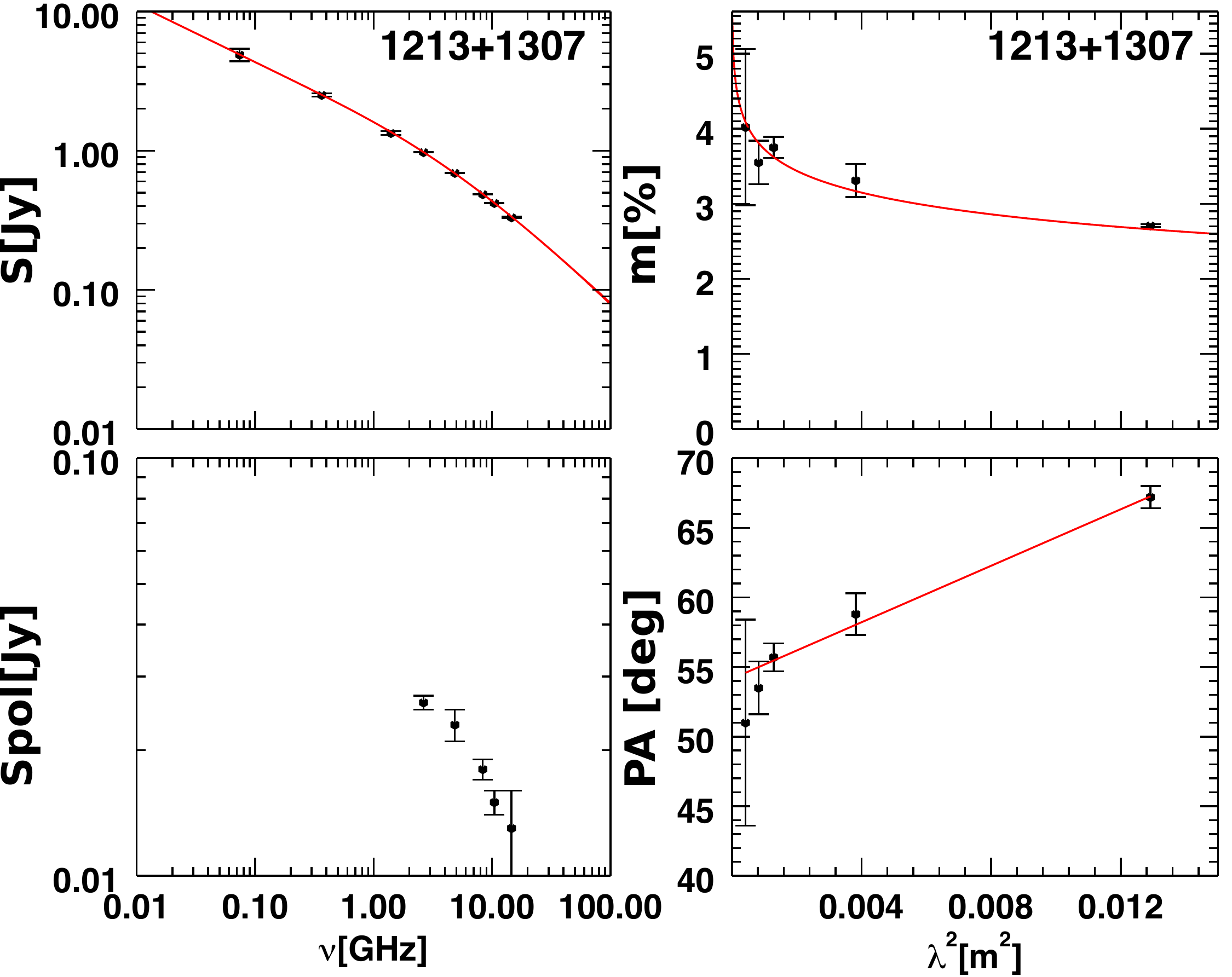}
\includegraphics[width=0.49\textwidth]{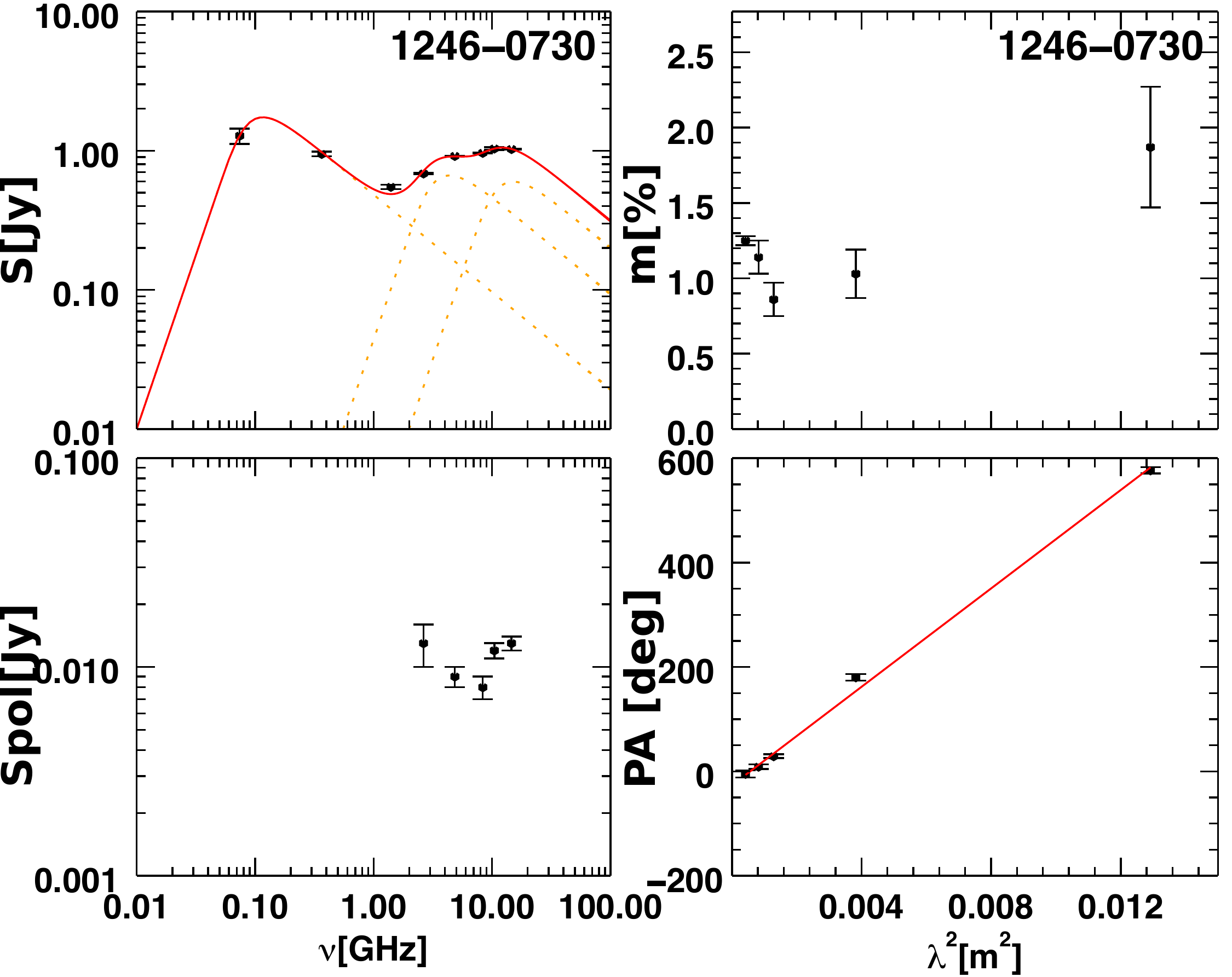}
\includegraphics[width=0.49\textwidth]{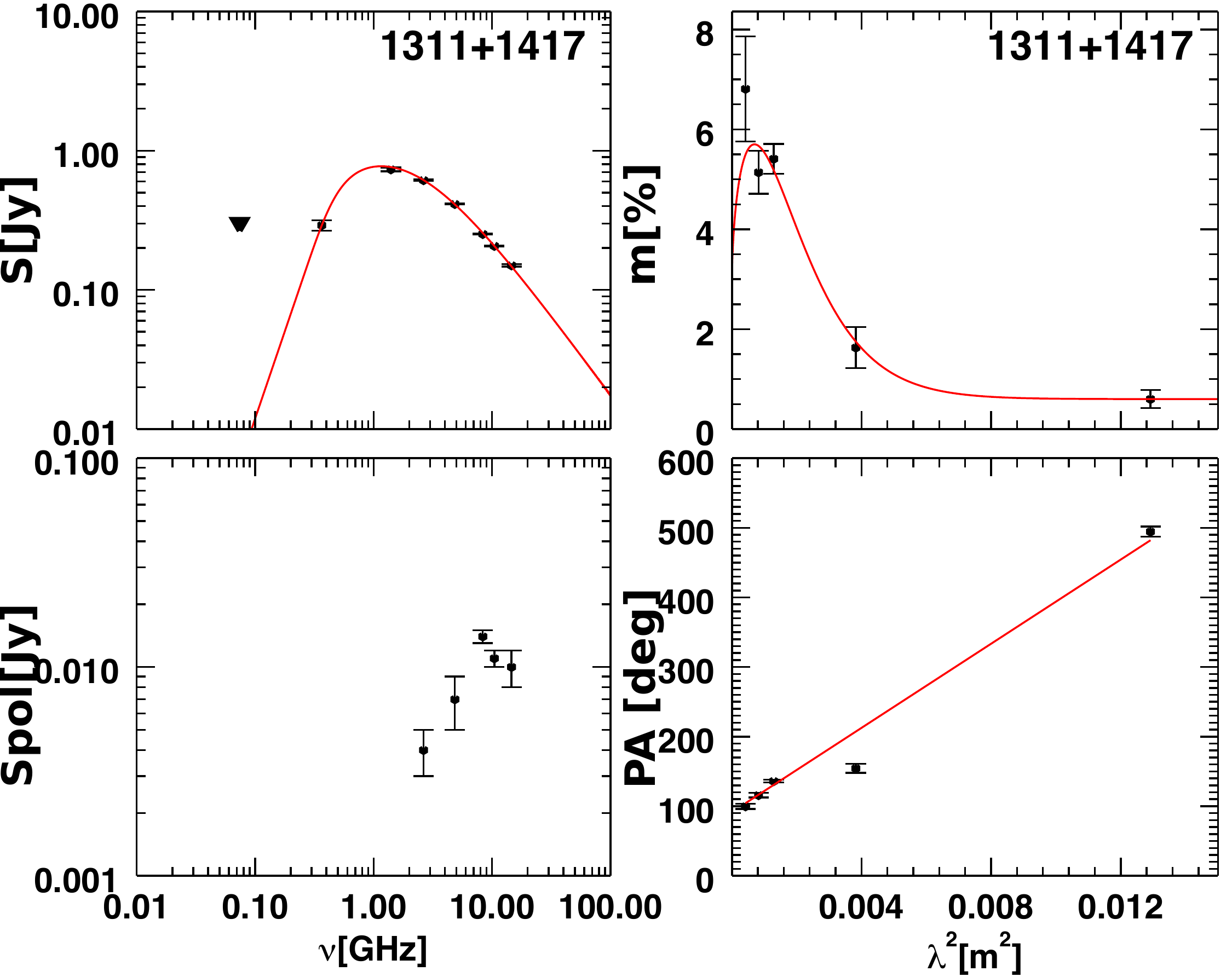}
\label{default}
\end{center}
\end{figure}

\begin{figure}[!h]
\begin{center}
\caption{Source 1312+5548, 1351+0830 and 1405+0415. For each source we present their SED $\textit{S}$[Jy], the polrization flux density $\textit{S$_{pol}$}$[Jy], the fractional polarization $\textit{m}$[\%] and the polarization angle PA[rad]. }
\includegraphics[width=0.49\textwidth]{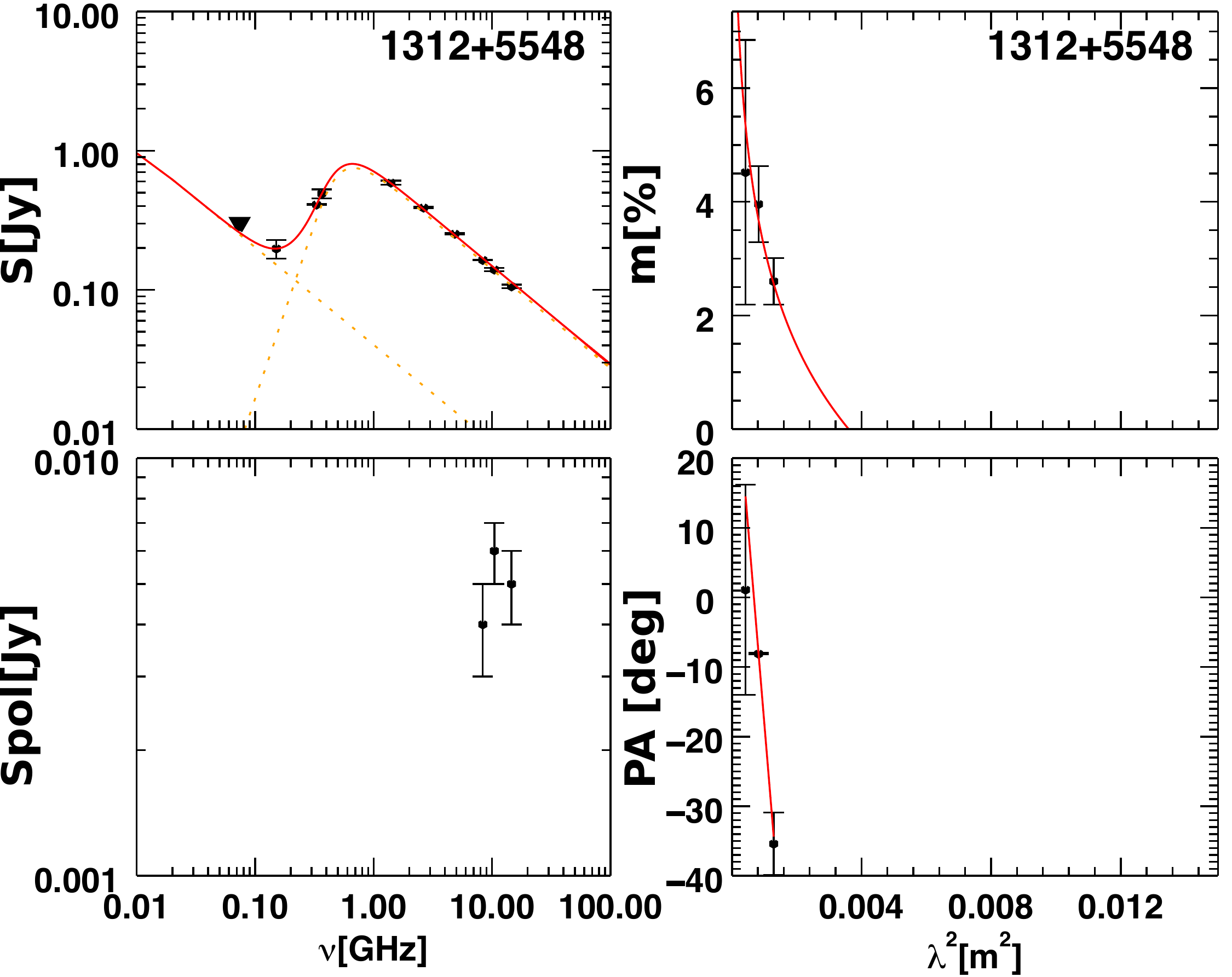}
\includegraphics[width=0.49\textwidth]{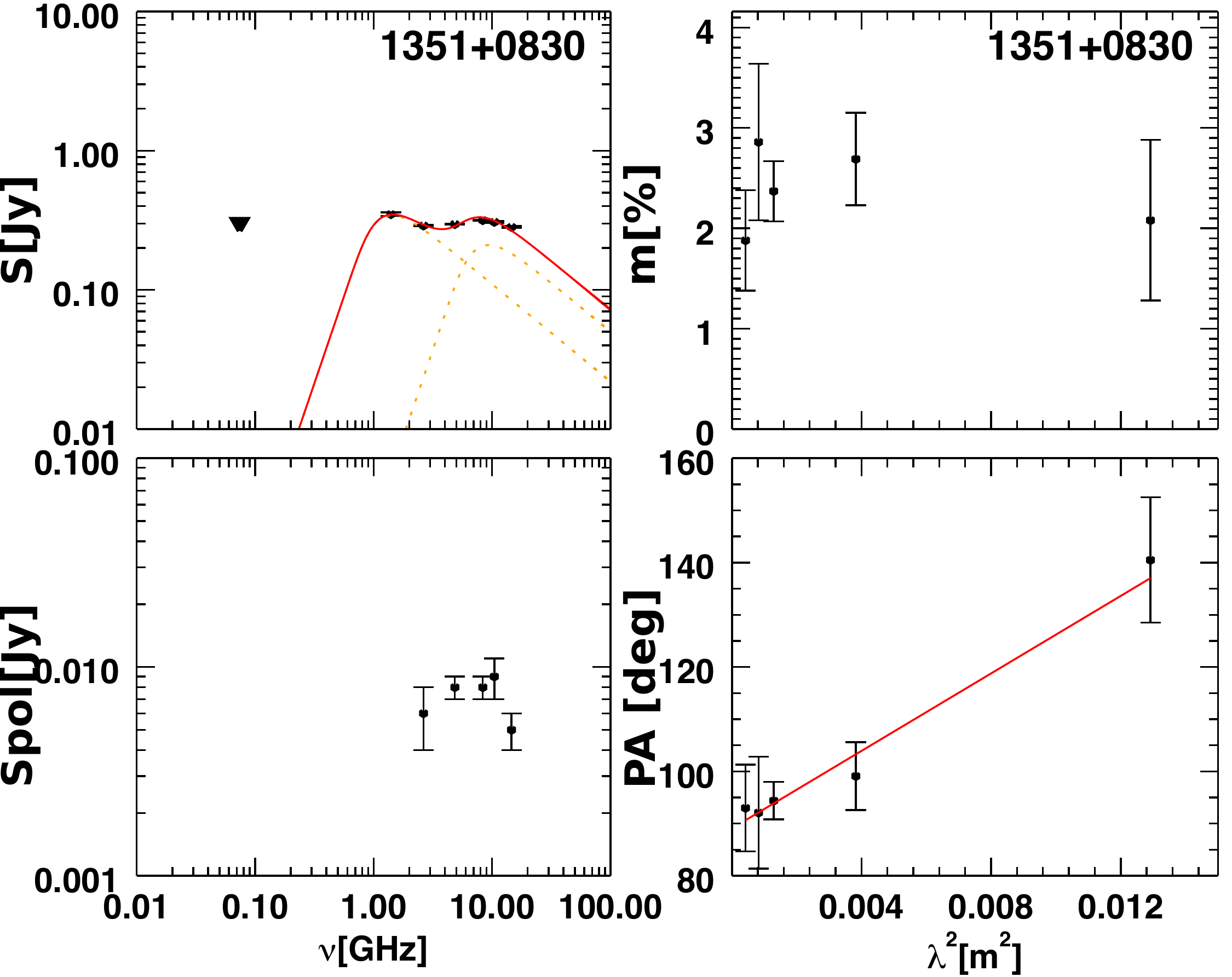}
\includegraphics[width=0.49\textwidth]{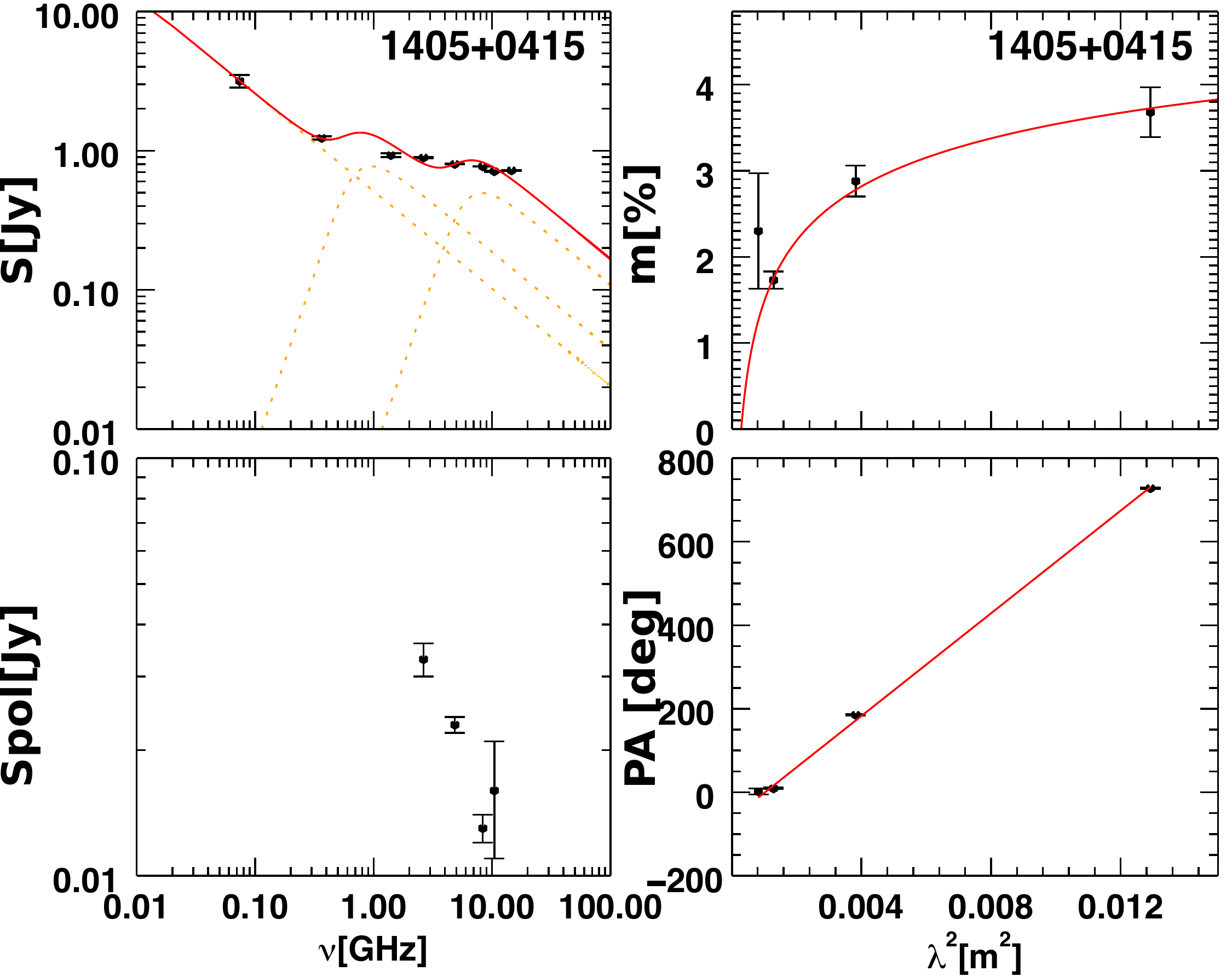}
\label{default}
\end{center}
\end{figure}

\begin{figure}[!h]
\begin{center}
\caption{Source 1435-0414, 1549+5038 and 1616+0459.  For each source we present their SED $\textit{S}$[Jy], the polrization flux density $\textit{S$_{pol}$}$[Jy], the fractional polarization $\textit{m}$[\%] and the polarization angle PA[rad].  }
\includegraphics[width=0.49\textwidth]{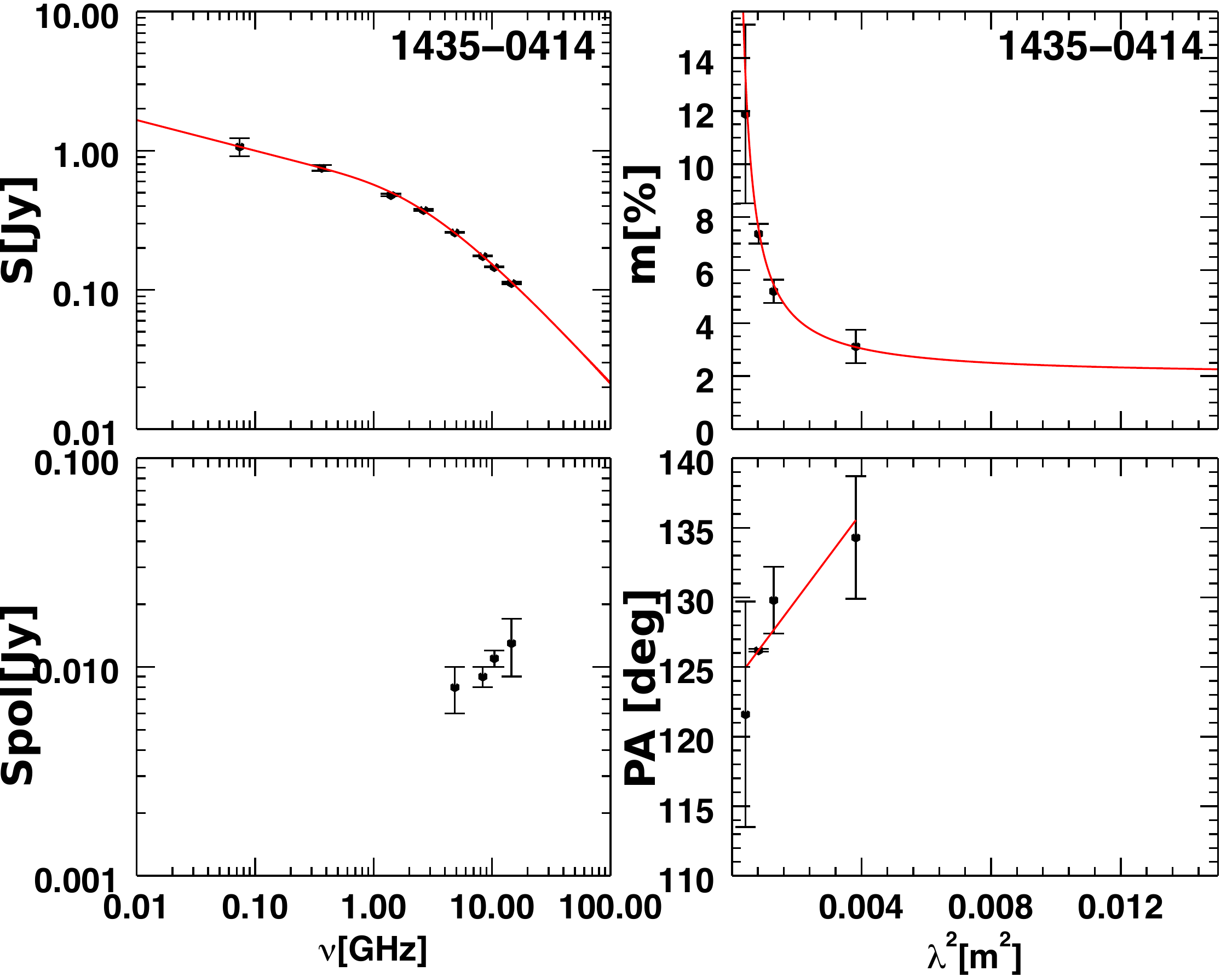}
\includegraphics[width=0.49\textwidth]{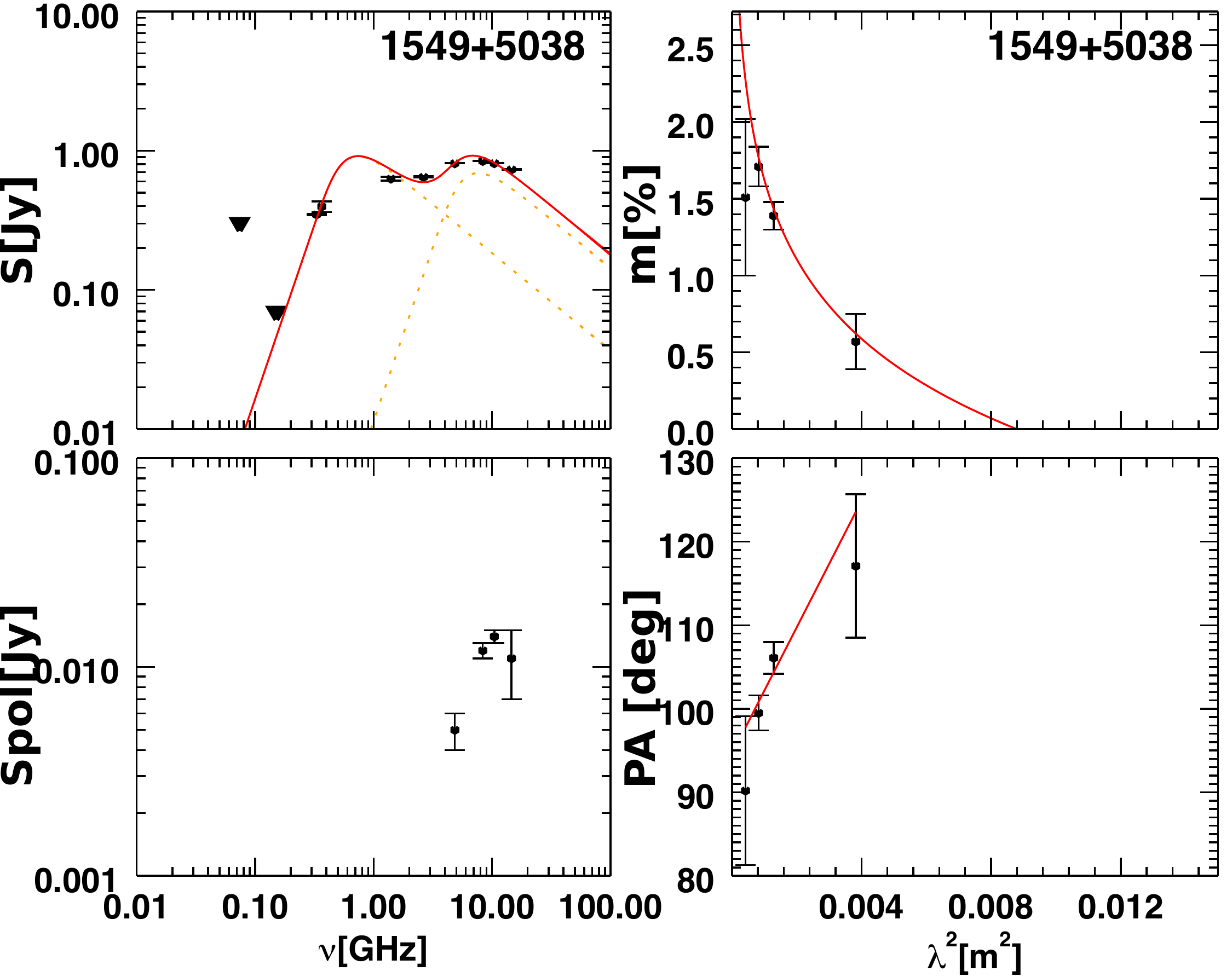}
\includegraphics[width=0.49\textwidth]{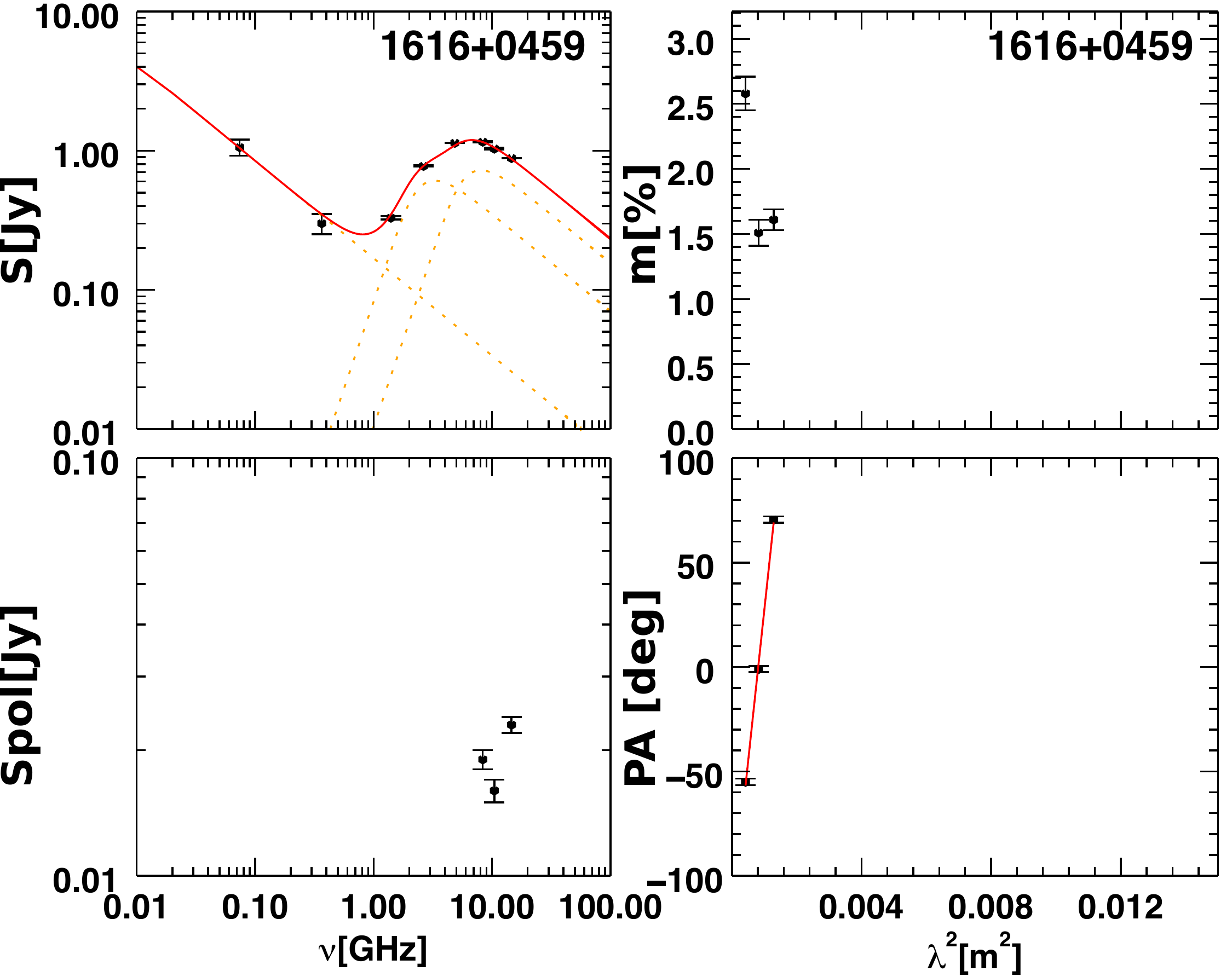}
\label{default}
\end{center}
\end{figure}

\begin{figure}[!h]
\begin{center}
\caption{Source 1616+2647, 1647+3752 and 1713+2813.  For each source we present their SED $\textit{S}$[Jy], the polrization flux density $\textit{S$_{pol}$}$[Jy], the fractional polarization $\textit{m}$[\%] and the polarization angle PA[rad]..  }
\includegraphics[width=0.49\textwidth]{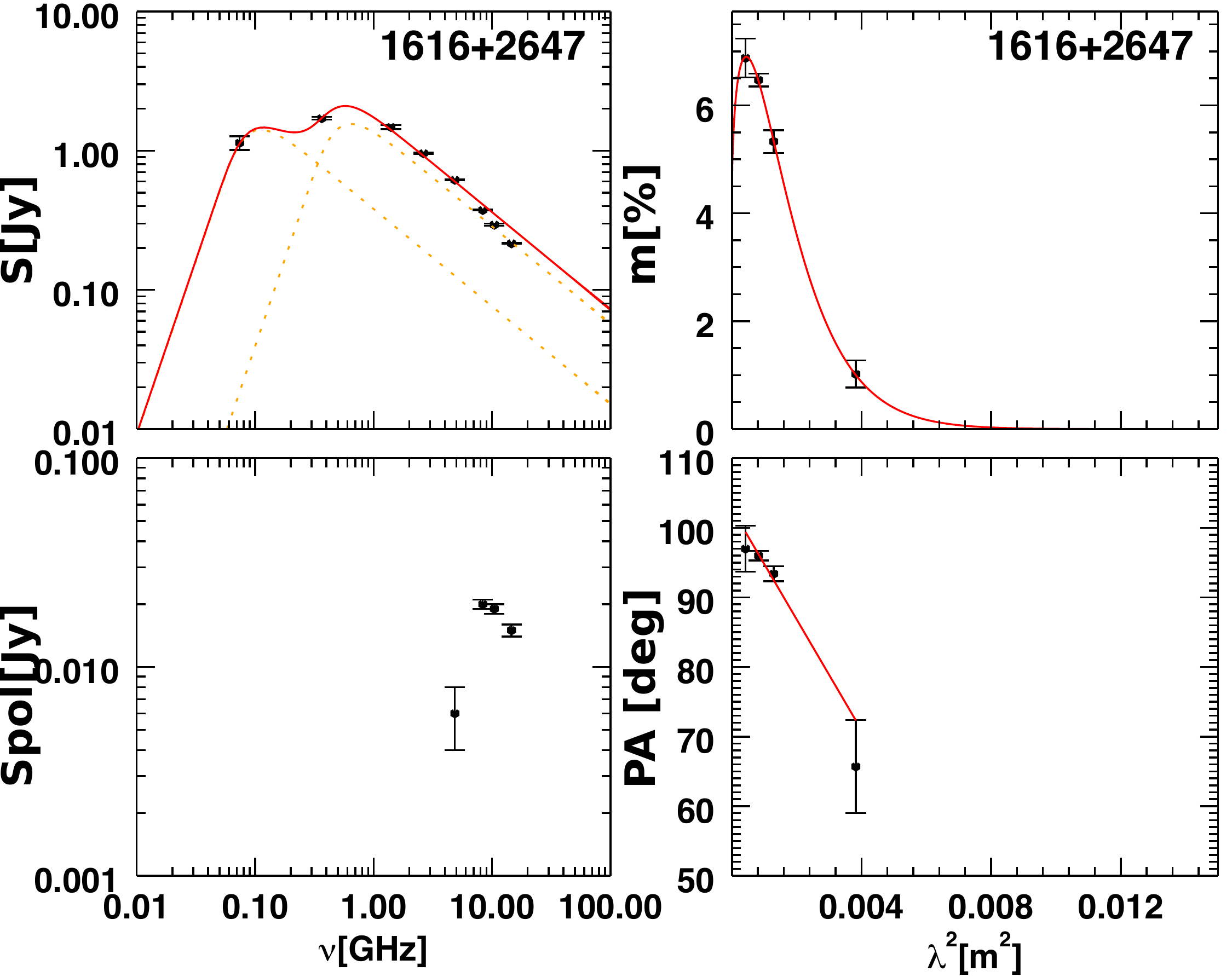}
\includegraphics[width=0.49\textwidth]{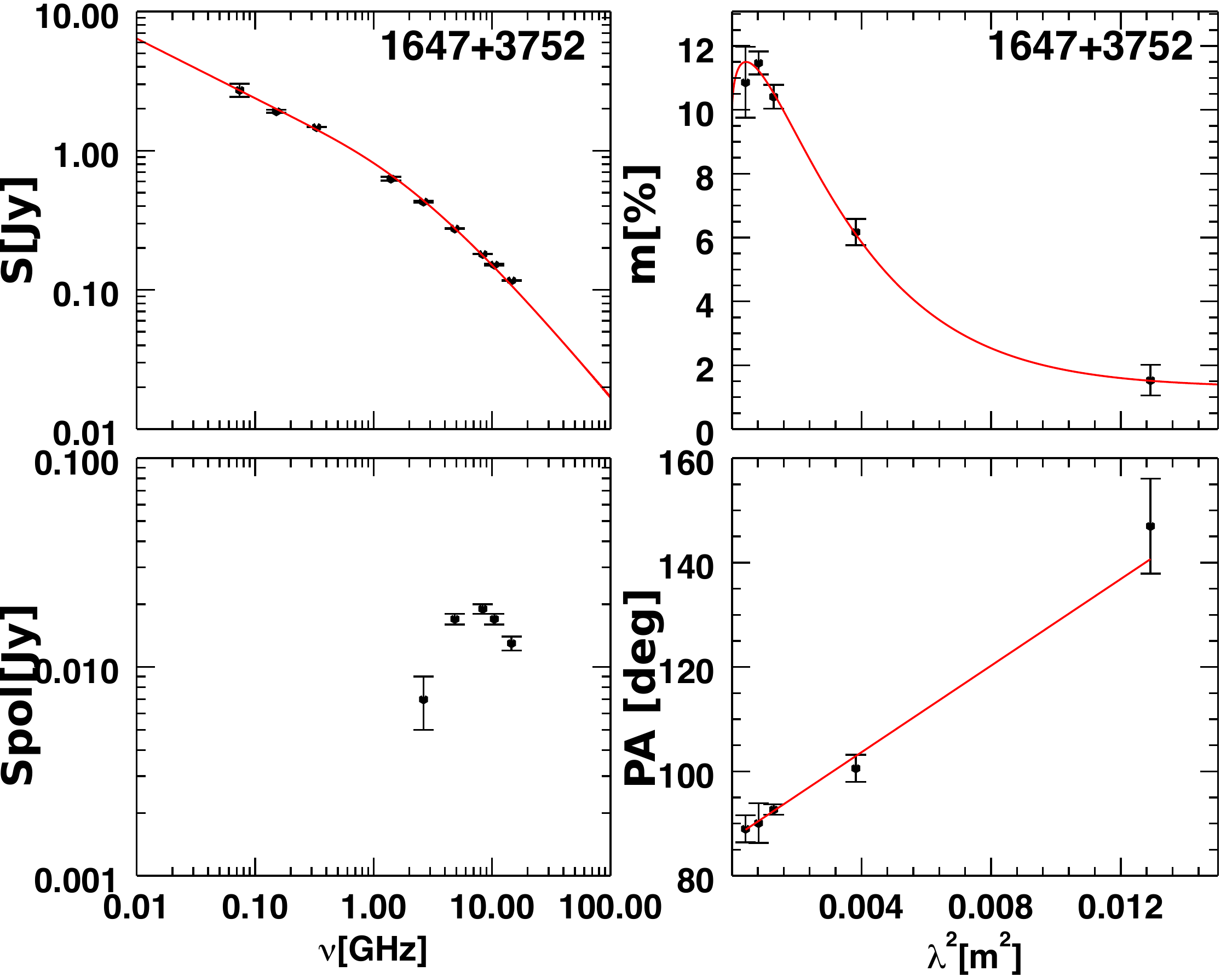}
\includegraphics[width=0.49\textwidth]{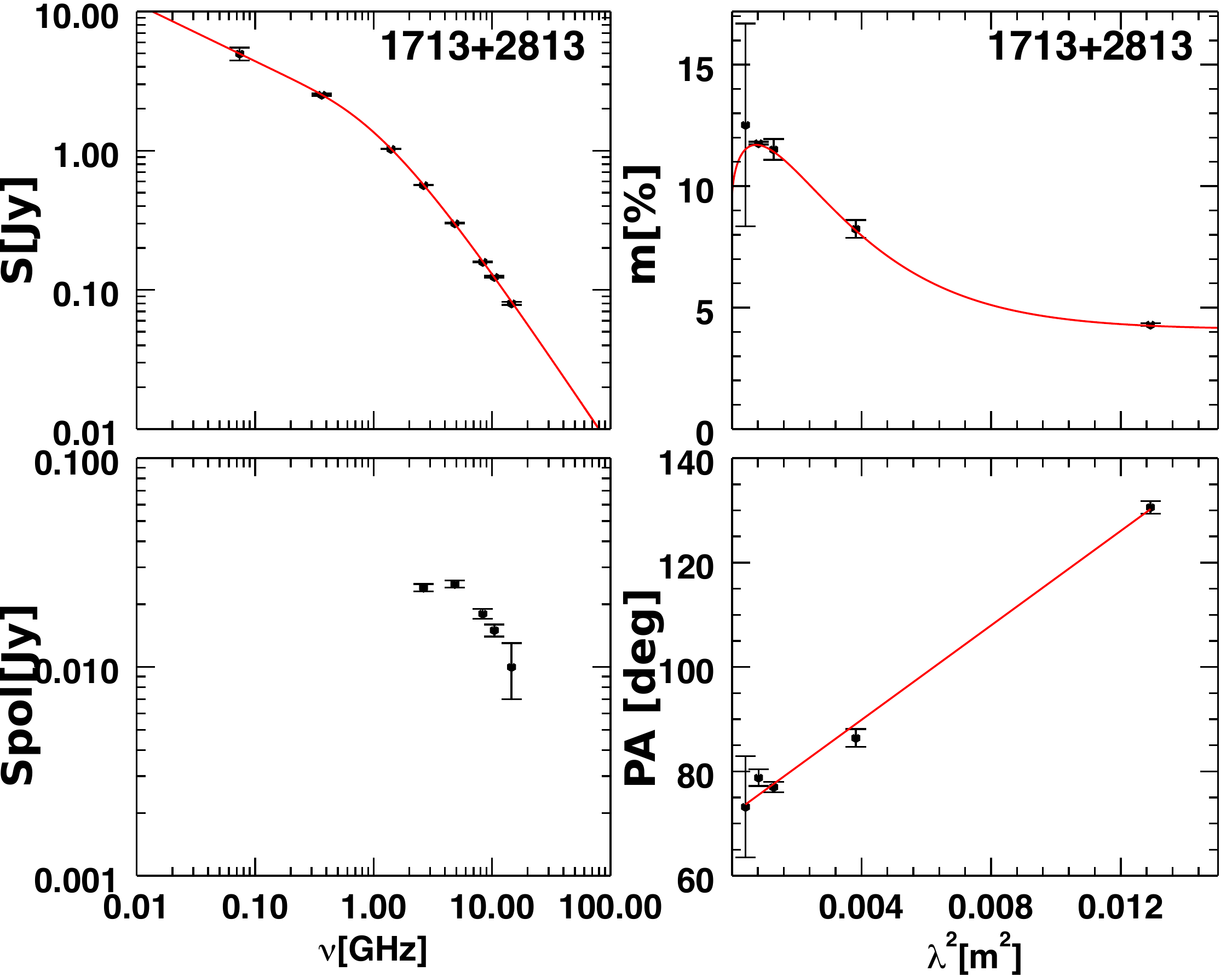}
\label{default}
\end{center}
\end{figure}

\begin{figure}[!h]
\begin{center}
\caption{Source 1723+3417, 2050+0407 and 2101+0341.  For each source we present their SED $\textit{S}$[Jy], the polrization flux density $\textit{S$_{pol}$}$[Jy], the fractional polarization $\textit{m}$[\%] and the polarization angle PA[rad]. }
\includegraphics[width=0.49\textwidth]{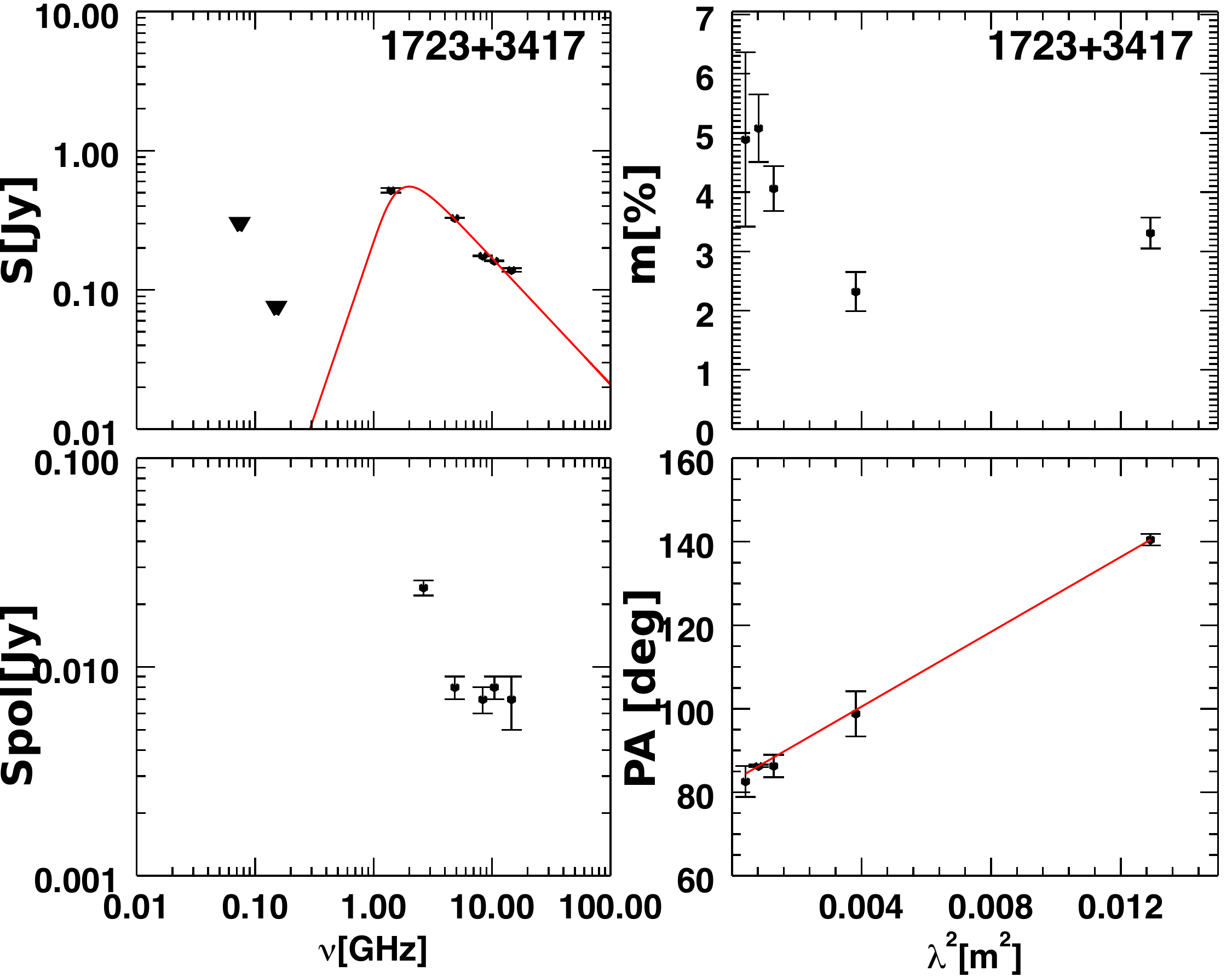}
\includegraphics[width=0.49\textwidth]{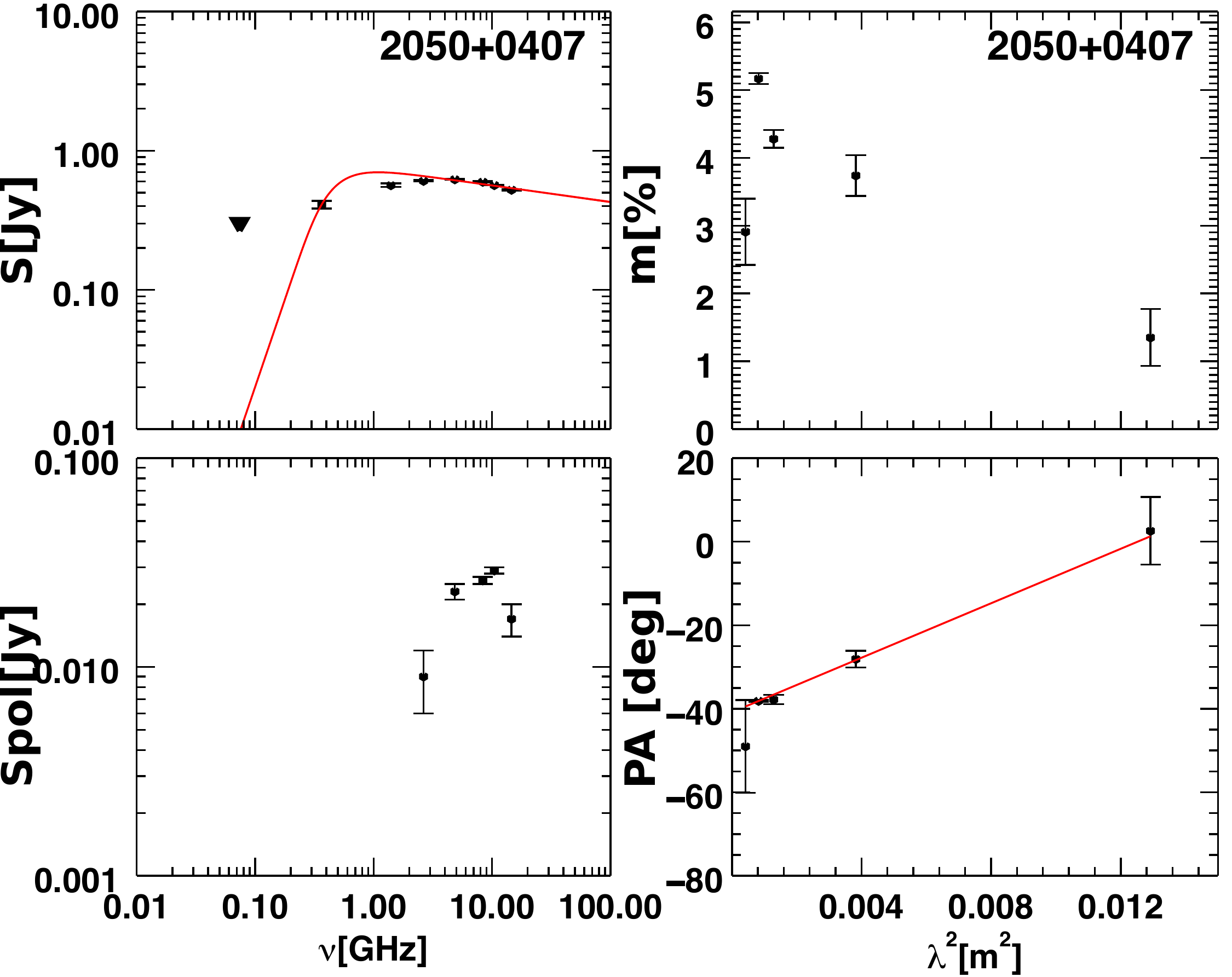}
\includegraphics[width=0.49\textwidth]{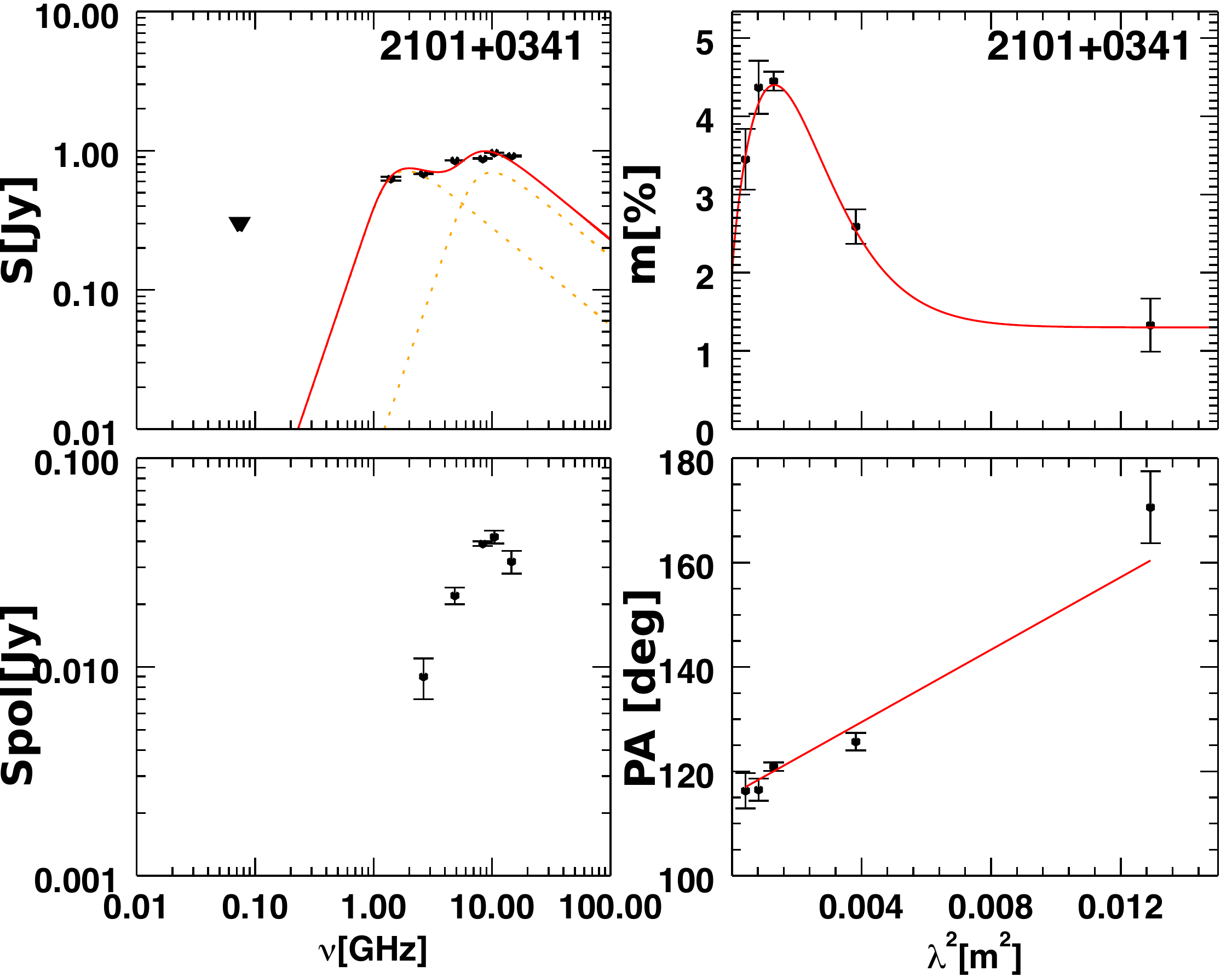}
\label{default}
\end{center}
\end{figure}

\begin{figure}[!h]
\begin{center}
\caption{Source 2147+0929, 2200+0708 and 2245+0324.  For each source we present their SED $\textit{S}$[Jy], the polrization flux density $\textit{S$_{pol}$}$[Jy], the fractional polarization $\textit{m}$[\%] and the polarization angle PA[rad]. }
\includegraphics[width=0.49\textwidth]{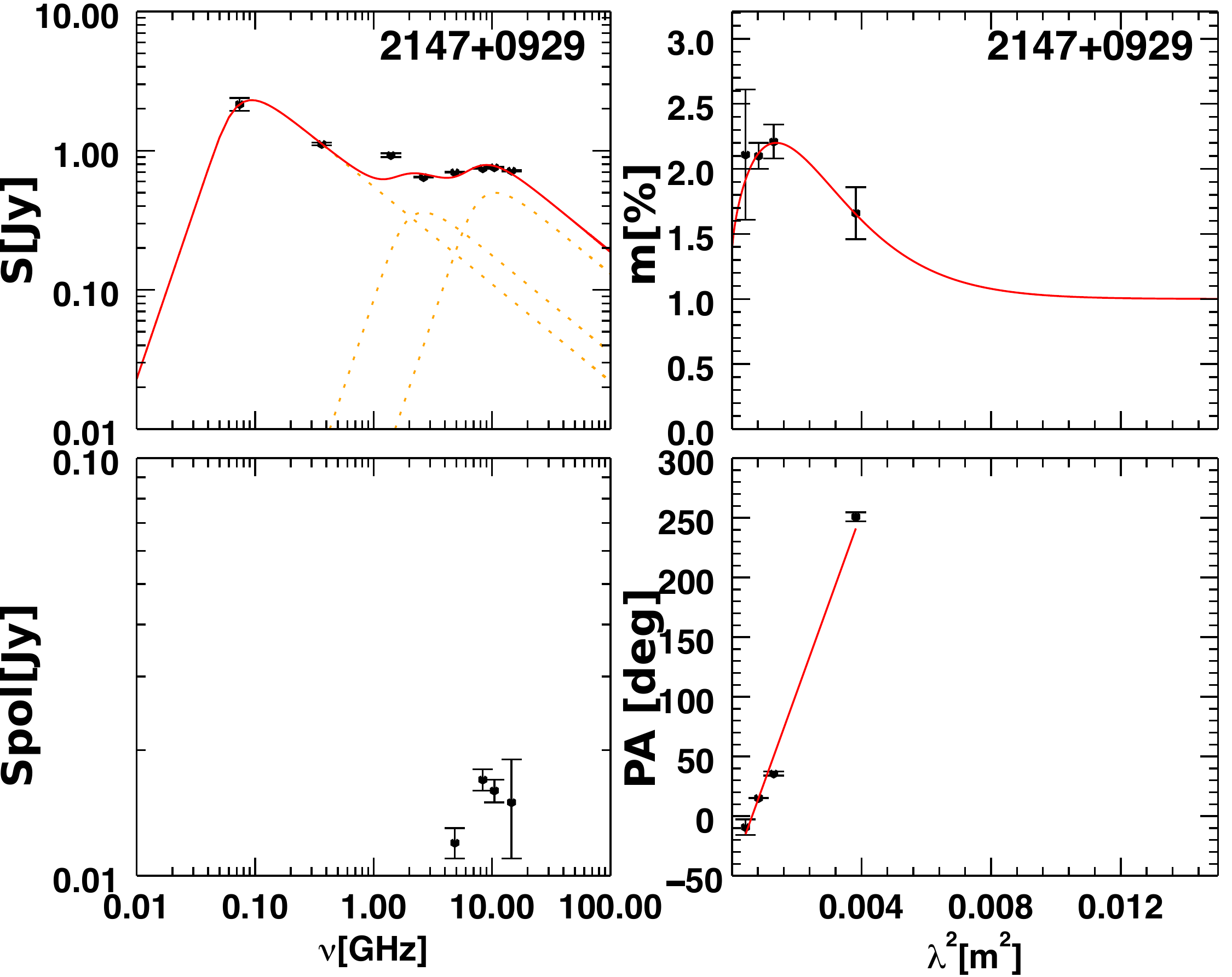}
\includegraphics[width=0.49\textwidth]{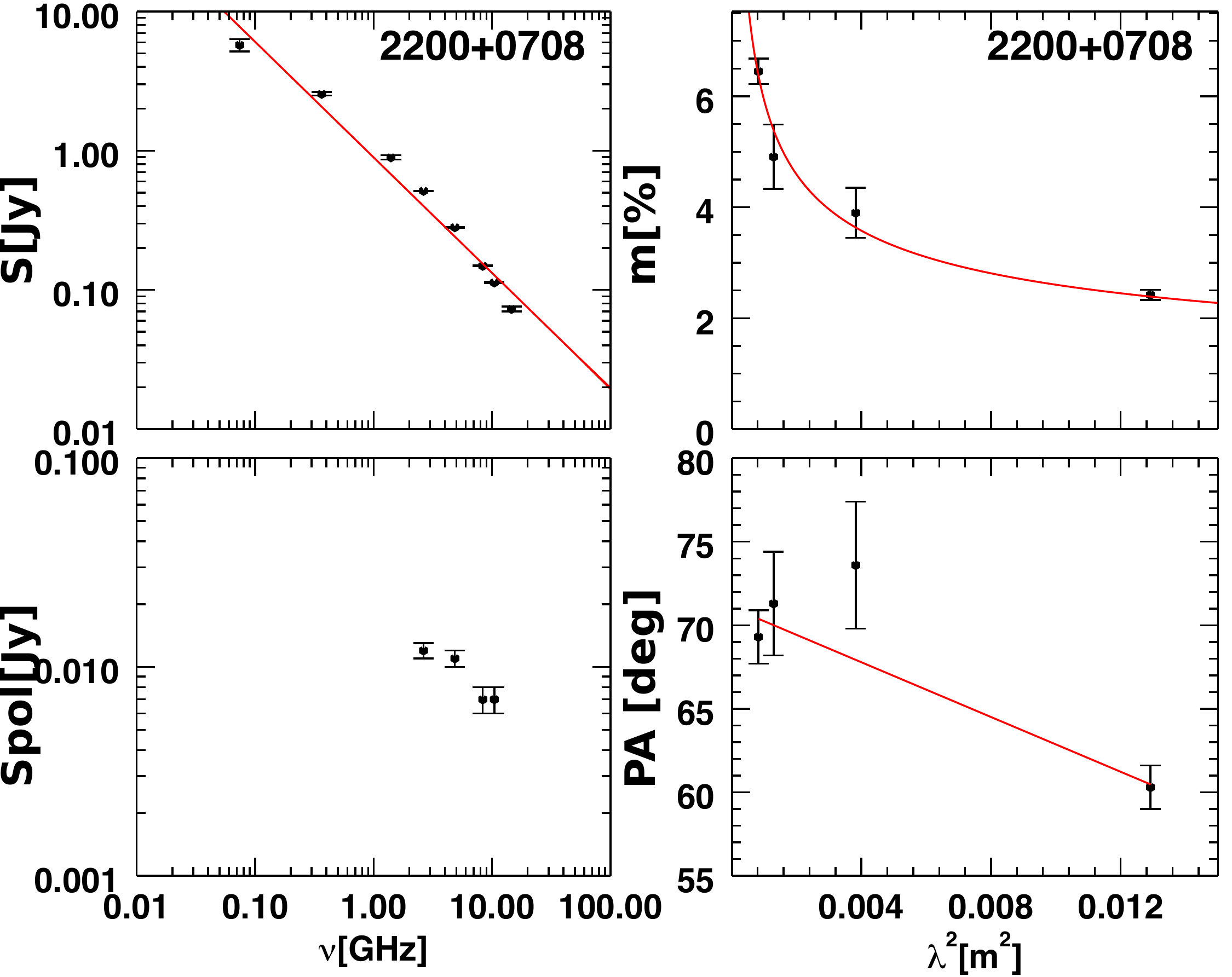}
\includegraphics[width=0.49\textwidth]{2245-eps-converted-to.pdf}
\label{default}
\end{center}
\end{figure}

\end{appendix}

\end{document}